\documentclass[aps,pre,amsmath,amsfonts,floatfix,superscriptaddress,nofootinbib]{revtex4}
\pdfoutput=1
\usepackage{mathrsfs} \usepackage{amssymb} \usepackage{graphicx,color} \usepackage{bbm} \usepackage{multirow}
\usepackage{bbm} \usepackage{mathrsfs} \usepackage{commath}\usepackage{tikz}
\usetikzlibrary{shapes,arrows,positioning,calc}\usepackage{stmaryrd}\usepackage{bm}

\font\msytw=msbm9 scaled\magstep1

\let\a=\alpha \let\b=\beta \let\g=\gamma \let\d=\delta
\let\e=\epsilon \let\z=\zeta \let\h=\eta 
\let\l=\lambda \let\m=\mu \let\n=\nu  \let\p=\pi
\let\s=\sigma \let\t=\tau \let\f=\varphi \let\c=\chi
   \let\G=\Gamma
\let\D=\Delta \let\Th=\Theta  
\let\Si=\Sigma   
\let\ee=\varepsilon \let\r=\rho \let\th=\theta \let\io=\infty

\def\ie{{i.e. }}\def\eg{{e.g. }}

\def\PP{{\cal P}} \def\VV{{\cal V}}
\def\CC{{\cal C}}\def\FF{{\cal F}} \def\HH{{\cal H}}
 
\def\RR{{\cal R}}\def\LL{{\cal L}}  \def\OO{{\cal O}}
\def\DD{{\cal D}}\def\GG{{\cal G}} \def\SS{{\cal S}}
\def\KK{{\cal K}}  
\def\ZZ{{\cal Z}}

\def\Re{{\rm Re}\,}

  \def\erf{\text{erf}}

\def\redv{\bar v}
\def\Dl{\D_{\rm liq}}
\def\sl{s_{\rm liq}}
\def\bth{\bar\th}

\def\to{\rightarrow} \def\la{\left\langle} \def\ra{\right\rangle}

\def\RRR{\hbox{\msytw R}}
\def\SSS{\hbox{\msytw S}}

\newcommand{\beq}{\begin{equation}} \newcommand{\eeq}{\end{equation}}
\newcommand{\wh}{\widehat} 
\newcommand{\Tr}{\text{Tr}}

\newcommand{\afunc}[1]{\operatorname{\mathsf{#1}}}
\def\DE{\afunc{D}}
\def\De{\mathrm D}
\def\de{\mathrm d}

\begin{document}

\title{
Statics and dynamics of infinite-dimensional liquids and glasses: \\
a parallel and compact derivation
} 

\author{Jorge Kurchan}
\affiliation{LPS,
\'Ecole Normale Sup\'erieure, UMR 8550 CNRS, 24 rue Lhomond, 75005 Paris, France}

\author{Thibaud Maimbourg}
\affiliation{LPT,
\'Ecole Normale Sup\'erieure, UMR 8549 CNRS, 24 rue Lhomond, 75005 Paris, France}

\author{Francesco Zamponi}
\affiliation{LPT,
\'Ecole Normale Sup\'erieure, UMR 8549 CNRS, 24 rue Lhomond, 75005 Paris, France}

\begin{abstract}
We provide a compact derivation of the static and dynamic equations for infinite-dimensional
particle systems in the liquid and glass phases. 
The static derivation is based on the introduction of an ``auxiliary'' disorder and the use of the replica method. The 
dynamic derivation is based on the general analogy between replicas and the supersymmetric formulation of
dynamics. We show that static and dynamic results are consistent, and follow the
Random First Order Transition scenario of mean field disordered glassy systems.
\end{abstract}

\maketitle

\tableofcontents

\clearpage

\section{Introduction}\label{sec:intro}

A major step in the understanding of glasses has been the ``Random First Order Transition'' (RFOT)
scenario~\cite{KW87,KT87,KW87b,KT87b,KT88,KT89,KTW89} (see~\cite{LW07,WL12,KT15} for recent reviews). 
It may be described in short as the 
assumption that a class of solvable disordered models -- most notably the spin-glass with $p$-spin interactions, $p>2$ -- are the 
mean-field representation (or metaphor) of fragile glasses. This development hinged on two main observations: the existence
in these models  of  a thermodynamic transition  essentially identical in nature to the Kauzmann ``entropy crisis'' scenario, 
and of a dynamic (pseudo) transition 
governed by the Mode-Coupling equations. That these two features, both of which had been  proposed previously,
 appear generically and with no parameter tuning in a vast family
of models, was considered a remarkable unifying fact.  
Several successes followed, including the understanding of the out of equilibrium (aging) dynamics~\cite{CK93}, 
the natural appearance of effective temperatures~\cite{BCKM97}, 
and the glassy rheology with generic shear-thinning properties~\cite{BBK00}.

In spite of these successes, a weak point remained which led to some skepticism: because there was no microscopic derivation starting from a particle system
-- at least none without uncontrolled approximations~\cite{BGS84,SSW85,KW87,KT89,MP96,CFP98,MP99,MP09,Go09,BJZ11}, 
the scenario was mostly an analogy
between systems with superficially little in common. 
This situation has changed 
thanks to the solution of  system of particles in the limit of large dimensions $d$,
an often used  remedy  to  the  absence  of  a  small  parameter \cite{1N,KW87,FRW85}.  
The entire RFOT scenario is recovered in this limit~\cite{PZ10,KPZ12,KPUZ13,CKPUZ13,RUYZ14,nature,MKZ15}, but now as 
an inevitable consequence of the exact solution, and not as a postulated analogy. 
The extra element that somehow closes the circle is that, in addition, there are simulations of particle systems
in dimensions from three to twelve~\cite{CIPZ11,CIPZ12,CCPZ12,CCPZ15}, where one can see the features extrapolating smoothly to large $d$, 
{\em and also best appreciate the limitations of the approach.}

The purpose of this paper is to present a parallel derivation of the thermodynamics and the dynamics of systems of particles
interacting through spherical potentials in dimension $d \rightarrow \infty$, simpler than the previously published ones~\cite{KPZ12,KPUZ13,CKPUZ13,RUYZ14,nature,MKZ15}.
The basic degrees of freedom are $N$ point particles of typical size $\sigma$ confined in a ``box''.  For our purposes, 
the simplest way to do this 
is to place each particle on the surface of 
a $(d+1)$-dimensional hypersphere
of radius $R\gg\sigma$, 
$x_i \in \RRR^{d+1}$ with  $x_i^2 = R^2$. The thermodynamic limit corresponds to $R \to\io$ with constant density.
With this choice rotational and translational invariances can be handled together. Moreover, the long-time limit
of the particles' mean-square displacement in the liquid phase can be computed easily.

The  ``thermodynamics'' we are aiming at is a partial one:
we explicitly exclude, as we shall discuss below, crystalline states. These certainly dominate the equilibrium measure of the condensed phase in
three dimensions, and the same might be the case in $d \rightarrow \infty$, see~\cite{TS10} for a discussion.
The reason why it is at all possible to separate amorphous from crystalline configurations is that it is expected, and it has to
be shown self-consistently, that these regions of phase space are strictly disconnected in the limit $d \rightarrow \infty$. 
In finite dimensions, the separation 
is not perfect and ultimately depends on the dynamic regime under study: the consensus is, however, that in glassy regimes
the formation of crystallites may usually be neglected, especially when $d>3$~\cite{SDST06,VCFC09}.
The study of dynamics poses no such problems of principle, as one may start from any chosen configuration and let the system decide of its
own evolution.
 
\begin{figure}
\begin{tabular}{cc}
 \includegraphics[width=7cm]{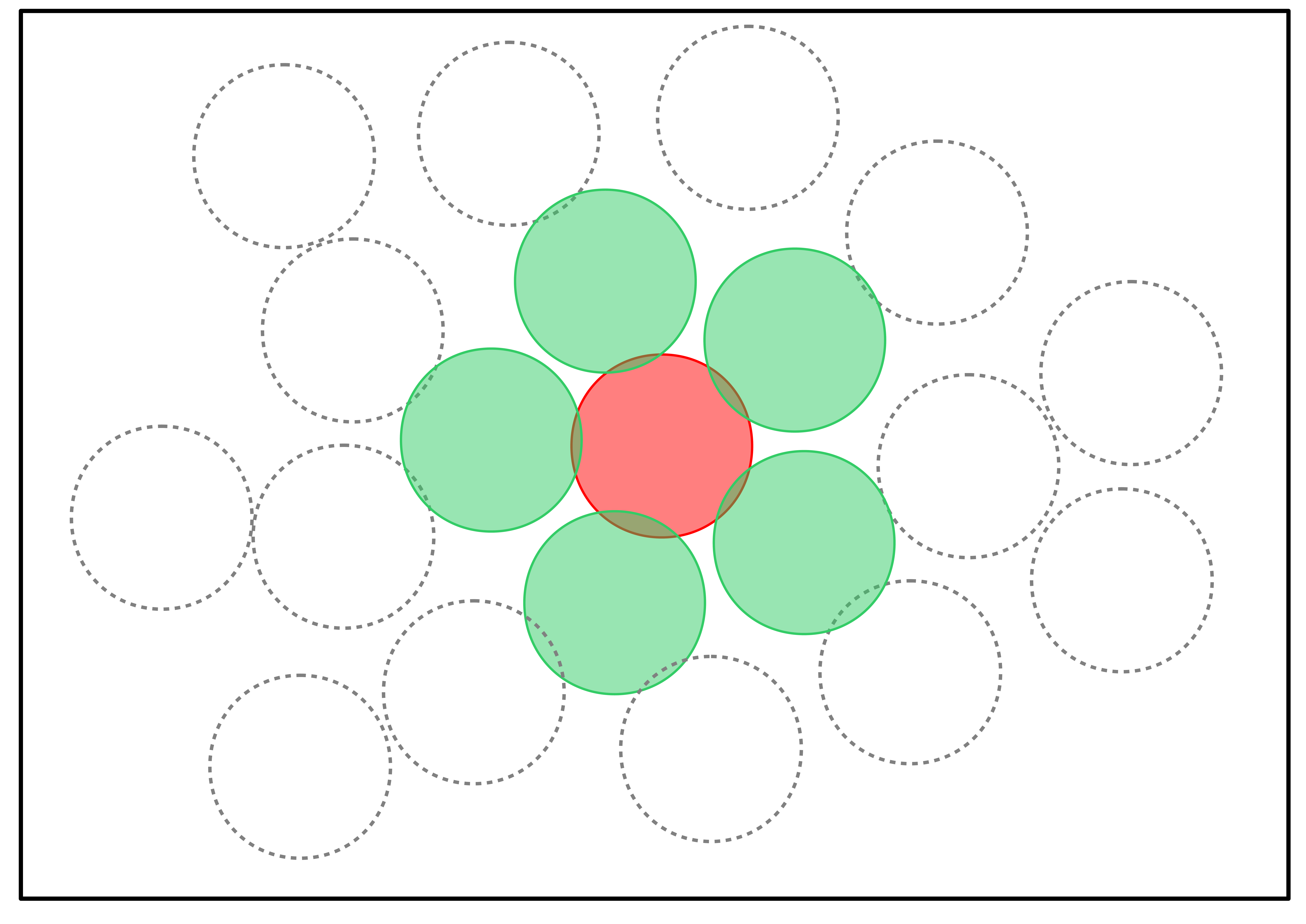} &
 \includegraphics[width=7cm]{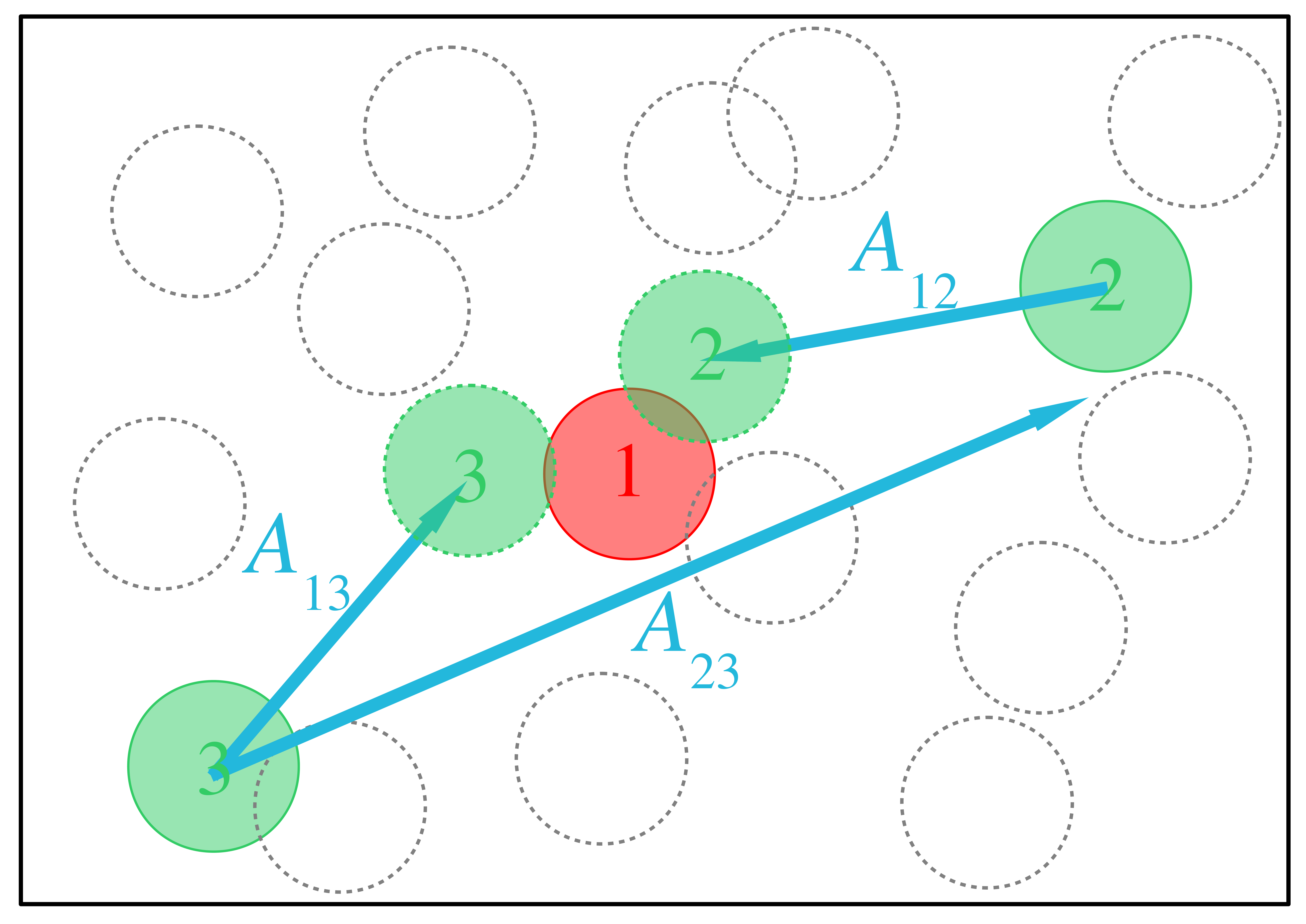} \\
 (a)&(b)
\end{tabular}
\caption{(a) Soft-Sphere system in $d\to\io$ dimensions (original model \ie $R\to\io$ limit for the spherical setting, and in absence random rotations $\RR_{ij}$). The red particle interacts with its green neighbours, which do not ``see'' each 
other (the most likely configuration is that they all are in orthogonal directions). The others (dashed gray) particles do not interact with the red one and again have a tree-like structure of contacts. (b) In the MK 
model in $d$ dimensions ($R\to\io$), random rotations become random shifts $A_{ij}$~\cite{MK11}. The red particle 1 interacts only with the green ones 2 and 3 owing to the shifts, the others (dashed gray) do not interact
with particle 1. 2 and 3 may interact as well if $|A_{12}+A_{13}+A_{23}|\sim \s$, which is very unlikely for high $d$ or for shifts that are of the
order of the linear size of the ``box''.}
\label{fig:MKHS}
\end{figure}

The reason why one expects the limit $d \rightarrow \infty$ to become simple was pointed out by Frisch et al.~\cite{FRW85,WRF87,FP99,EF88} 
thirty years ago, and it is common to many other fields of physics, e.g. ferromagnetic systems~\cite{GY91} or strongly correlated electrons~\cite{GKKR96}. 
Consider a particle 1 interacting with, amongst others, two particles 2 and 3; what can we say about the interaction between 2 and 3?
In order that 2 and 3 interact, if the forces have finite range, we need 1,2,3 in ``contact'' (defined by the range of interaction) with each other, forming a closed chain \begin{tikzpicture}[baseline={([yshift=-.5ex]current bounding box.center)}] 
\draw[fill=white] (0:0) circle (0.2) node {2};\draw[fill=white] (60:0.4) circle (0.2) node {1};\draw[fill=white] (0:0.4) circle (0.2) node {3};\end{tikzpicture}. 
Now, the number of configurations for which 2-1-3 form an open chain \begin{tikzpicture}[baseline={([yshift=-.5ex]current bounding box.center)}] 
\draw[fill=white] (0:0) circle (0.2) node {2};\draw[fill=white] (30:0.4) circle (0.2) node {1};\draw[fill=white] (0:0.692) circle (0.2) node {3};\end{tikzpicture} in $d=\infty$ is 
overwhelmingly larger than those in which 2-1-3 close a chain: we conclude that in $d \rightarrow \infty$, 2 and 3 may be considered non-interacting, unless there is an underlying (very special) order, see~Fig.\ref{fig:MKHS}a.
This simple observation led to the exact solution of liquid equilibrium in $d \rightarrow \infty$~\cite{FRW85,WRF87,FP99}.
Using this idea, and adapting them to the glassy regime following the proposal of Kirkpatrick and Wolynes~\cite{KW87},
the exact solution was extended to liquid dynamics~\cite{MKZ15} and glass thermodynamics~\cite{PZ10,KPZ12,KPUZ13,CKPUZ13}, as we will 
describe here.
Note, however, that contrary to Ref.~\cite{EF88}, we will not make use of a collision expansion to solve the dynamics:
in fact, in the glassy regime there are multiple collisions between a particle and its neighbors, which makes a collision expansion
ineffective; also, we will not be restricted to hard spheres but will also consider soft potentials, for which the notion of collision is not
well defined.

There is another useful way of imposing a similar situation, valid in any dimension. Consider a system with interaction potential
 \beq
H_{\rm normal} = \sum_{i<j} v(|x_i -  x_j |) \ ,
\label{original}
\eeq
 in any dimension. We will refer to this as the ``normal'' system.
Replace it now by a different system, this one with ``randomly shifted'' interactions
\beq
H_{\rm MK} = \sum_{i<j} v(|x_i - \RR_{ij} x_j |) \ ,
\label{MK}
\eeq
where $\RR_{ij}$ is a ``shift'' (in our case a rotation on the $d+1$-dimensional hypersphere) chosen randomly for each pair of particles once and for all.
We refer to this as the ``MK'' system, because it was studied extensively in~\cite{MK11}.
 If the size of these
shifts is large\footnote{Or if the spatial dimension is large, see Appendix~\ref{app:MKHS}.}, for example of the order of the box itself, then we may repeat the argument above and conclude that the chances that particles $2$ and
$3$, both interacting with $1$, interact between themselves are negligible~\cite{MK11}, see~Fig.\ref{fig:MKHS}b.
This model has been studied in finite dimensions, 
and it is definitely much closer to mean-field behaviour than the normal system, although
it is not clear what is the exact nature of its
glass transition, if even there is a sharp one:
we will not discuss this issue here and we refer the reader to~\cite{MK11,CCJPZ13} for further details.

A point which is clear is that the limit $d \rightarrow \infty$ of the MK model (\ref{MK}) and the normal model (\ref{original}) 
should coincide exactly (see Appendix~\ref{app:MKHS}) in all disordered phases --
liquid and glass -- but not in a possible crystalline phase, which would be suppressed in (\ref{MK}).
In this paper we use this as a trick to simplify the derivation of the equations for the dynamics and glassy thermodynamics
of (\ref{original}): the introduction of a disorder that is {\em a posteriori} irrelevant helps to justify and simplify the derivation.
The same technique in various forms has been used in glassy systems (see~\cite{Mo95} and references therein).
Note that here we focus only on the derivation of the equations, rather than on the extraction of physical results from them, which
have already been discussed in other papers~\cite{PZ10,KPZ12,KPUZ13,CKPUZ13,nature,RUYZ14,MKZ15}.

The structure of the paper is the following.
In Section~\ref{sec:II}, we recapitulate the basic definitions, and we derive the free energy as a functional of the single-particle
density. We show that this functional contains only the ideal-gas term plus a mean-field density-density interaction. In Section~\ref{sec:III},
we show that in the limit $d\to\io$, thanks to rotational invariance, one can evaluate all integrals involving
the single-particle density through a saddle point. We thus obtain the free energy as a simple function of a matrix $\hat\D$ 
that encodes the mean square displacement
between different replicas in the thermodynamics or, by an analogy, different times in the dynamics. In Section~\ref{sec:hierarchical}, we consider a special choice of $\hat\D$, corresponding to
Parisi's hierarchical ansatz, and we show that in this case we reproduce previous results for glassy thermodynamics~\cite{PZ10,KPZ12,KPUZ13,CKPUZ13,nature}.
In Section~\ref{sec:V} we write a general equation for the matrix $\hat\D$, without assuming that it is a hierarchical matrix, and we show that this
equation has the form of a Mode-Coupling equation~\cite{Go09}, controlled by a memory kernel for which we give a microscopic expression in terms of force-force
or stress-stress correlations. 
In Section~\ref{sec:dynamics} we exploit the general analogy between the supersymmetric formulation of Langevin dynamics
and the replica method to obtain dynamical equations for the model, thus reproducing the results of~\cite{MKZ15}. 
In Section~\ref{sec:VII} we show that (glassy) thermodynamics and (glassy) dynamics
give consistent results, in compliance with the general Random First Order Transition picture. Finally, we draw our conclusions.
For the convenience of the reader, in Appendix~\ref{app:formulae} we provide a list of the most recurrent mathematical definitions and notations.
The other Appendices contain some details of the calculations that are omitted from the main text.

\section{Setting of the problem}
\label{sec:II}

\subsection{Definition of the model}
\label{sec:definition}

Let us recapitulate here the precise definition of the system we wish to investigate.

\begin{itemize}
\item
The basic degrees of freedom are $N$ point particles. 
Each particle lives on the surface of 
a $(d+1)$-dimensional hypersphere
of radius $R$ (which we call $\SSS$), 
hence its coordinate is a point $x_i \in \RRR^{d+1}$ with the constraint $x_i^2 = R^2$.
The volume of this space is $V = \text{vol}(\SSS) = \Omega_{d+1} R^d$.
This is just a very convenient choice of ``boundary conditions''
for particles enclosed in a finite volume $V$. In fact, for $R\to\io$ we recover a system
defined on a flat (Euclidean) and infinite space $\RRR^d$, while at finite $R$ the global
rotational invariance in $\RRR^{d+1}$ encode both the rotations and the translational symmetries of
the $d$-dimensional Euclidean problem.

\item 
We wish then to consider {\it first} the thermodynamic limit where $R\to\io$ with constant $N/V$, in which the model becomes equivalent to the
usual definition in a $d$-dimensional Euclidean periodic cubic volume, and {\it then} the limit $d\to\io$,
where the model is exactly solvable.

\item
Each particle pair interact through 
a potential $v(|x_i - \RR_{ij} x_j |)$, where  
$\RR x$ is a uniformly distributed random rotation of point $x$ on the sphere.
Here $|x| = \left( \sum_{\mu=1}^{d+1} x_\mu^2 \right)^{1/2}$ is the modulus of the vector $x$ in $\RRR^{d+1}$, or in other words
$|x-y|$ is the Euclidean distance\footnote{For a central potential we could have defined the model with the great-circle distance (geodesic distance) on $\mathbb{S}$ instead of the Euclidean one, however in the large $R$ limit these two coincides up to irrelevant $\OO(1/R^2)$ corrections. 
For convenience we choose to work with the Euclidean distance.} between points $x$ and $y$ in $\RRR^{d+1}$.
The total potential energy is thus (we drop the ``MK'' suffix for simplicity)
\beq
H = \sum_{i<j} v(|x_i - \RR_{ij} x_j |) \ .
\eeq
As discussed in the introduction, the random rotations $\RR_{ij}$ are introduced for pedagogical convenience. In the limit $d\to\io$, they become irrelevant,
and the model becomes equivalent to a standard model where all the $\RR_{ij}$ are equal to the identity. We discuss this in more details 
in Section~\ref{sec:RR} and in Appendix~\ref{app:MKHS}.

\item
In order to have a proper limit $d\to\io$, we 
consider a 
class of inter-particle potentials at temperature $T=1/\b$ (here $k_{\rm B}=1$), 
such that
\beq\label{eq:redv}
\underset{d\to\io}{\lim}\, v(r) = \redv(h) \ , \hskip20pt h = d (r - \s)/\s \ , \hskip20pt r = \s (1 + h/d) \ ,
\eeq
where $\redv(h)$ is a finite function of $h$, and $\s$ is a reference value for the particle size. The scaling $r = \s (1 + h/d)$ is physically related to the fact that interactions 
are dominated by neighbouring particles that are typically almost touching and whose positions are fluctuating with amplitude $O(1/d)$~\cite{KW87}. 
This means, together with 
the scaling of density given by the packing fraction $\varphi=O(d/2^d)$, as shown below, that a particle interacts with $O(d)$ neighbours, consistently with the mean-field behaviour~\cite{GY91}.
 Concrete examples are:
\begin{itemize}
\item Hard Spheres with $e^{-\b v(r)} = \th(r - \s) = \th(h) = e^{-\b \redv(h)}$.
\item Soft Harmonic Spheres with $v(r) = \e d^2 (r/\s - 1)^2 \th(\s - r) = \e h^2 \th(-h) = \redv (h)$.
\item Soft Spheres with $v(r) = \e (\s/r)^{\a d} \to \e e^{-\a h} = \redv (h)$.
\item Lennard-Jones with $v(r) = \e \left[ (\s/r)^{4 d} - (\s/r)^{2 d} \right] \to \e [e^{-4 h} - e^{-2 h} ] = \redv(h)$.
\end{itemize}
Note that this is the natural generalisation of potentials such as the Lennard-Jones one to $d>3$, because
in any case one has to impose $v(r) \ll r^{-d-1}$ at large $r$ to keep the interaction short ranged and a finite second virial coefficient.
In many cases we will specialize to the Hard Sphere potential for concreteness, but all the main results we obtain in the paper
apply to a generic potential.

\end{itemize}
The main definitions are summarized in Appendix~\ref{eqapp:liquid}.

\subsection{The role of random rotations}
\label{sec:RR}

Let us comment on the choice of introducing the random rotations $\RR_{ij}$ in the interaction potential. As we will see in the following,
there are a few reasons for that choice:
\begin{enumerate}
\item 
all contributions to the free energy of the system involving three particles or more vanish, both in the statics and in the dynamics;
\item the crystalline state cannot exist in presence of random rotations, so we can focus on the amorphous liquid and glass states;
\item the presence of quenched disorder allows one to treat the thermodynamic problem by introducing replicas in a straightforward 
way.
\end{enumerate}
The first result, when $d\to\io$, is also true in absence of the $\RR_{ij}$; the proof is based on the fact that as we will see in the following, the typical mean squared
displacement of particles is always of the order of $1/d$~\cite{KW87}. Therefore, particle trajectories form a ``cloud'' of typical size $1/d$ while the distance
between them is of order 1. One can then apply the arguments of Frisch et al.~\cite{FRW85,WRF87,FP99}: they used a virial expansion of the entropy in powers of
the liquid density, and showed that in $d\to\io$ only the first correction to the ideal gas, a two-particle virial term, survives (as if it were a Van der Waals gas).
Note also that the random shifts disappear from the two-particle virial term. 
The second result is not true in absence of random rotations: there might be a crystal state. However, as shown in~\cite{SDST06,VCFC09}, 
crystallization is strongly suppressed in $d>3$. Thus, the liquid and glass states are metastable but have an extremely large lifetime, that
is expected to diverge when $d\to\io$. Finally, concerning the third point, in absence of quenched disorder one can still use replicas
within the Monasson~\cite{Mo95} or Franz-Parisi~\cite{FP95} schemes to describe glassy states. For particle systems in $d\to\io$
this has been done in~\cite{PZ10,KPZ12} and \cite{RUYZ14}, respectively. This only require minor modifications of the replica scheme
(see Appendix~\ref{app:equivD} for a discussion),
and no modification to the dynamics.
We conclude that the presence of the random rotations $\RR_{ij}$
is irrelevant in $d\to\io$, and that the MK model with random rotations is equivalent to the normal model with $\RR_{ij}$ equal to the identity.
Additional details 
can be found in Appendix~\ref{app:MKHS}. 
The results presented in the following therefore hold also for a normal particle model without random rotations.

\subsection{Replicated partition function}\label{sub:repf}

For simplicity we focus in this section on the Hard Sphere potential, but the derivation can be easily extended
to a generic potential, as mentioned in Section~\ref{sec:definition}.
We denote by $\de\RR$ the uniform measure over rotations, and by an overbar the average over it, i.e. over all the random rotations.
To compute the average of the free energy over the 
random rotations, we apply the so-called replica trick by considering the $n$-times replicated partition function
and use the relation $\overline{\log Z} = \underset{n\to 0}{\lim}\, \partial_n \overline{Z^n}$~\cite{MPV87}.
We denote by $\bar x = (x^1,\cdots,x^n) \in \SSS^n$ the coordinate of a replicated particle, 
and by $\bar X = (X_1, \cdots, X_n )$ a full replicated configuration of the system.
Let us define
\beq\begin{split}
&\c(\bar x_i,\bar x_j) = \prod_{a=1}^n e^{-\b v(|x_i^a - x_j^a|)} = \prod_{a=1}^n \th( | x_i^a- x_j^a | - \s) \ , \\
&\bar \chi(\bar x,\bar y) =\int \de \RR \,   \chi(\bar x, \RR \bar y) 
= \int \de \RR  \, \prod_{a=1}^n \th( | x^a - \RR y^a | - \s)
\ .
\end{split}\eeq
We have
\beq\label{eq:olZ}
\begin{split}
\overline{Z^n} &= \overline{\int \de\bar X  \,\prod_{a=1}^n e^{-\b H[X_a]}} =
\int \de\bar X  \,  \overline{\prod_{i < j } \chi(\bar x_i,\RR_{ij}\bar x_j) } =
\int \de\bar X  \,  \prod_{i<j} \int \de\RR \, \chi(\bar x_i,\RR \bar x_j)  
=  \int \de\bar X \, e^{ \sum_{i<j} 
\log  \bar\chi(\bar x_i,\bar x_j)} \ ,
\end{split}\eeq
where we recall that the overline denotes the average over the $N(N-1)/2$ independent random rotations $\RR_{ij}$.
For an arbitrary point $x \in \SSS$, $\VV_d(\s)/V = \int \de\RR\, \th(\s - | x - \RR x |)$ is the fraction of volume excluded by a particle of radius $\s$
on $\SSS$.
Simple geometrical considerations allow one to bound the function $\bar \chi(\bar x,\bar y)$ from above and below.
In fact, the value of $\bar\chi(\bar x, \bar y)$ is obtained by taking the $n$ particles described by $\bar y$, rotating all of them by the same random rotation $\RR$, and computing the probability
that none of the rotated spheres overlap with the corresponding particle in $\bar x$.
Clearly the value of $\bar\chi$ is maximal when $\bar y = \RR_0 \bar x$ for some $\RR_0$, because in this case one minimizes the number of excluded rotations.
We can choose $\RR_0$ to be the identity, in such a
way that $\bar y = \bar x$, without loss of generality. In that case we have
\beq\label{eq:app77745}
\bar\chi(\bar x, \bar y) \leqslant \bar\chi(\bar x,\bar x) = \int \de\RR  \, \prod_{a=1}^n \th( | x^a- \RR x^a | - \s)= 
1-\frac{\VV_d(\s)}{V} \ .
\eeq
Similarly, the value of $\bar\chi$ is minimal if $\bar y$ is chosen in such a way that, for any rotation $\RR$, at most one of the particles in $\RR \bar y$ is in overlap with the corresponding particle of $\bar x$.
Indeed, in this way one maximizes the number of excluded rotations.
Using this we have
\beq
\bar\chi(\bar x,\bar y) \geqslant 1-n\frac{\VV_d(\s)}{V} \ ,
\eeq
because the integrand is 1 except in $n$ distinct regions where the rotation brings one of the $\bar y$ particles in overlap with one of the $\bar x$. 
For generic configurations we thus get
\beq
1-n\frac{\VV_d(\s)}{V} \leqslant \bar\chi(\bar x,\bar y) \leqslant  1-\frac{\VV_d(\s)}{V} \ .
\eeq
Hence, defining the Mayer function
\beq\label{eq:fxy}
f(\bar x,\bar y) = \chi(\bar x,\bar y)  - 1 
= -1 + \prod_{a=1}^n \th( | x^a-y^a | - \s) 
\ ,
\hskip40pt
\bar f(\bar x,\bar y) =\bar\chi(\bar x,\bar y) -1 =\int \de\RR  \,  f(\bar x, \RR \bar y) \ ,
\eeq
we deduce that $\bar f \propto \VV_d(\s) / V$ is small in the thermodynamic limit, and in Eq.~\eqref{eq:olZ} we
can expand 
$\log  \bar\chi(\bar x_i,\bar x_j) = \log [ 1 + \bar f(\bar x_i,\bar x_j) ] \sim \bar f(\bar x_i,\bar x_j)$.
Introducing the order parameter (density of replicated configurations)
\beq\label{eq:rhodef}
\r(\bar x) =  \frac1N \sum_i \d(\bar x-\bar x_i) \ ,
\eeq
we thus have
\beq
 \sum_{i<j} 
\log  \bar\chi(\bar x_i,\bar x_j) \sim  \sum_{i<j}  \bar f(\bar x_i,\bar x_j)
= \frac{N^2}2  \int \de\bar x \de\bar y \, \r(\bar x)\r(\bar y) 
 \bar f(\bar x,\bar y) - \frac{N}2 \int \de\bar x  \,\r(\bar x) \bar f(\bar x,\bar x) \ .
\eeq
Note that from Eq.~\eqref{eq:app77745} we have $\bar f(\bar x,\bar x) = -\VV_d(\s) / V$
and thus $-\frac{N}2 \int \de\bar x \, \r(\bar x) \bar f(\bar x,\bar x) = N \VV_d(\s)/(2V)$ is a constant.
In the following, we do not keep track explicitly of all the multiplicative constants in the partition function. We will fix this at the end 
of the computation in Section~\ref{sec:disting}, so this term will be dropped. Note also that the constant is finite in the thermodynamic limit and therefore
it is subdominant with respect to the extensive terms of the free energy.
Inserting a delta function for $\r(\bar x)$ in Eq.~\eqref{eq:olZ} and representing it
as a Fourier integral over $\wh\r(\bar x)$, we obtain
\beq\label{eq:app45223}
\begin{split}
\overline{ Z^n} & \propto \int \de\bar x_i \, \int_{\r,\wh \r}  e^{N\int \de\bar x \, \r(\bar x) \wh \r(\bar x)
- \sum_i \wh \r(\bar x_i) + \frac{N^2}{2}  \int \de\bar x \de\bar y \, \r(\bar x)\r(\bar y) 
 \bar f(\bar x,\bar y) - \frac{N}2 \int \de\bar x  \,\r(\bar x) \bar f(\bar x,\bar x)
 }  \\ &\propto
\int_{\r,\wh \r} e^{N \left\{\int \de\bar x  \,\r(\bar x) \wh \r(\bar x)
+ \log\int \de\bar x e^{-\wh\r(\bar x)} + \frac{N}{2}  \int \de\bar x \de\bar y \, \r(\bar x)\r(\bar y) 
\bar f(\bar x,\bar y) \right\} } = \int_{\r,\wh \r} e^{N \SS(\r,\wh\r)}
 \ .
\end{split}\eeq
The last integral 
can be evaluated by the saddle point method by
optimizing $\SS$, which represents the ``free entropy'' functional\footnote{ ``Free entropy'' is the logarithm of the canonical partition function. 
For Hard Spheres, it is the same as the entropy because the partition function is temperature-independent.} at fixed $\r,\wh\r$. 
We will simply refer to it as ``entropy'' in the following.
The saddle point equations for $\wh \r$ is
\beq\label{eq:SPRHOHAT} \begin{split}
\r(\bar x) & = \frac{e^{-\wh \r(\bar x)}}{\int \de\bar y\, e^{-\wh \r(\bar y)}} \ , \\ 
\end{split}\eeq
which is very simple and is compatible with the normalization of $\r(\bar x)$.
Note that the original integral over $\wh\r$ was on the imaginary axis, but the saddle-point lies on the real axis as shown explicitly by
Eq.~\eqref{eq:SPRHOHAT} because $\r$ must be real-valued.
We can use this equation and substitute it in the entropy, then we get:
\beq\begin{split}
\label{eq:repent}
\SS(\r) &= - \int \de\bar x \, \r(\bar x) \log \r(\bar x) + \frac{N}{2} \int \de\bar x \de\bar y \, \r(\bar x)\r(\bar y) 
\bar f(\bar x,\bar y) \\
 &= - \int \de\bar x \, \r(\bar x) \log \r(\bar x) + \frac{N}{2}  \int \de\bar x \de\bar y \, \r(\bar x)\r(\bar y) 
\int \de\RR  \,f(\bar x, \RR \bar y) \\
 &= - \int \de\bar x  \,\r(\bar x) \log \r(\bar x) + \frac{N}{2}  \int \de\RR \de\bar x \de\bar y  \,\r(\bar x)\r(\RR^{-1}\bar y) 
 f(\bar x, \bar y) \\
 &= - \int \de\bar x \, \r(\bar x) \log \r(\bar x) + \frac{N}{2}  \int \de\bar x \de\bar y \, \r(\bar x)\r(\bar y) 
f(\bar x, \bar y)  \ ,
\end{split}\eeq
where in the last step we assumed that $\r(\bar y)$ is rotationally invariant, hence $\r(\RR^{-1}\bar y) = \r(\bar y)$,
and
\beq\label{eq:logZn}
\frac1N \log \overline{Z^n} = \max_{\r} \SS(\r) + C_n \ ,
\eeq
where the additive constant $C_n$ comes from
the proportionality constant in Eq.~\eqref{eq:app45223}. We will see in next Section~\ref{sec:disting} that $C_n=0$.
Let us emphasize once again that, as discussed in Section~\ref{sec:RR}, 
the last line in Eq.~\eqref{eq:repent} holds also in absence of random shifts in the limit $d\to\io$.
In fact, it is the usual starting point of replica computations for hard spheres in large dimensions~\cite{PZ10,KPZ12}.
The derivation presented in this section has the
advantage that it does not require to introduce the virial expansion, so it is more compact.
Note that here we normalized $\r(\bar x)$ to $\int \de \bar x \r(\bar x) = 1$ while in previous works~\cite{PZ10,KPZ12}
the standard normalization of liquid theory, $\int \de \bar x \r(\bar x) = N$, was used.

\subsection{Liquid phase (and a problem with distinguishability)}
\label{sec:disting}

As a first check we derive the entropy in the liquid phase of Hard Spheres. This corresponds to having independent and uniformly distributed particles over the
sphere, so $\r(\bar x) = V^{-n}$. Then we have, neglecting the constant $C_n$,
\beq\label{eq:SSliq}
\begin{split}
\SS(\r) &= n \log V + \frac{N}2  \int \frac{\de\bar x}{V^n}\frac{\de\bar y}{V^n} \, f(\bar x,\bar y)  = n \log V + \frac{N}2 \left[ - 1 +  \left( \int \frac{\de x}V \frac{\de y}V  \, \th( | x-y | - \s) \right)^n \right] \\
&= n \log V + \frac{N}2 \left[ -1 + \left( 1 - \frac{\VV_d(\s)}V \right)^n \right]  
\sim  n \log V - \frac{N n}2 \frac{\VV_d(\s)}V = n \sl \ ,
\end{split}\eeq
where
\beq\label{eq:Sliq}
\sl = \log V - \frac{N \VV_d(\s)}{2V} = \log V - \frac{2^d \f}2 \ ,
\eeq
and $\f = N \VV_d(\s)/(2^d V)$ is the packing fraction in the large $R$ limit.
We therefore recover the desired result, that the replicated entropy is given by $n$ times the liquid entropy if
replicas are decorrelated~\cite{MPV87}. This also shows that the constant $C_n$ in Eq.~\eqref{eq:logZn} is equal to zero.

Note that for the liquid entropy 
we obtain almost the same results that has been obtained by Frisch and Percus~\cite{FP99} for standard
$d$-dimensional hard spheres when $d\to\io$, which is
\beq
\sl^{\rm HS} =1 -  \log(N/V) - \frac{2^d \f}2 \ .
\eeq
In fact, we have $\sl = \sl^{\rm HS} -1 + \log N$, hence 
\beq
Z_{\rm liq} \sim e^{N \sl} \sim Z^{\rm HS}_{\rm liq} e^{N \log N - N} \sim Z^{\rm HS}_{\rm liq} N! \ .
\eeq
This factor of $N!$ is due to the fact that in the MK model particles are distinguishable, while in the HS model they are not.
One could correct by dividing (artificially) the MK partition function by $N!$~\cite{MK11}, but in any case
this factor is irrelevant for dynamics: it only affects the location of the Kauzmann point~\cite{MK11}.

\subsection{Pair correlation function}

Note that Eq.~\eqref{eq:olZ} can be written as
\beq
\overline{Z^n} =  \overline{\int \de\bar X \,  e^{-\frac\b2 \sum_{a,i\neq j} v(|x_i^a - \RR_{ij} x_j^a |)  }}
= \overline{\int \de\bar X  \, e^{-\frac\b2 \int \de\bar x \de\bar y  \,\r^{(2)}(\bar x, \bar y) V(\bar x, \bar y)  }}
\eeq
where
\beq
\r^{(2)}(\bar x, \bar y) = \sum_{i \neq j} \d(\bar x - \bar x_i) \d(\bar y - \RR_{ij}\bar x_j) \ , 
\hskip20pt
V(\bar x,\bar y) = \sum_{a} v(|x^a - y^a |) \ .
\eeq
Therefore we obtain
\beq\label{eq:rho2}
\begin{split}
\r(\bar x,\bar y) &=
 \overline{\la \r^{(2)}(\bar x,\bar y) \ra} = 
 \overline{\frac{1}{Z^n} 
 \int \de\bar X \, e^{-\b \sum_{a} H[X_a]  }   \r^{(2)}(\bar x,\bar y)    } \\
 &\sim \frac{1}{\overline{Z^n}} 
 \overline{\int \de\bar X \, e^{-\b \sum_{a} H[X_a]  }   \r^{(2)}(\bar x,\bar y)    }
 = - 2 T \frac{\d \log \overline{Z^n}}{\d V(\bar x, \bar y)} = N^2 \r(\bar x)\r(\bar y) \c(\bar x,\bar y) \ .
\end{split}\eeq
The equality between the first and second lines in Eq.~\eqref{eq:rho2} holds for $n\to 0 $ in which we are interested eventually, 
because $Z^n \to 1$ in that limit.

From the knowledge of $\r(\bar x,\bar y)$ we can compute the averages of several interesting observables.
Consider for example an observable of the non-replicated system of the form
\beq
\OO = \sum_{i < j} \OO(x_i, \RR_{ij}x_j) \ .
\eeq
We have
\beq\begin{split}
\frac12 \int \de\bar x\de\bar y \, \OO(x^1,y^1) \r(\bar x,\bar y) &= \frac{1}{2\overline{Z^n}} 
 \overline{\int \de\bar X \,  e^{-\b \sum_{a} H[X_a]  } \sum_{i\neq j} \OO(x_i^1,\RR_{ij} x_j^1)
 } \\
& =\frac{1}{\overline{Z^n}} 
 \overline{Z^{n-1} \int \de X  \,e^{-\b H[X]  } \sum_{i < j} \OO(x_i,\RR_{ij} x_j)
 }
\underset{n\to 0}{\longrightarrow} \overline{\la \OO \ra} \ .
\end{split}\eeq
As for the free energy, the calculation presented in this section has been done for the MK model with random rotations
for simplicity; however it also holds for the normal potential without random rotations, as a result of keeping the lowest order virial expansion of two-point correlation functions~\cite{hansen,MH61}.

To conclude, let us write explicitly the result for the liquid phase where $\r(\bar x) = V^{-n}$, and specializing to a finite-ranged potential 
(of interaction range $\s$) for simplicity.
We have
\beq\begin{split}
\overline{\la \OO\ra} &= \frac12 \int \de\bar x\de\bar y\, \OO(x^1,y^1) N^2 V^{-2n} \c(\bar x,\bar y)
= \frac12 \left(\frac{N}{V}\right)^2 \int \de x \de y\, \OO(x,y) e^{-\b v(|x-y|) } \left( \frac1{V^2} \int \de x \de y\, e^{-\b v(|x-y|) } \right)^{n-1} \\
&= \frac{N^2}{2V^2} \int \de x \de y\, \OO(x,y) e^{-\b v(|x-y|) } O\left(\left( \frac{V - \VV_d(\s)}{V}  \right)^{n-1}\right)
\underset{V\to\io}{\longrightarrow} \frac{ N^2}{2V^2} \int \de x \de y\, \OO(x,y) e^{-\b v(|x-y|) } \ ,
\end{split}\eeq
which is the correct result and corresponds to a liquid pair correlation $g(r) = e^{-\b v(r)}$, which is the leading term in the virial
expansion~\cite{hansen, MH61} and thus gives the exact result concerning the original model in the limit $d\to\io$, as well as for the MK model in all dimensions~\cite{MK11}.

\subsection{Summary of the results}

Let us summarize the results obtained in this section:
\begin{enumerate}
\item
The free energy functional has a simple form, composed by
two terms, the ideal gas and a simple mean field density-density interaction:
\beq
\SS(\r) = \SS_{\rm IG}(\r) + \SS_{\rm int}(\r) = - \int \de\bar x  \,\r(\bar x) \log \r(\bar x) + \frac{N}{2}  \int \de\bar x \de\bar y \, \r(\bar x)\r(\bar y) 
f(\bar x, \bar y) \ ,
\hskip20pt
\frac1N \log \overline{Z^n} = \max_{\r} \SS(\r) \ .
\eeq
Here, for a generic potential, $\SS$ is given by $-\b$ times the free energy; it is also sometimes called ``free entropy'' (we nevertheless refer to it as ``entropy'' in the following).
\item
By definition of $\r(\bar x)$, Eq.~\eqref{eq:rhodef}, averages of one-particle quantities can be written as
\beq
\OO = \sum_i \OO(x_i)  \hskip20pt
\Rightarrow
\hskip20pt
 \overline{\la \OO \ra} = \int \de\bar x \, \r(\bar x) \OO(x^1) \ .
\eeq
\item
Two-particle quantities can be written as
\beq\label{eq:twoO}
\OO = \sum_{i<j} \OO(x_i, \RR_{ij} x_j)  \hskip20pt
\Rightarrow
\hskip20pt
 \overline{\la \OO \ra} = \frac12 \int \de\bar x \de\bar y  \,\r(\bar x,\bar y) \OO(x^1,y^1) \ ,
 \hskip20pt
 \r(\bar x,\bar y) = N^2 \r(\bar x)\r(\bar y) \chi(\bar x,\bar y) \ .
\eeq
Note that the random shifts in the definition of $\OO$ have to be included for the MK model, while they should not be included
for the normal particle system.
\end{enumerate}
Correlations involving more than two particles are factorized in terms of one- and two-particle correlations,
as discussed in~\cite{PZ10,MK11,YZ14}.

\section{Rotational invariance and large dimensional limit}
\label{sec:III}

In this section we show how to take into account rotational invariance in order to solve exactly the limit $d\to\io$.
Before proceeding, let us recall that we also wish to take the thermodynamic limit $R\to\io$. 
In some cases the order of the two limits is irrelevant, but when relevant, according to Section~\ref{sec:definition},
we should take the $R\to\io$ limit first. In other words, we should consider that $R/d$ is a large quantity.

In a few words, the strategy we will use in this section is the following.
Due to rotational invariance on the hypersphere, the density of replicated configurations
$\rho(\bar x)$ can only depend on the matrix of the scalar products $q_{ab} = x_a \cdot x_b$,
or more physically, on the matrix of
mean square displacements
between replicas (recall that $q_{aa} = x_a^2 = R^2$):
\beq\label{eq:DEq}
\DE_{ab} = (x_a - x_b)^2 = 2 R^2 - 2 q_{ab} \ ,
\hskip20pt
q_{ab} = x_a \cdot x_b \ .
\eeq
These definitions are summarized in Appendix~\ref{eqapp:MSD}.
We can thus make a change of variables in the integration over $\de\bar x$
to $q_{ab}$ or $\DE_{ab}$, integrating out the irrelevant degrees of freedom.
We will see that, roughly speaking, the change of variables gives for density averages:
\beq
\int \de\bar x  \; \bullet \;  \rho(\bar x) \rightarrow \int \de \hat q \; \bullet \; e^{\frac{d}2 \log \det \hat q + d  \, \Omega(\hat q)}  \ ,
\eeq 
where the factor $e^{\frac{d}2 \log \det \hat q}$ is the Jacobian of the transformation, and one can show that
$\rho(\hat q)= e^{d \, \Omega(\hat q)}$ where $\Omega(\hat q)$ is finite\footnote{
This is shown explicitly in Eq.(65) of~\cite{KPZ12}, recalling that in the relevant regime $2^d\f = d \wh\f$ with finite $\wh\f$ and
that $\FF$, as defined in~\cite{KPZ12}, is a finite function.
}
for large $d$~\cite{KPZ12}. 
The appearance of the dimension in the exponent leads to a narrowing of fluctuations of correlations, when $d\to\io$, and saddle-point
evaluation becomes exact~\cite{Fy02,FS07,MV09}. 
In this way we will obtain an exact expression of $\SS(\r)$ in terms of the matrix $\hat \DE$.
In the rest of this section we will make these ideas mathematically precise.

\subsection{One-particle integrals: normalization of the density and ideal gas term}

As a preliminary remark, $V = \Omega_{d+1} R^d$ being the surface of the $d+1$-dimensional
hypersphere, we have:
\beq
\int_{\RRR^{d+1}} \de\bar x\, \prod_{a=1}^n \d(x_a^2 - R^2) = \left[ \Omega_{d+1} \int_0^\io \de r \, r^d \d(r^2 - R^2)  \right]^n
= \left[ \Omega_{d+1} \frac{R^d}{2R}  \right]^n
= \left[  \frac{V}{2R} \right]^n = \frac{1}{(2R)^n} \int_V \de\bar x \ ,
\eeq
and therefore, defining $D\bar x = \de\bar x \prod_{a=1}^n \d(x_a^2 - R^2)$, we have
\beq\label{eq:repent_RRR}
\SS(\r) = - (2R)^n \int_{\RRR^{d+1}} D \bar x\, \r(\bar x) \log \r(\bar x) + \frac{N}{2} (2R)^{2n} \int_{\RRR^{d+1}} D\bar x D\bar y\, \r(\bar x)\r(\bar y) 
 f(\bar x, \bar y)  \ .
\eeq
We now use rotational invariance to deduce that the density $\r(\bar x)$ must depend only on $q_{ab} = x_a \cdot x_b$, 
or equivalently on $\DE_{ab} = (x_a - x_b)^2$. 
For a rotationally invariant function we have (Appendix~\ref{app:A}):
\beq\label{eq:J}
\begin{split}
\int_{\RRR^{d+1}} D\bar x \, f(\bar x) & = 
C_{n+1}^{d+1} \int \prod_{a<b}^{1,n} \de q_{ab} \, (\det\hat q)^{\frac{d-n}2} f(\hat q) \\
&= 
C_{n+1}^{d+1} (-2)^{\frac{-n(n-1)}2} \int  \prod_{a<b}^{1,n} \de \DE_{ab} \, 
e^{\frac{d-n}2 \left[ \log\det(-\hat \DE/2) + \log(1-2 R^2 v^{\rm T} \hat \DE^{-1} v) \right] } f(\hat \DE) \ ,
\end{split}\eeq
where $v = (1 ,\cdots, 1)$, by definition $q_{aa} = R^2$ and $\DE_{aa}=0$, and
\beq\label{eq:C}
C_{n+1}^{d+1} = 2^{-n} \Omega_{d+1} \Omega_{d} \cdots \Omega_{d-n+2} \sim e^{\frac{d}2 n \log(2 \pi e/d) } \ .
\eeq

The density is normalized as
\beq\begin{split}
\label{eq:rhonorm}
1 = \int_V \de\bar x \,\r(\bar x) 
&= (2R)^n  C_{n+1}^{d+1} \int \prod_{a<b}^{1,n} \de q_{ab} \, (\det\hat q)^{\frac{d-n}2} \r(\hat q)  \\
&=(2R)^n C_{n+1}^{d+1} (-2)^{\frac{n(n-1)}2} \int  \prod_{a<b}^{1,n} \de \DE_{ab} \, 
e^{\frac{d-n}2 \left[ \log\det(-\hat \DE/2) + \log(1-2 R^2 v^{\rm T} \hat \DE^{-1} v) \right]} \r(\hat \DE) \ .
\end{split}\eeq
We write $\r(\hat \DE) = e^{d \, \Omega(\hat \DE)}$, and we take a saddle point in $d$ assuming that $\Omega$ is finite for large $d$. 
Taking the logarithm
of the Eq.~\eqref{eq:rhonorm}, at leading order for $d\to\io$, using Eq.~\eqref{eq:C}, we have
\beq\label{eq:34}
0 = \frac{d}2 n \log(2 \pi e/d) + \frac{d}2 \left[ \log\det(-\hat \DE_{\rm sp}/2) + \log(1-2 R^2 v^{\rm T} \hat \DE_{\rm sp}^{-1} v) \right] 
+ d \, \Omega(\hat \DE_{\rm sp}) \ .
\eeq
Note that Eq.~\eqref{eq:34} holds only for the matrix
$\hat\DE_{\rm sp}$ that maximizes the exponent in Eq.~\eqref{eq:rhonorm}, and not for generic values of $\hat\DE$: in other words,
Eq.~\eqref{eq:34} does not give the full shape of $\Omega(\hat \DE)$ but only its value at $\hat \DE = \hat \DE_{\rm sp}$.
For a rotationally invariant observable $\OO(\hat\DE)$ that is not exponential in $d$, the average over $\r(\bar x)$ is dominated
by the same value $\hat\DE_{\rm sp}$ and
we have
\beq\label{eq:avO}
\int_V \de\bar x\, \r(\bar x) \OO(\bar x) = \OO(\hat \DE_{\rm sp}) \ .
\eeq
In particular, the first term in Eq.~\eqref{eq:repent_RRR} (the ideal gas term) is the average of $\log \r(\bar x)$, which by hypothesis
is not exponential in $d$. Thus we can apply Eq.~\eqref{eq:avO} and we obtain
\beq\label{eq:SIG}
\SS_{\rm IG} = - \log \r(\hat \DE_{\rm sp}) =  - d \Omega(\hat \DE_{\rm sp}) = \frac{d}2 n \log(2 \pi e/d) + 
\frac{d}2  \left[ \log\det(-\hat \DE_{\rm sp}/2) + \log(1-2 R^2 v^{\rm T} \hat \DE_{\rm sp}^{-1} v) \right]
 \ .
\eeq

\subsection{Two-particle integrals: the interaction term}

For two-particle integrals, the exact calculation of the Jacobian of the change of variables is more difficult, so 
we will use a slightly different procedure where we compute the Jacobian by a saddle point in $d$.
This procedure is simpler but the price to pay is that we cannot keep track easily of all the normalization constants\footnote{This is why we
did not use this procedure for the ideal gas term, where the normalization is crucial.}.
We will compute the normalization constant only at the end, and for the moment all the proportionality factors will be neglected.

\subsubsection{Change of variables}

We consider a 
generic function $f$ that depends only on the distances between pairs of atoms in two replicas, $ | x_a - y_a |$, and
a two-particle integral of the form
\beq\label{eq:int1}
\begin{split}
 I_f &=\frac{N}{2}  \int_V \de\bar x \de\bar y\, \r(\bar x)\r(\bar y) 
f(\bar x, \bar y) \propto \int_{\RRR^{d+1}} D\bar x D\bar y\, \r(\bar x)\r(\bar y) 
 f(\{| x_a - y_a | \}) \\ &=
 \int \de\hat q^x \de\hat q^y \de\bar\omega\,
 \r(\hat q^x) \r(\hat q^y) f( \sqrt{\bar\omega} ) K(\hat q^x,\hat q^y,\bar\omega) \ .
\end{split}\eeq
Here $q^x_{ab} = x_a \cdot x_b$ and $q^y_{ab} = y_a \cdot y_b$ are symmetric matrices such that
$q^x_{aa} = q^y_{aa} = R^2$, hence $\de\hat q^{x,y} = \prod_{a<b} \de q^{x,y}_{ab}$, while $\bar\omega = (\omega_1, \cdots, \omega_n)$ with
$\omega_a = (x_a - y_a)^2$. With an abuse of notation we defined $\sqrt{\bar\omega} = (\sqrt{\omega_1}, \cdots ,\sqrt{\omega_n})$.
Therefore
\beq\label{eq:int2}
\begin{split}
K(\hat q^x,\hat q^y,\bar\omega) & = \int \de\bar x \de\bar y\, \prod_{a\leqslant b}^{1,n} \d(x_a \cdot x_b - q^x_{ab})\d(y_a \cdot y_b - q^y_{ab})
\prod_{a=1}^n \d(\omega_a - (x_a - y_a)^2 )  \\
& \propto \int \de\bar x \de\bar y \de\hat \l^x \de\hat \l^y \de\bar \mu \,
e^{ \sum_{ab}^{1,n} (\l^x_{ab} q^x_{ab} -  \l^x_{ab} x_a \cdot x_b +\l^y_{ab} q^y_{ab} -  \l^y_{ab} y_a \cdot y_b )
+ \sum_{a=1}^n (\mu_a \omega_a - \mu_a (x_a - y_a)^2 ) } \\
& \propto \int \de\hat \l^x \de\hat \l^y \de\bar\mu \, \exp\left\{\Tr( \hat\l^x \hat q^x + \hat\l^y \hat q^y ) +  \bar\mu^T \bar \omega - \frac{d}2 \log \det 
\begin{pmatrix}
\hat \l^x + \hat\mu & - \hat\mu \\
-\hat\mu  & \hat \l^y + \hat \mu 
\end{pmatrix}
\right\} \ ,
\end{split}\eeq
where $\bar\mu$ has been written for convenience as
a diagonal $n\times n$ matrix $\hat\m$ with $\mu_{ab} = \mu_a \d_{ab}$. In the above derivation we performed the following steps:
\begin{enumerate}
\item
In the second line, we introduced a Fourier representation of the delta functions by integrating over
$\hat \l^x, \hat \l^y, \bar\mu$. Note that because the delta functions are introduced for $a\leqslant b$, the matrix $\hat \l^x$ (and similarly $\hat\l^y$)
has as independent elements the ones for $a\leqslant b$ only. Correspondingly $\de\hat \l^{x} = \prod_{a\leqslant b} \de\l^{x}_{ab}$, differently from
the matrices~$\hat q^x$.
\item
The integrals over $\hat \l^x, \hat \l^y, \bar\mu$ should be done on the imaginary axis.
However, we are going to compute $K$ by a saddle point and
we anticipate that the saddle point is on the real axis, so we can equivalently treat them as real variables~\cite{CC05}.
Then the integral over $\bar x,\bar y$ is a simple Gaussian integral and gives the determinant term.
\end{enumerate}
In Eqs.~\eqref{eq:int1} and \eqref{eq:int2} there is clearly a symmetry $x \leftrightarrow y$ and it is very unreasonable that this symmetry is broken
at the saddle point. Hence we assume that $\hat q^x = \hat q^y = \hat q$ and $\hat\l^x = \hat\l^y = \hat\l$ at the saddle point. 
Using
\beq
\det 
\begin{pmatrix}
\hat \l + \hat\mu & - \hat\mu \\
-\hat\mu  & \hat \l + \hat \mu 
\end{pmatrix}
=\det (\hat \l) \, \det(\hat \l + 2 \hat  \mu) 
=\det (\hat \l)^2 \, \det(1 + 2 \hat \l^{-1} \hat  \mu) 
\ ,
\eeq
we have
\beq\label{eq:K}
K(\hat q,\hat q,\bar\omega) \propto \int \de\bar\mu \exp\left\{
2 \Tr ( \hat q \hat \l ) + \bar\m^T \bar\omega
- d \log \det\hat \l - \frac{d}2 \log \det(1 + 2 \hat \l^{-1} \hat  \mu) 
\right\} \ ,
\eeq
where $\hat \l$ (and $\bar\mu$, but we postpone its saddle-point evaluation) are determined by maximizing the exponent.
We now make a simplifying assumption\footnote{This assumption is not necessary~\cite{MKZ15} but it simplifies the derivation.} (to be checked {\it a posteriori}), 
namely that the last term in the expression above is not proportional to $d$ and therefore does not affect
the saddle point on $\hat\l$. Maximizing the exponent with respect to $\hat\l$ we obtain the relation
\beq
\hat q - \frac{d}2 \hat\l^{-1} =0  \  ,
\eeq
and therefore
\beq
K(\hat q,\hat q,\bar\omega) \propto  \int \de\bar\mu \exp\left\{
\bar\m^T \bar\omega
+ d \log \det\hat q - \frac{d}2 \log \det \left(1 + \frac{4}d \hat q \hat  \mu \right) 
\right\} \ ,
\eeq
and
\beq
I_f \propto 
\int \de\hat q \de\bar\omega \de\bar\mu \,
 \r(\hat q)^2   f(\sqrt{\bar\omega}  ) \,
 e^{ \bar\m^T \bar\omega
+ d \log \det\hat q - \frac{d}2 \log \det \left(1 + \frac{4}d \hat q \hat  \mu \right) } \ .
\eeq
Under the assumption that the last term is not exponential in $d$, $\hat q$ is determined
by maximizing
$ \left( \r(\hat q) e^{  \frac{d}2 \log \det\hat q } \right)^2$, which is exactly the same factor
that determines the saddle point value of $\hat q$ in the ideal gas term, see Eq.~\eqref{eq:rhonorm}.
We therefore assume from now on that $\hat q$ is equal to this saddle point value (without adding explicitly
the suffix ``sp'' to $\hat q$ for notational convenience). 
We know from Eq.~\eqref{eq:rhonorm} that at this saddle point value $  \r(\hat q) e^{  \frac{d}2 \log \det\hat q } $
is a constant. We obtain
\beq\label{eq:int3}
I_f \propto
\int  \de\bar\omega \de\bar\mu \,
 f(\sqrt{\bar\omega}  ) \,
 e^{ \bar\m^T \bar\omega
 - \frac{d}2 \log \det \left(1 + \frac{4}d \hat q \hat  \mu \right) } \ .
\eeq

\subsubsection{Scaling of the mean square displacement}

We now change variables by introducing the mean square displacement $\DE_{ab}$, Eq.~\eqref{eq:DEq},
with $q_{ab} = R^2 - \DE_{ab}/2$.
Our crucial assumption is that
$\DE_{ab}= \s^2 \D_{ab}/d$, with $\D_{ab}$ remaining finite for $d\to\io$.
This assumption is based on the scaling that has been already found in~\cite{PZ10,CKPUZ13,KW87}, and
we will check {\it a posteriori} that it is the only possible choice to obtain a meaningful scaling for $d\to\io$.
In matrix form we have
\beq
\hat q = R^2 v v^{\rm T} - \frac12 \hat \DE = R^2 v v^{\rm T} - \frac{\s^2}{2d} \hat \D \ .
\eeq
In Eq.~\eqref{eq:int3} we have
\beq\begin{split}
& \log \det \left(1 + \frac{4}d \hat q \hat  \mu \right) =\log \det \left(1 + \frac{4 R^2}d v v^{\rm T} \hat \mu - \frac{2\s^2}{d^2} \hat \D \hat \m \right) \\
&= \log \det \left(1 + \frac{4 R^2}d v v^{\rm T} \hat \mu \right)
+ \log \det \left[1 - \frac1{1 + \frac{4 R^2}d v v^{\rm T} \hat \mu} \frac{2\s^2}{d^2} \hat \D \hat \m \right] \ .
\end{split}\eeq
Using the cyclic properties of the trace we have
\beq
\log \det \left(1 + \frac{4 R^2}d v v^{\rm T} \hat \mu \right) =\Tr \log\left(1 + \frac{4 R^2}d v v^{\rm T} \hat \mu \right)
= \log \left(1 + \frac{4 R^2}d v^{\rm T} \hat \mu v \right) \ .
\eeq
Similarly, we have
\beq\begin{split}
 \frac1{1 + \frac{4 R^2}d v v^{\rm T} \hat \mu}  &= 1 + \sum_{n=1}^\io \left( -\frac{4 R^2}{d} \right)^n ( v v^{\rm T} \hat \mu )^n 
= 1 -\frac{4 R^2}{d} v v^{\rm T} \hat \mu \sum_{n=0}^\io \left( -\frac{4 R^2}{d} \right)^{n} ( v^{\rm T} \hat \mu v )^n
\\ &= 1 - \frac{4R^2/d}{  1+ (4 R^2/d) v^{\rm T} \hat \mu v}  v v^{\rm T}\hat \mu 
\to 1 - \frac{ v v^{\rm T}\hat \mu}{   v^{\rm T} \hat \mu v}  \ ,
\end{split}\eeq
where the last result holds for $R^2/d$ large. Using these results we obtain for large $R^2/d$
\beq
I_f \propto
\int  \de\bar\omega  \de\bar\mu \,
   f(  \sqrt{\bar\omega} ) \,
 e^{ \bar\m^T \bar\omega
 - \frac{d}2 
 \log \left(v^{\rm T} \hat \mu v \right)
 - \frac{d}2   \Tr \log \left[1 -  \left( 1 - \frac{ v v^{\rm T}\hat \mu}{   v^{\rm T} \hat \mu v} \right)
\frac{2\s^2}{d^2} \hat \D \hat \m \right] 
  } \ .
\eeq
The last term can be expanded for large $d$, we have
\beq
\frac{d}2   \Tr \log \left[1 -  \left( 1 - \frac{ v v^{\rm T}\hat \mu}{   v^{\rm T} \hat \mu v} \right)
\frac{2\s^2}{d^2} \hat \D \hat \m \right] \sim
- \frac{\s^2}{d}   \Tr \left[   \left( 1 - \frac{ v v^{\rm T}\hat \mu}{   v^{\rm T} \hat \mu v} \right) \hat \D \hat \m  \right]
=  \frac{\s^2}{d}   \frac{ \Tr (  v v^{\rm T}\hat \mu \hat \D \hat \m ) }{   v^{\rm T} \hat \mu v} 
= \frac{\s^2}{d}   \frac{ v^{\rm T}\hat \mu \hat \D \hat \m v}{   v^{\rm T} \hat \mu v}  \ ,
\eeq
where we used that $\Tr (\hat \D \hat \m ) = \sum_{a=1}^n \mu_a \D_{aa}=0$ because $\D_{aa}=0$.
Finally we obtain
\beq
I_f \propto \int  \de\bar\omega \, \de\bar\mu \,
  f( \sqrt{\bar\omega}  ) \,
 e^{ \sum_a \m_a \omega_a
 - \frac{d}2 
 \log \left(\sum_a \mu_a \right)
 -  \frac{\s^2}{d}   \frac{ \sum_{ab} \mu_a \D_{ab}  \m_b }{   \sum_a \mu_a}
  } \ .
\eeq
We now make a change of variable, $\omega_a = \s^2 (1 + 2 h_a/d)$ and $\mu_a = d g_a/\s^2$.
We have $\sqrt{\omega_a} \sim \s (1 + h_a/d)$
 and
we get
\beq\label{eq:shiftD}
\begin{split}
I_f & \propto 
- \int  \de\bar h \, \de\bar g \, f( \{ \s(1+h_a/d) \} )
   \,
 e^{ d\sum_a g_a + 2\sum_a h_a g_a
 - \frac{d}2 
 \log \left(\sum_a g_a \right)
 -   \frac{ \sum_{ab} g_a \D_{ab}  g_b }{   \sum_a g_a}
  }  \\
   & \propto - \int  \de\bar h \, \de\bar g \, \de x \, \de\l \,  f( \{ \s(1+h_a/d) \} )
      \,
 e^{ \l \left(x - \sum_a g_a\right) + 
  d\, x + 2\sum_a h_a g_a
 - \frac{d}2 
 \log x
 -   \frac{1}x \sum_{ab} g_a \D_{ab}  g_b 
  } \ ,
\end{split}\eeq
where we introduced a delta function of $x = \sum_a g_a$ through the integral representation (rotated on the real axis).
The integral over $x$ can be done via a saddle point because of the presence of a factor $d$ in front of the exponential.
For the saddle point over $x$ we can neglect the last term, and we obtain $1 = 1/(2x)$ hence $x=1/2$. 
The integral over $g_a$ is Gaussian, giving
\beq\label{eq:int4}
\begin{split}
I_f & = - \frac{ \CC}2 \int  \de\bar h \, \de\l \,  f( \{ \s(1+h_a/d) \} )
   \,
 e^{ \frac12 \l 
 +  \frac{1}2 \sum_{ab} (h_a - \l/2) (\D^{-1})_{ab}  (h_b - \l/2) 
  } \\
  &= -  \CC \int  \de\bar h \, \de\l \, f( \{ \s(1+ h_a/d + \l/d ) \} )   \,
 e^{  \l 
 +  \frac{1}2 \bar h^T \hat\D^{-1} \bar h
  } \ .
 \end{split}\eeq
Note that the crucial assumption made for the saddle point in Eq.~\eqref{eq:K} has now been checked self-consistently: the terms that were neglected are not exponential\footnote{Only multiplicative constants resulting from these terms are exponential in $d$. They are computed in the following in an easier way.} in $d$.

The proportionality constant $\CC$ does not depend upon the choice of $f$. Hence we can choose a test function\footnote{
Note that the choice of the test function is not completely arbitrary. In particular it should satisfy the properties of the
Mayer function $f$, that we used to derive the entropy, such as $\bar f \sim \VV_d(\s) / V$. Making a choice that does not respect these properties would lead to absurd results.
}
$f(\{ | x_a - y_a | \}) = \th(\s - |x_1 - y_1|)$. Recall that $\int \de\bar x \r(\bar x) =1$, and $\int \de x_2 \cdots \de x_m \r(\bar x) = 1/V$
because it must be a constant due to translational invariance. With the test function $f$ we obtain
\beq
I_f = 
\frac{N}{2}\int \de\bar x \de\bar y \r(\bar x)\r(\bar y) 
 \th(\s - |x_1 - y_1|) = \frac{N}{2V^2} \int \de x_1 dy_1 \th(\s - |x_1 - y_1|) = \frac{N \VV_d(\s)}{2V} 
 = \frac{2^d\f}2 \ .
\eeq
From Eq.~\eqref{eq:int4} we obtain instead (recalling that $\D_{11}=0$)
\beq
I_f 
= -  \CC \int  \de \bar h \, \de \l \, \th( - h_1 - \l )   \,
 e^{  \l 
 +  \frac{1}2 \bar h^T \hat\D^{-1} \bar h
  }  = -  \CC \int  \de \bar h \, 
   e^{ -h_1 
 +  \frac{1}2  \bar h^T \hat\D^{-1} \bar h
  }
  = - \CC (2\pi)^{n/2} \sqrt{\det(-\hat\D)} 
\eeq
Comparing these two expressions we obtain
$\CC = - \frac{2^d \f}{2} \frac{1}{(2\pi)^{n/2} \sqrt{\det(-\hat\D)}}$ which leads to the result:
\beq\label{eq:SSint_final} \begin{split}
I_f  &
  = \frac{2^d \f}{2}   \int  \DD_{-\hat\D} \bar h \, \de \l \, e^{  \l   }  f( \{ \s(1+ h_a/d + \l/d ) \} )   \ ,
  \hskip30pt
 \DD_{\hat\D} \bar h  = \de \bar h \frac{1}{(2\pi)^{n/2} \sqrt{\det(\hat\D)}} e^{ - \frac{1}2 \bar h^T \hat\D^{-1} \bar h} \ .
\end{split}\eeq
An important remark is that the measure $\DD_{-\hat\D} \bar h$ defined in Eq.~\eqref{eq:SSint_final}
cannot really be considered as a Gaussian measure over the $h_a$.
In fact, one has $\la h_a h_b \ra = - \D_{ab}$ which clearly makes no sense, because $\la h_a^2 \ra = -\D_{aa}=0$ which implies
that actually all the $h_a=0$. A related problem is that $\Tr \hat\D = \sum_a \D_{aa} =0$, hence $\hat\D$ has both positive and
negative eigenvalues, which makes the Gaussian integral ill-defined.
However, these problems can bypassed by considering $\DD_{-\hat\D} \bar h$ as an abstract measure, and the prescription to compute
integrals of functions of $h_a$ is that $\la h_a \ra=0$, $\la h_a h_b \ra = -\D_{ab}$, and higher moments are computed using the Wick rule
for Gaussian integrals. 

The problem can be fixed by a change of variables\footnote{
This discussion could have been hidden by introducing the shift $\hat\D \to \hat \D-A v v^{\rm T} $ directly in~\eqref{eq:shiftD}. 
However equations like~\eqref{eq:crucialrelation} and~\eqref{eq:FFdef6} will be needed in the following to simplify computations. 
The important point is that there exist well-chosen values of
$A$ that makes the expression of $\FF$ in Eq.~\eqref{eq:FFdef6} well-defined; Eq.~\eqref{eq:FFdef1} can be seen as an analytic continuation to $A=0$.
}. Let us define the function
\beq\label{eq:FFdef1}
\FF(\hat \D) =  - \int \DD_{-\hat\D} \bar h \, \de \l \,  e^{  \l   }  f( \{ \s(1+ h_a/d + \l/d ) \} )  \ .
\eeq
Here $\hat\D$ is (minus) the matrix of correlations of the Gaussian measure of the $h_a$.
Then, if we wish to compute $\FF(-A v v^{\rm T}  + \hat \D)$, we have $\la h_a h_b \ra = A - \D_{ab}$.
Equivalently, we can write $h_a = h'_a + H$, where
$h_a$ and $H$ are uncorrelated Gaussian variables 
with zero mean, such that $\la  H^2 \ra = A$ and $\la h'_a h'_b \ra = - \D_{ab}$.
We thus have
\beq\label{eq:crucialrelation}
\begin{split}
\FF(-A v v^{\rm T}  + \hat \D) &= - \int  \DD_{-\hat\D} \bar h' \, \de H \frac{e^{-\frac{H^2}{2 A}}}{\sqrt{2\pi A}} \,
\de\l \,  e^{  \l   }  f( \{ \s(1+ h'_a/d + H/d+  \l/d ) \} )  \\
&= - \int  \DD_{-\hat\D} \bar h \, \de H \frac{e^{-\frac{H^2}{2 A}}}{\sqrt{2\pi A}} \,
\de\l \,  e^{  \l - H  }  f( \{ \s(1+ h_a/d + \l/d ) \} )  \\
&= - e^{A/2} \int \DD_{-\hat\D} \bar h \, \de\l \,  e^{  \l  }  f( \{ \s(1+ h_a/d + \l/d ) \} )   = e^{A/2} \FF(\hat \D)
\ .
\end{split}\eeq
This shows that $\FF(\hat \D) = e^{-A/2} \FF(-A v v^{\rm T}  + \hat \D) $ for arbitrary $A$ and leads to
our final result:
\beq\label{eq:FFdef6}
\begin{split}
I_f &= -\frac{2^d \f}{2}  \FF(\hat \D) \ ,
\hskip20pt
\FF(\hat \D) = -e^{-A/2} \int  \DD_{A v v^{\rm T} - \hat\D} \bar h \, \de\l \,  e^{  \l   }  f( \{ \s(1+ h_a/d + \l/d ) \} ) \ . \\
\end{split}\eeq
We will see that $A$ can be chosen conveniently to have a well defined Gaussian measure, and 
simplify these expressions in concrete cases.

\subsubsection{Mayer function}

Let us now specialize to the case in which $f$ is the replicated Mayer function defined in Eq.~\eqref{eq:fxy}
and make contact with previous results~\cite{CKPUZ13}.
 Then using Eq.~\eqref{eq:redv} we get
\beq
f( \{ \s(1+ h_a/d + \l/d ) \} ) = - 1 + \prod_{a=1}^n e^{-\b v[ \s(1+ h_a/d + \l/d ) ]} = - 1 + \prod_{a=1}^n e^{-\b\redv( h_a + \l )} \ .
\eeq
and thus
\beq\label{eq:FFdef2}
\FF(\hat \D) = e^{-A/2} 
\int \de\l \, 
 e^{  \l   } \, \left\{
  1 -  \int \DD_{A v v^{\rm T} -\hat\D} \bar h \prod_{a=1}^n e^{-\b\redv( h_a + \l ) }
   \right\} 
 \ ,
\eeq
One can show that for any function $f(\{h_a\})$ the following relation holds\footnote{
The proof is obtained by performing, on the left hand side, 
a Taylor expansion of the function $f(\{h_a\})$ and using the Wick rule, while on the right hand side expanding 
the exponential. For example, at the lowest order, one obtains
\beq
\int \DD_{-\hat\D} \bar h \, f(\{h_a\})  = f(\{ 0 \}) - \frac12 \sum_{ab}^{1,n} \D_{ab} \frac{\partial^2 f}{\partial h_a\partial h_b}(\{ 0 \}) + \cdots = 
\left.
\exp\left(-\frac{1}{2}\sum_{ab}^{1,n} \Delta_{ab}\frac{\partial^2}{\partial h_a\partial h_b}\right) f(\{h_a\})  
 \right|_{\{h_a=0\}} \ .
\eeq
} (here we choose $A=0$ for simplicity):
\beq
\int \DD_{-\hat\D} \bar h \, f(\{h_a\})  =  
\left.
\exp\left(-\frac{1}{2}\sum_{ab}^{1,n} \Delta_{ab}\frac{\partial^2}{\partial h_a\partial h_b}\right) f(\{h_a\})  
 \right|_{\{h_a=0\}} \ .
\eeq
Using this we obtain
\beq\label{eq:FFpaperIII}
\begin{split}
\FF(\hat \D) &= 
\int \de\l \, 
 e^{  \l   } \, \left\{
  1 -  \left.
\exp\left(-\frac{1}{2}\sum_{ab}^{1,n} \Delta_{ab}\frac{\partial^2}{\partial h_a\partial h_b}\right) \prod_{a=1}^n e^{-\b\redv( h_a + \l ) }
 \right|_{\{h_a=0\}}
 \right\} \\
&= 
\int \de h \, 
 e^{  h   } \, \left\{
  1 -  \left.
\exp\left(-\frac{1}{2}\sum_{ab}^{1,n} \Delta_{ab}\frac{\partial^2}{\partial h_a\partial h_b}\right) \prod_{a=1}^n e^{-\b\redv( h_a ) }
 \right|_{\{h_a=h\}}
 \right\} \\
&= 
\int \de h \,e^{  h   }\frac{\de}{\de h} 
 \left\{\left.
\exp\left(-\frac{1}{2}\sum_{ab}^{1,n} \Delta_{ab}\frac{\partial^2}{\partial h_a\partial h_b}\right) \prod_{a=1}^n e^{-\b\redv( h_a ) }
 \right|_{\{h_a=h\}}
 \right\} 
 \ .
\end{split}\eeq
which coincides with the result obtained
in~\cite[Eq.~(15)]{CKPUZ13}
(the last line is obtained by an integration by parts).

\subsection{Summary of the results}

Let us summarize the main results of this section.
\begin{enumerate}
\item
We have shown that for a generic function $f(\{|x_a - y_a|\})$ that is not exponential in $d$ we have, from
Eq.~\eqref{eq:FFdef6}:
\beq\label{eq:Iffinal}
\begin{split}
I_f &=\frac{N}{2}  \int_V \de\bar x \de\bar y \r(\bar x)\r(\bar y) 
f(\bar x, \bar y)
= -\frac{2^d \f}{2}  \FF(\hat \D) \ , \\
\FF(\hat \D) &= -e^{-A/2} \int  \DD_{A v v^{\rm T} -\hat\D} \bar h \, \de \l \,  e^{  \l   }  f( \{ \s(1+ h_a/d + \l/d ) \} ) \ , \\
\end{split}\eeq
where $\DD_{A v v^{\rm T} -\hat\D} \bar h$ is a Gaussian measure 
with $\la h_a h_b \ra = A - \D_{ab}$, as
defined in Eq.~\eqref{eq:SSint_final},
and $A$ is an arbitrary constant. Here $\hat\D$ is the saddle point matrix defined in Eq.~\eqref{eq:34}.
Other equivalent expressions for $\FF(\hat\D)$, namely Eqs.~\eqref{eq:FFdef2}
and \eqref{eq:FFpaperIII}, have been derived in the special case in which $f$ is the replicated Mayer function
defined in Eq.~\eqref{eq:fxy}. Eq.~\eqref{eq:FFpaperIII} reproduces the previous result of~\cite{CKPUZ13}.
Using Eq.~\eqref{eq:twoO}, this result can be used to compute the averages of two-particle rotationally invariant observables.
\item
Our second result is an expression of the entropy in terms of the saddle-point scaled mean square displacement matrix 
$\hat \D = d \hat \DE/\s^2$ and the scaled density $\wh \f = 2^d \f/d$.
The ideal gas term is given by Eq.~\eqref{eq:SIG}.
For the interaction term we use Eq.~\eqref{eq:Iffinal}.
We obtain\footnote{It may seem that this expression of the entropy is ill-defined (imaginary) because the logarithms might be evaluated at negative values. Indeed, 
since $\Tr\hat\D=0$, $\hat\D$ has both positive and negative eigenvalues; thus, $\det(-\hat \D)$ and $1- \frac{2 d R^2}{\s^2} v^{\rm T} \hat \D^{-1} v $ might be negative. However, remember that in the 
end one wishes to take the limit $n\to0$. In this limit, these expressions are regularized: one can check from Appendix~\ref{app:hierarchical} that $\hat \D$ is negative definite.
Another option, which will be used in sections~\ref{sec:V} and~\ref{sec:dynamics}, is to express $\SS$ in terms of $\hat Q\equiv\Dl v v^{\rm T} -\hat\D$, which is positive definite.}
\beq\label{eq:SSfinal}
\SS(\hat \D) = \frac{d}2 n \log( \pi e \s^2/d^2) + 
\frac{d}2  \log\det(-\hat \D) + \frac{d}2 \log\left(1- \frac{2 d R^2}{\s^2} v^{\rm T} \hat \D^{-1} v \right) -  \frac{d}{2} \wh\f \FF(\hat\D)
 \ ,
\eeq 
where for $\FF(\hat\D)$ we have three expressions: Eqs.~\eqref{eq:FFdef2},~\eqref{eq:FFpaperIII} and~\eqref{eq:Iffinal}.

\item
The matrices $\hat\D$ or $\hat\DE$ should be determined by solving the $d\to\io$ saddle-point condition, i.e. by
maximizing the terms that are exponential in $d$ in Eq.~\eqref{eq:rhonorm}.
The problem is that
we have never derived explicitly the form of $\Omega(\hat\DE)$. However, one can show that $\hat\D$ can be equivalently determined by
maximizing the final result for the entropy, Eq.~\eqref{eq:SSfinal}, which is quite intuitive (a formal proof can be found in~\cite{KPZ12,KPUZ13,MKZ15}). Indeed, the thermodynamic limit saddle-point equation is $\d\SS/\d\rho=0$.
In infinite $d$, $\SS$ depends on $\rho$ only through the saddle-point value of $\Omega(\hat\Delta_{\rm sp})$, which is known in terms of $\hat \Delta_{\rm sp}$ via~\eqref{eq:34}. 
Therefore $\d\SS/\d\rho=0$ is equivalent to $\de\SS/\de\hat\D_{\rm sp}=\hat 0$ where we have expressed $\SS$ only in terms of the saddle-point value $\hat\Delta_{\rm sp}$ as in~\eqref{eq:SSfinal}. 
This condition fully determines the saddle-point value $\Delta_{\rm sp}$ and is thus equivalent to the $d\to\io$ saddle-point equation.

\end{enumerate}
Thanks to these results, 
we can express both the free energy and two-particle correlations in terms of the matrix $\hat\D$. Our next task
is to determine explicitly this matrix.

\section{Hierarchical matrices and replica symmetry breaking}
\label{sec:hierarchical}

In this section we show that Eq.~\eqref{eq:SSfinal} reproduces all the correct results in the different thermodynamic phases of the system,
where the matrix $\hat\D$ is a hierarchical matrix~\cite{MPV87}. 
We will focus on Hard Spheres for simplicity, and to make contact with previous results~\cite{KPZ12,KPUZ13,CKPUZ13}.
In particular we will show that

\begin{enumerate}

\item For a general form of the matrix $\D_{ab}$, Eq.~\eqref{eq:SSfinal} coincides with the result obtained in~\cite{KPZ12,KPUZ13,CKPUZ13}
through a quite different derivation. For the interaction term, this has been shown in Eq.~\eqref{eq:FFpaperIII}. For the ideal gas term,
this is shown in Appendix~\ref{app:equivD}.

\item In the liquid phase, we expect that all replicas are uncorrelated. Hence for $a\neq b$, $x_a \cdot x_b=0$ and 
$\D_{ab} = d (x_a - x_b)^2/\s^2 = 2 d R^2/\s^2 \equiv \Dl$.
Consistently we will show that in this phase
the matrix $\hat\D$ is replica symmetric (RS) with
$\D_{ab} =  \D_0 (1-\d_{ab})$ and $\D_0 = \Dl$. Furthermore, $\SS = n \sl$ as expected from Section~\ref{sec:disting}.
These results are discussed in Section~\ref{sec:liquid}.

\item In the glass phase, where $\hat\D$ is a hierarchical replica symmetry breaking (RSB) matrix~\cite{MPV87,Mo95,MP09,CC05}, 
from Eq.~\eqref{eq:SSfinal} we can derive the expression
of $s = \underset{n\to 0}{\lim}\, \SS/n$ and again we find the same results as in~\cite{CKPUZ13} for the 1RSB, 2RSB, $\cdots$, fullRSB cases.
For pedagogical reasons we first discuss 
the 1RSB computation (Section~\ref{sec:1RSB}) and then the general $k$RSB computation (Section~\ref{sec:fRSB}).

\end{enumerate}
Some useful mathematical properties of hierarchical RSB matrices are discussed in Appendix~\ref{app:hierarchical};
we will also use the notations for Gaussian integrals defined in Appendix~\ref{app:Gauss}.

\subsection{Liquid (replica symmetric) phase}
\label{sec:liquid}

The liquid phase is described by a replica symmetric matrix $\hat \D = \D_0 (v v^{\rm T}- I)$.
As an example for $n=3$ we have
\begin{equation}
\label{eq:DeltaRSmaintext}
\hat\D=
\left(
\begin{array}{ccc}
0 & \D_0 & \D_0 \\
\D_0 & 0 & \D_0 \\
\D_0 & \D_0 & 0 \\
\end{array}
\right) \  .
\end{equation}
Using this ansatz amounts to assume that in the liquid phase (moderate density), 
the free energy landscape describing the system as a function of the mean-square displacement matrix $\hat \Delta$ has a minimum, 
corresponding to the stable thermodynamic phase, having the form given by Eq.~\eqref{eq:DeltaRSmaintext}. 
Dynamically, this means that the time-dependent mean-square displacement has a single plateau at long times, corresponding to $\D_0$,
see Fig.~\ref{fig:delta}.
We can compute $\SS(\Delta_0)$ and find the stable value of $\Delta_0$, 
as one would compute the magnetization of the paramagnetic phase at high temperature as the minimum of the free energy of the Curie-Weiss ferromagnet model.
We define $\Dl = 2 d R^2/\s^2$. Note that $\Dl \to\io$ in the thermodynamic limit.
We wish to show that $\D_0 = \D_{\rm liq}$ and recover Eq.~\eqref{eq:Sliq}.

\subsubsection{Replica symmetric entropy}

We start from Eq.~\eqref{eq:SSfinal} and we plug in the RS form of $\hat\D$.
Eq.~\eqref{eq:RSdeltasum} implies that $1 - \frac{2 d R^2}{\s^2} v^{\rm T} \hat \D^{-1} v = 1 - \frac{\D_{\rm liq}}{\D_0} \frac{n}{n-1}$.
Using also Eq.~\eqref{eq:RSdeltadet}, the ideal gas term in Eq.~\eqref{eq:SSfinal} becomes
\beq\label{eq:SIG_liquid}
\SS_{\rm IG} = \frac{d}2 n \log \left( \frac{ \pi e \s^2 \D_0}{d^2} \right) + \frac{d}2 \log(1-n) + \frac{d}2 \log\left(1 - \frac{\D_{\rm liq}}{\D_0} \frac{n}{n-1} \right) \ .
\eeq
For the interaction part, using Eq.~\eqref{eq:crucialrelation} and the representation in Eq.~\eqref{eq:FFdef2}, we have
\beq\begin{split}
\FF( \D_0 v v^{\rm T} - \D_0 I)  &= e^{-\D_0/2} \FF(-\D_0 I) = e^{-\D_0/2} \int \de \l \, e^\l \left\{ 1 - \frac{1}{(2\pi \D_0)^{n/2}} \int_{-\l}^\io \de h_a e^{-\sum_a \frac{h_a^2}{2\D_0}} \right\} \\
&= e^{-\D_0/2} \int \de \l \, e^\l \left\{ 1 -  \left( \int_{-\l}^\io \DD_{\D_0} h \right)^n  \right\} 
= e^{-\D_0/2} \int \de \l \, e^\l \left\{ 1 -  \Th\left( \frac{\l}{\sqrt{2\D_0} } \right)^n  \right\}
\ .
\end{split}\eeq
Note that through an integration by parts and a change of variables, we obtain
\beq
\FF(\D_0) = n \int \DD\l \, \Th\left(\frac{\sqrt{\D_0} - \l}{\sqrt{2}} \right)^{n-1} \ ,
\eeq
where the function $\Th(x)$ is defined in Appendix~\ref{app:formulae}.
This is
exactly the replica-symmetric result for $\FF$ obtained in~\cite[Eq.~(40)]{KPUZ13}.
In the limit $n\to 0$ we obtain
\beq\label{eq:sRS}
s_{\rm RS}(\D_0) = \underset{n\to 0}{\lim}\, \frac{\SS(\hat\D_{\rm RS})}n = 
\frac{d}2  \log \left( \frac{ \pi  \s^2 \D_0}{d^2} \right)  + \frac{d}2  \frac{\D_{\rm liq}}{\D_0} - \frac{d}2 \wh\f \int \DD\l \, \Th\left(\frac{\sqrt{\D_0} - \l}{\sqrt{2}} \right)^{-1} \ . 
\eeq

\subsubsection{Saddle point equation}
\label{sec:SPrepRS}

The saddle point equation for $\D_0$ is obtained by taking the derivative of Eq.~\eqref{eq:sRS}. 
We expect that $\D_0 = \D_{\rm liq}$ and we are thus interested in the case where $\D_0$ is large.
For $\D_0  \to \io$, we have
\beq\begin{split}
& \Th\left(\frac{\sqrt{\D_0} - \l}{\sqrt{2}} \right) \sim 1 - \frac{e^{- \frac12 (\sqrt{\D_0} - \l)^2}}{\sqrt{2\pi} (\sqrt{\D_0} - \l)} \ , \\
& \frac1n \FF(\D_0) \sim 1 + \int \DD\l \frac{e^{- \frac12 (\sqrt{\D_0} - \l)^2}}{\sqrt{2\pi} (\sqrt{\D_0} - \l)}
= 1 + e^{-\D_0/4} \int \frac{\de\l}{2\pi} \frac{e^{- (\sqrt{\D_0}/2 - \l)^2}}{\sqrt{\D_0} - \l}
\sim 1 + \sqrt{\frac{1}{\pi \D_0}} e^{-\D_0/4} \ .
\end{split}\eeq
We conclude that $\FF(\hat\D)/n \to 1$ with corrections exponentially small in $\D_0$.
It follows that the interaction term is a constant for large $\D_0$, and its derivative vanishes exponentially.
We therefore obtain
\beq\label{eq:75}
0 = \frac{\partial s_{\rm RS}}{\partial \D_0} \propto
 \frac{1}{\D_0} - \frac{\D_{\rm liq}}{\D_0^2}  + \OO( e^{-\D_0/4}/\sqrt{\D_0} ) \ ,
\eeq
which is solved by $\D_0 = \D_{\rm liq}$ in the limit $R\to\io$ where $\D_{\rm liq} \to\io$.

\subsubsection{Thermodynamic entropy}

Plugging the result $\D_0 = \D_{\rm liq} =  2 d R^2/\s^2$ in Eq.~\eqref{eq:SIG_liquid} we obtain
\beq
\SS_{\rm IG} = \frac{d}2 n \log(2 \pi e/d) + d n \log  R \sim  n \log( \Omega_{d+1} R^d)  = n \log V
 \ .
\eeq
Hence, recalling that $\FF(\D_0\to\io) \to 1$, Eq.~\eqref{eq:SSfinal} becomes
\beq\label{eq:nsliq}
\SS_{\rm RS} = n \left( \log V - \frac{d}{2} \wh\f \right) = n \SS_{\rm liq} \ .
\eeq
and we recover Eq.~\eqref{eq:Sliq}: the replicated entropy is given by $n$ times the liquid entropy\footnote{This might seem surprising since our scaling hypotheses in the derivation 
of Section~\ref{sec:III} constrain replicas within the same configuration to be close and two particles to be almost at contact, which is not the case in the liquid phase. An explanation for this fact 
is given in Appendix~\ref{app:MKHS}.} if replicas are decorrelated~\cite{MPV87}.

\begin{figure}[t]
 \includegraphics[width=10cm]{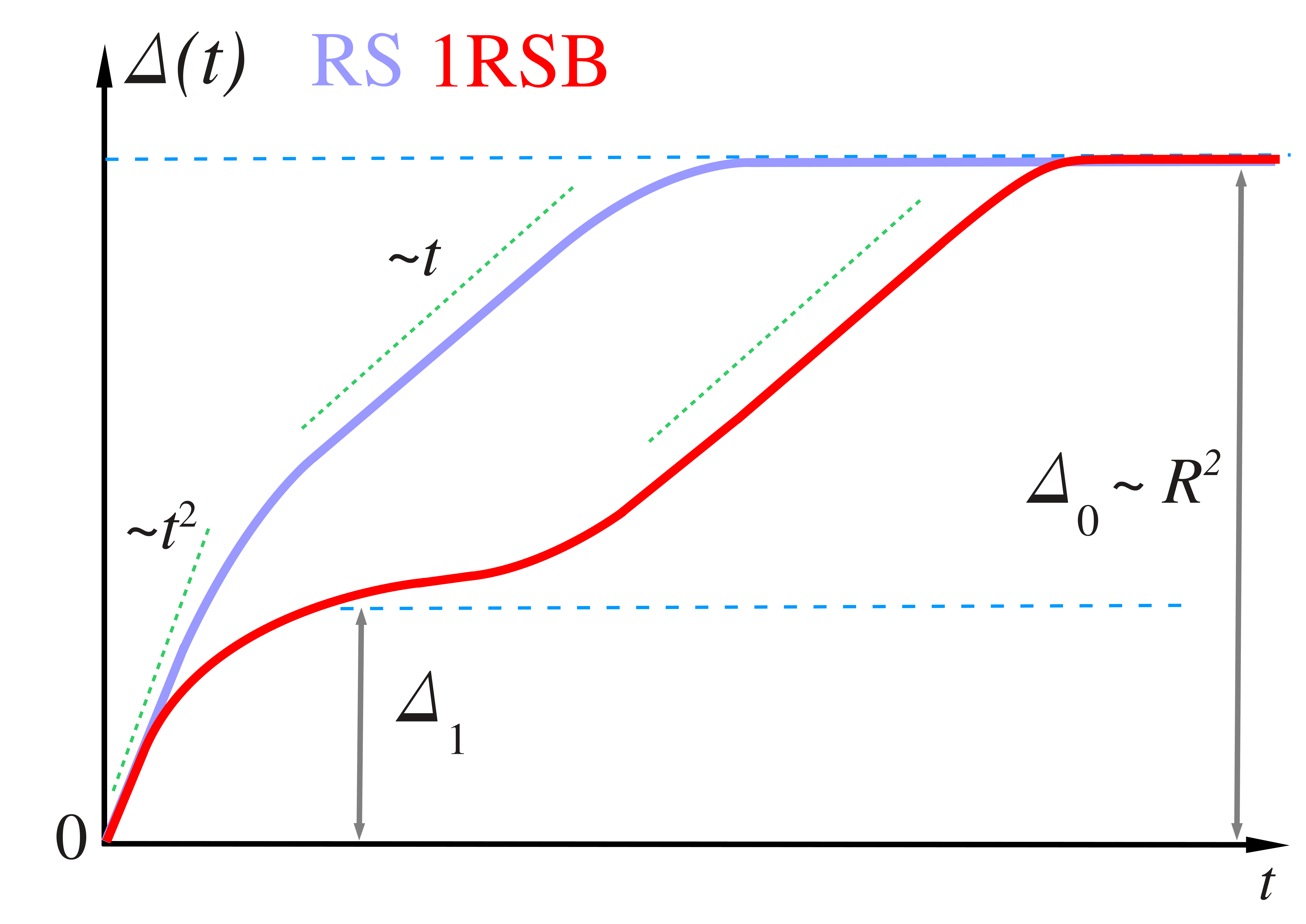}
 \caption{Interpretation of the replica symmetric (RS) and one-step replica-symmetry-breaking (1RSB) hierarchical matrices in terms of the corresponding dynamical quantity, the scaled mean-square displacement (MSD) $\D(t)$. 
 In the liquid phase where the RS solution is stable, the MSD displays the usual ballistic (for inertial dynamics) then diffusive regimes, saturating at the volume of the ``box'' represented by $\D_0=\Dl$. In the 1RSB glass phase, 
 the diffusive regime is replaced by an infinite plateau measured by the parameter $\D_1$, related to the size of the cage. Before reaching this liquid-glass transition, the plateau develops as a crossover 
 between ballistic and diffusive behaviours.}
 \label{fig:delta}
\end{figure}

\subsection{The 1RSB glass phase}
\label{sec:1RSB}

We now repeat the same procedure for a 1RSB matrix that describes the glass phase in the vicinity of the liquid phase~\cite{PZ10,KPUZ13},
and we show that we recover the results of~\cite{PZ10}.
The properties of 1RSB matrices~\cite{MPV87}, 
that are parametrized by $n$ and by an additional integer $m$ and by elements $\D_0$ and $\D_1$, 
are derived in
Appendix~\ref{app:1RSBformula}.
As an example, for $m=3$ one has (with $n/m$ blocks):
\begin{equation}\label{eq:ex1RSB}
\hat\D=\left(
\begin{array}{ccc}
\left(
\begin{array}{ccc}
0 & \D_1 & \D_1 \\
\D_1 & 0 & \D_1 \\
\D_1 & \D_1 & 0 \\
\end{array}
\right) & &\D_0 \\
&\ddots&\\
\D_0 &  &
\left(
\begin{array}{ccc}
0 & \D_1 & \D_1 \\
\D_1 & 0 & \D_1 \\
\D_1 & \D_1 & 0 \\
\end{array}
\right) \\
\end{array}
\right) \ .
\end{equation}
This amounts to assume that, 
at high enough density the free energy landscape develops another minimum while the liquid one becomes unstable, somewhat similarly to what happens 
in the low temperature phase of the Curie-Weiss model, in a ``direction'' given by the 1RSB ansatz. 
Dynamically, this means we assume that the time-dependent mean-square displacement has a plateau at intermediate times, corresponding
to $\D_1$, followed by the true long-time plateau corresponding to $\D_0$, as in Fig.~\ref{fig:delta}.
The new parameter $\Delta_1$ thus represents the typical size of a cage (scaled by $1/d$), 
\ie the amplitude of particles vibrations around an amorphous lattice. It is also sometimes called ``non-ergodicity factor'' because it
signals the breaking of ergodicity in the liquid phase:
the set of liquid configurations which were previously solution of the problem is now split into many disconnected 
clusters of glassy configurations~\cite{CC05}.

\subsubsection{1RSB entropy}\label{sub:1RSBent}

We start by computing $\FF(\hat\D_{\rm 1RSB})$. 
Using Eqs.~\eqref{eq:crucialrelation}, \eqref{eq:FFdef2},
\eqref{eq:1RSBdelta}, \eqref{eq:1RSBinv2} and \eqref{eq:1RSBdeltadet2},
and defining $\D_{\rm B} = [m \D_0 + (1-m) \D_1]/\D_1$,
we have
\beq\begin{split}
\FF(\hat\D_{\rm 1RSB}) & = e^{-\D_0/2} \FF[  (\D_1 - \D_0) \hat I^m - \D_1 \hat I^1 ] \\
 & = e^{-\D_0/2} 
\int \de \l \, e^\l \left\{
1 - \frac{1}{(2\pi)^{n/2} \sqrt{\det[  (\D_0 - \D_1) \hat I^m + \D_1 \hat I^1 ]}} \int_{-\l}^\io \de h_a e^{\frac12 \left[ \frac{\D_0 - \D_1}{\D_{\rm B} \D_1^2} \sum_B \left(\sum_{a \in B} h_a \right)^2
- \frac{1}{\D_1} \sum_a h_a^2  \right]}
\right\} \\
& = e^{-\D_0/2} \int \de \l \, e^\l \left\{
1 - \frac{1}{ \D_{\rm B}^{\frac{n}{2m}} } \int_{-\l}^\io \DD_{\D_1} h_a \prod_B \int \DD z_B
e^{\sqrt{ \frac{\D_0 - \D_1}{\D_{\rm B} \D_1^2} } z_B \sum_{a \in B} h_a  }
\right\} \\
&= e^{-\D_0/2} \int \de \l \, e^\l \left\{
1 - \left[ \frac{1}{ \sqrt{\D_{\rm B}}}  \int \DD z \left( \int_{-\l}^\io \DD_{\D_1} h \, 
e^{\sqrt{ \frac{\D_0 - \D_1}{\D_{\rm B} \D_1^2} } z h } \right)^m \right]^{\frac{n}{m}}
\right\} \\
& =e^{-\D_0/2} \int \de \l \, e^\l \left\{
1 - \left[ \int \DD_{\D_0- \D_1}  z \,  \Th\left( \frac{\l-z}{\sqrt{2\D_1}} \right)^m \right]^{\frac{n}{m}}
\right\} \ ,
\end{split}\eeq
where the last equality can be proven by a series of changes of variable on $z$ and $\l$.
Therefore we obtain, setting $h\equiv\l$ to recover the notations of previous results,
\beq
\begin{split}
\underset{n\to 0}{\lim}\, \frac{\FF(\hat\D_{\rm 1RSB})}n & = -\frac{1}m e^{-\D_0/2} \int \de h \, e^h \log 
\left[ \int \DD_{\D_0- \D_1}  z \,  \Th\left( \frac{h-z}{\sqrt{2\D_1}} \right)^m \right] \ .
\end{split}\eeq
Finally, in Eq.~\eqref{eq:SSfinal} we plug Eqs.~\eqref{eq:1RSBdeltasum} and \eqref{eq:1RSBdeltadet} and we obtain
\beq\label{eq:s1RSBcomplete}
\begin{split}
s_{\rm 1RSB}(\D_0,\D_1,m) &= \underset{n\to 0}{\lim}\, \frac{\SS(\hat\D_{\rm 1RSB})}n = 
\frac{d}2 \log( \pi e \s^2/d^2)  \\
&+\frac{d}2 \left[ \frac{m-1}m \log \D_1 + \frac1m \log (m \D_0 + (1-m) \D_1) - \frac{\D_0}{m\D_0 + (1-m)\D_1} \right] \\
&+
\frac{d}2  \frac{\D_{\rm liq}}{m \D_0 + (1-m) \D_1} +  \frac{d}{2m}\wh\f  e^{-\D_0/2} \int \de h \, e^h \log 
\left[ \int \DD_{\D_0- \D_1}  z \,  \Th\left( \frac{h-z}{\sqrt{2\D_1}} \right)^m \right]  \ . 
\end{split}\eeq

\subsubsection{Saddle point equations}

The reasoning is similar to the RS one: we conjecture that $\D_0$ is very large at the saddle point level,
and we thus expand the entropy for large $\D_0$.
We have for the interaction term
\beq\begin{split}
-m \underset{n\to 0}{\lim}\, \frac{\FF(\hat\D_{\rm 1RSB})}n &= e^{-\D_0/2}  \int \de h \, e^h \log 
\left\{ 1 + \int \de z \g_{\D_0- \D_1}(h-z) \,  \left[ \Th\left( \frac{z}{\sqrt{2\D_1}} \right)^m - 1 \right] \right\} \\
& = e^{-\D_0/2}  \int \de h \, e^h \log 
\left\{ 1 + \int \de z \g_{\D_0- \D_1}(h-z) \, f(z) \right\} \ ,
\end{split}\eeq
where $f(z) = \Th\left( \frac{z}{\sqrt{2\D_1}} \right)^m - 1$ decays quickly to zero for $z\to\io$ and to -1 for $z\to -\io$.
As in the RS case, the integral over $h$ is dominated by large values of $h$, where $\g_{\D_0- \D_1} \star f(h)$ is small,
we can thus expand the logarithm and we obtain, at the leading order for large $\D_0$,
\beq\label{eq:1RSBasy}
\begin{split}
-m \underset{n\to 0}{\lim}\, \frac{\FF(\hat\D_{\rm 1RSB})}n & =   \int \de h \, e^{h-\D_0/2}
 \int \de z \g_{\D_0- \D_1}(h-z) \, f(z) 
 -\frac12   \int \de h \, e^{h-\D_0/2}
\left( \int \de z \g_{\D_0- \D_1}(h-z) \, f(z)  \right)^2 + \cdots \\
&= e^{-\D_1/2} \int \de z \, e^z \, f(z) 
- \frac12 e^{-\D_0/4} \int \de z\de z' \frac{ e^{-\frac{(z-z')^2}{4 \D_0}  } }{\sqrt{4 \pi \D_0}}
 e^{\frac{z+z'}2} f(z) f(z')  + \cdots \\
 &= e^{-\D_1/2} \int \de z \, e^z \, \left[ \Th\left( \frac{z}{\sqrt{2\D_1}} \right)^m - 1 \right]  + \OO( e^{-\D_0/4}/\sqrt{\D_0} )
 \ ,
\end{split}\eeq
where we see that the corrections have the same scaling than in the RS case.

Combining Eqs.~\eqref{eq:s1RSBcomplete} and \eqref{eq:1RSBasy} we obtain:
\beq
0 = \frac{\partial s_{\rm 1RSB}}{\partial \D_0} \propto  
\frac{\D_0 - \Dl}{[m \D_0 + (1-m)\D_1]^2}
  + \OO( e^{-\D_0/4}/\sqrt{\D_0} ) 
\ ,
\eeq
which is again solved by $\D_0 = \D_{\rm liq}$ in the limit $R\to\io$ where $\D_{\rm liq} \to\io$.

\subsubsection{Thermodynamic entropy}
\label{sec:1RSB_C}

Plugging the result $\D_0 = \D_{\rm liq} =  2 d R^2/\s^2$ in Eq.~\eqref{eq:s1RSBcomplete} and using Eq.~\eqref{eq:1RSBasy} we obtain
\beq\begin{split}
s_{\rm 1RSB}&(\D_1,m) = 
\frac{d}2 \log( \pi e \s^2/d^2)  
+\frac{d}2 \left[ \frac{m-1}m \log \D_1 + \frac1m \log (m \D_{\rm liq} )  \right] 
 +  \frac{d}{2m}\wh\f  e^{-\D_1/2} \int \de z \, e^z \, \left[ \Th\left( \frac{z}{\sqrt{2\D_1}} \right)^m - 1 \right] \\
&= \frac1m \left\{
\log V +
\frac{d}2  (m-1) \log \left( \frac{\pi e \s^2 \D_1}{d^2}  \right)
+ \frac{d}2 \log m   
 +  \frac{d}{2}\wh\f   \int \de z \, e^z \, \left[ \Th\left( \frac{z + \D_1/2}{\sqrt{2\D_1}} \right)^m - 1 \right] 
 \right\} \ ,
\end{split}\eeq
which is exactly\footnote{With two small differences. First, in~\cite{PZ10} the entropy of $m$ replicas was computed, while here we divided the entropy
by $n$, hence we computed the entropy per replica. This explains the additional factor $1/m$ in front of the entropy. For a more detailed discussion, see Appendix~\ref{app:equivD}. Second, we should keep in mind
that to obtain the correct result in absence of random rotations 
we should take into account that particles are identical, which introduces an additional factor of $N!$, see
Sec.~\ref{sec:disting}.
} 
the result derived in~\cite[Eq.~(50)]{PZ10}.
Taking the derivative with respect to $\D_1$ and the limit $m\to 1$ one obtains the equation
\beq\label{eq:D1}
\frac1{\D_1} = \frac{\wh\f}2 \int \DD\h \frac{e^{-\frac12(\h +\sqrt{\D_1})^2}}{\sqrt{2\pi \D_1}} \frac{1}{\Th[ (\h+\sqrt{\D_1})/2]} \ ,
\eeq
that gives the cage parameter $\D_1$ on the equilibrium line~\cite{PZ10} and will be useful for future comparison with the
dynamic result.

\subsection{The fullRSB glass phase}
\label{sec:fRSB}

Finally, we consider here full hierarchical replica matrices that describe the Gardner phase, and we show that in this case we obtain
the same results as~\cite{CKPUZ13}. FullRSB matrices are obtained by iterating the procedure
that brings from RS to 1RSB. To save space, we assume here that the reader is familiar with this construction, that can be found in
several textbooks, e.g.~\cite{MPV87} -- see also Appendix~\ref{app:fRSBformula} and Ref.~\cite{CKPUZ13}. 
A physical discussion of the properties of this phase 
can be found for general systems in~\cite{MPV87}, and in the specific case of particle systems in~\cite{KPUZ13,CKPUZ13,nature}.

\subsubsection{FullRSB entropy}

We have
\beq
s_{\rm fRSB} = s_{\rm IG} + s_{\rm int} \ .
\eeq
For the ideal gas term,
plugging Eqs.~\eqref{eq:fRSBdeltadet} and \eqref{eq:fRSBdeltasum} in
Eq.~\eqref{eq:SSfinal}, we have
\beq\label{eq:SIGx}
s_{\rm IG} = \underset{n\to 0}{\lim}\, \frac1n \SS_{\rm IG} =
 \frac{d}2  \log( \pi e \s^2/d^2) 
 + \frac{d}2 \left[
 \log\left( \la \D \ra \right)  - \int_0^1 \frac{\de x}{x^2}\, \log\left(
1 + \frac{ [\D](x)}{\la \D \ra}
\right) + \frac{\D_{\rm liq} - \D(0)}{\la \D \ra} 
 \right] \ .
\eeq
For the interaction term, we start from Eq.~\eqref{eq:FFpaperIII}, which coincides with the result of~\cite{CKPUZ13}. 
We can then follow the derivation of~\cite{CKPUZ13}, with a slight modification.
In fact here we have $n$ replicas and we wish to take the limit $n\to 0$. The function $\D(x)$ is
parametrized as in~\cite{CKPUZ13} but with non-zero $\D(0) = \D_0$ in the interval $[n,m_0]=[0,m]$.
Adapting the results of~\cite[Eq.~(42)-(46)]{CKPUZ13} to take into account this modification, and using the same
notations, we obtain 
\beq\begin{split}
g(1,h) &=  \g_{\D_k} \star \th(h) = \Th\left( \frac{h}{\sqrt{2\D_k}} \right)  \ , \\
g(m_{i},h) & = \g_{\D_i - \D_{i+1}} \star  g(m_{i+1},h)^{\frac{m_{i}}{m_{i+1}}} \ ,
\hskip50pt i = 0 \cdots k-1 \ ,
\\
\mathcal F(\hat \Delta) 
&=e^{- \frac{ \D_0}{2}}  \int_{-\infty}^\infty \de h\, e^{h} 
  \left\{ 1- g(m_0,h)^{\frac{n}{m_0}} \right\} \ .
\end{split}\eeq
Therefore the interaction term becomes (recall that $m_0=m$):
\beq\label{eq:sintfRSB}
\begin{split}
s_{\rm int}& = -\frac{d}2 \wh\f
\underset{n\to 0}{\lim}\, \frac1n \mathcal F(\hat \Delta) 
= \frac{d}{2 m_0} \wh\f e^{- \frac{ \D_0}{2}}  \int_{-\infty}^\infty \de h\, e^{h} \log g(m_0,h) \\
&= \frac{d}{2 m_0} \wh\f e^{- \frac{ \D_0}{2}}  \int_{-\infty}^\infty \de h\, e^{h} \log\left[
\g_{\D_0 - \D_{1}} \star  g(m_{1},h)^{\frac{m_{0}}{m_{1}}} \right] \\
&= \frac{d}{2 m} \wh\f e^{- \frac{ \D_0}{2}}  \int_{-\infty}^\infty \de h\, e^{h} \log\left\{ 1 +
\int \de z\, 
\g_{\D_0 - \D_{1}}(h-z) \left[ g(m_{1},z)^{\frac{m}{m_{1}}} - 1 \right]
\right\} \ .
\end{split}\eeq

\subsubsection{Saddle point equations}

Like in the previous discussions,
 we conjecture that at the saddle point level $\D(x) = \D(0) = \D_0 \to\io$ for $0< x <m$, 
 while $\D(x)$ remains finite for $R\to\io$ when $m<x<1$.
In the ideal gas term,
we have $[\D](x)=0$ for $0<x<m$, and $\la \D \ra = m \D_0 + \int_m^1 \de x \D(x)
= m \D_0 + \la \D \ra_m$. We can also write
\beq
\la \D \ra + [\D](x) = x \D(x) + \int_x^1 \de y\,\D(y) \ ,
\eeq
which remains therefore finite for $m<x<1$.
Then we get at the leading order for $\D_0 \to\io$
\beq\label{eq:SIGfull} 
\begin{split}
s_{\rm IG} &= 
 \frac{d}2  \log( \pi e \s^2 /d^2) 
 + \frac{d}2 \left[
 \frac1m \log\left( m \D_0 \right)  - \int_m^1 \frac{\de x}{x^2}\, \log\left(
x \D(x) + \int_x^1 \de y\,\D(y)
\right) + \frac{\D_{\rm liq} - \D_0}{m \D_0}
 \right] \ . 
\end{split}\eeq
In the interaction term,
for $\D_0 \to \io$,
the integral over $h$ in Eq.~\eqref{eq:sintfRSB} is
dominated by large values of $h$. At large $h$, we have that
$\int \de z\,  \g_{\D_0 - \D_{1}}(h-z) \left[ g(m_{1},z)^{\frac{m}{m_{1}}} - 1 \right]$ is small
so we can expand the logarithm and we obtain
\beq\label{eq:sintfull}
\begin{split}
s_{\rm int} &= \frac{d}{2 m} \wh\f e^{- \frac{\wh \D_0}{2}}  \int_{-\infty}^\infty \de h\, e^{h} \int \de z\, 
\g_{\wh\D_0 - \wh\D_{1}}(h-z) \left[ g(m_{1},z)^{\frac{m}{m_{1}}} - 1 \right] \\
&= -\frac{d}{2 m} \wh\f e^{- \frac{\wh \D_1}{2}}  
 \int \de z 
\, e^z \, \left[ 1 - g(m_{1},z)^{\frac{m}{m_{1}}} \right]
\ .
\end{split}\eeq
Because the interaction term has a finite limit for $\D_0\to\io$, its derivative with respect to $\D_0$ must go to zero in that limit.
Then we have, at the leading order in $\D_0$
\beq
\frac{\partial s_{\rm fRSB}}{\partial \D_0} = \frac{\partial s_{\rm IG}}{\partial \D_0} = \frac{1}{m\D_0} - \frac{\D_{\rm liq}}{m \D_0^2} = 0 \ ,
\eeq
which implies that $\D_0  = \D_{\rm liq} = 2 d R^2/\s^2$.

\subsubsection{Thermodynamic entropy}

Plugging the result $\D_0  = \D_{\rm liq} = 2 d R^2/\s^2$ in Eq.~\eqref{eq:SIGfull}, we get
\beq \begin{split}
s_{\rm IG} &= 
 \frac{d}2  \log( \pi e \s^2 /d^2) 
 + \frac{d}2 \left[
 \frac1m \log\left( 2 m d R^2/\s^2 \right)  - \int_m^1 \frac{\de x}{x^2}\, \log\left(
x \D(x) + \int_x^1 \de y\,\D(y)
\right) \right] \\
 &=
 \frac1m \left[
\log V +   \frac{d}2 (m-1) \log( \pi e \s^2/d^2) + \frac{d}2 \log m 
- \frac{dm}2  \int_m^1 \frac{\de x}{x^2}\, \log\left(
x \D(x) + \int_x^1 \de y\,\D(y)
\right)
 \right]
 \ .
\end{split}\eeq
and adding the interaction term given in Eq.~\eqref{eq:sintfull} we obtain
\beq \begin{split}
s_{\rm fRSB} 
 &=
 \frac1m \Big\{
\log V +   \frac{d}2 (m-1) \log( \pi e \s^2/d^2) + \frac{d}2 \log m 
- \frac{dm}2  \int_m^1 \frac{\de x}{x^2}\, \log\left(
x \D(x) + \int_x^1 \de y\,\D(y)
\right)  \\
&- \frac{d}{2} \wh\f e^{- \frac{\wh \D_1}{2}}  
 \int \de z 
\, e^z \, \left[ 1 - g(m_{1},z)^{\frac{m}{m_{1}}} \right]
 \Big\}
 \ .
\end{split}\eeq
This is exactly\footnote{With the same small difference already noted for the 1RSB case.} the result reported in~\cite[Sec.~3.4]{CKPUZ13}.

\subsection{Relation with previous work}

Having explained the mathematical structure of the mean-squared displacement matrix $\D_{ab}$ in the different phases of the system,
let us give some additional comments on the relation with previous work.
Note that for a dynamic calculation (see Section~\ref{sec:dynamics} below), this
is just the mean-squared displacement in time of a particle, averaged over particles (Fig.~\ref{fig:delta}). For the static  calculation, 
 the formalism leads to considering distances between different replicas. As usual in the replica trick, the total number of replicas tends to zero
 to take the average over the disorder.
 In the case in which the system is solved by a 1RSB ansatz as in Eq.~\eqref{eq:ex1RSB},
 the replicas are grouped in ``blocks'', and all the
 replicas of a block may be pictured as constituting  a ``molecule''~\cite{MP99}, albeit with non-integer number of elements.
 If, as it happens at the highest densities or lowest temperature, the ansatz is fullRSB, then one may see the system as being 
 made of molecules, and molecules of molecules, and so on~\cite{CKPUZ13}. It must be however born in mind that this is  an evocative way of 
 picturing Parisi's ultrametric solution, and it involves no extra assumption.  

In previous work that used the replica scheme~\cite{MP99,PZ10,RUYZ14}, 
the problem was simplified by using the so-called Monasson~\cite{Mo95} or Franz-Parisi~\cite{FP95}
approaches. In these approaches, which are particularly efficient for systems without quenched disorder, one couples the replicas to
a reference system, in such a way that the replicas are always correlated. Mathematically, this corresponds to eliminating the outermost block
of the ultrametric ansatz, corresponding to the element $\D_0$ in Eq.~\eqref{eq:ex1RSB}. 
This decoupling is explicitly seen in Appendix~\ref{app:equivD}.
The problem is simplified because then all
the elements of the replicated matrix $\D_{ab}$ remain finite in the termodynamic limit: particles remain confined into ``molecules'' and one
can use molecular liquid methods to solve the problem~\cite{MP99,PZ10}.

This approach is however not efficient if one wishes to study the dynamics in the liquid phase: 
in fact, the value of $\D_0$ corresponds to the long-time limit of the mean square displacement in the liquid phase (Fig.~\ref{fig:delta}). 
Therefore, if one wishes to establish clearly the parallel between the static and dynamic treatments, one needs to keep the outermost
block in the replica structure. However, this corresponds to decorrelated replicas that have therefore a diverging mean square displacement
in the thermodynamic limit. In fact, we found above that $\D_0 \sim \Dl \to\io$ in the thermodynamic limit.

The advantage of the present derivation is that it makes no assumptions about the existence of molecules, and it allows one to treat a general
structure of $\D_{ab}$ including finite or diverging matrix elements. In this way we can at the same time reproduce previous results, and extend
them to include a complete relation with long-time dynamics in the liquid phase.

\section{Saddle point equation for the order parameter}
\label{sec:V}

In this section we will derive and discuss the equation for the order parameter $\hat\D$ without making any assumption on its
structure. While this is not very interesting for thermodynamics, where we already know that $\hat\D$ is a hierarchical matrix
(Section~\ref{sec:hierarchical}), it is interesting for dynamics. We will obtain the following results.
\begin{enumerate}
\item
The saddle point equation for $\hat\D$ can be written in a form that has the same algebraic structure of a Mode-Coupling (MCT) equation,
Eq.~\eqref{eq:MCTreplica2}, and involves a memory kernel $\hat M$ (Section~\ref{sec:MCTrep}).
\item The kernel $\hat M$ that enters in the MCT equation has a microscopic interpretation in terms of a force-force correlation
or stress-stress correlation between replicas, and gives the shear modulus of the glass (Section~\ref{sec:VD}).
\item
In the MCT equation
we will introduce a Lagrange multiplier to enforce the spherical constraint, and we will get its expression from replicas,
Eq.~\eqref{eq:mustatic}. This will be useful for later comparison with dynamics.
\item We will show that the MCT equation, plugging a 1RSB structure for $\hat\D$ and taking the limit $m\to1$, in which $\Delta_1$ corresponds 
to the equilibrium non-ergodicity factor~\cite{CC05}, gives the same equation as the 1RSB computation of Section~\ref{sec:hierarchical} (Section~\ref{sec:VC}).
\end{enumerate}
These results
will be compared with the dynamical results of Section~\ref{sec:dynamics}.

\subsection{Derivation of the saddle point equation}
\label{sec:MCTrep}

Before deriving the saddle point equation we write the replicated entropy in Eq.~\eqref{eq:SSfinal} using the representation
in Eq.~\eqref{eq:Iffinal} with $A=\Dl$. We obtain, neglecting irrelevant constant terms in the entropy
\beq\label{eq:SSint_final2}
\begin{split}
\SS(\hat \D) & =
\frac{d}2  \log\det(-\hat \D) + \frac{d}2 \log\left(1- \D_{\rm liq} v^{\rm T} \hat \D^{-1} v \right) 
- \frac{d\wh\f}{2} \FF(\hat\D) \ , \\
\FF(\hat\D) &= - e^{-\D_{\rm liq}/2}  \int  \DD_{\Dl v v^{\rm T} - \hat \D} \bar h \, \Psi(\bar h)  
 \ ,
 \hskip20pt
 \Psi(\bar h)  = \int \de\l \, e^\l \, \left(- 1 +  e^{-\b\sum_{a=1}^n \redv( h_a + \l )} \right) \ .
\end{split}\eeq
The advantage of this formulation is that the correlations of the $h_a$ are well defined. In fact, 
$\langle h_a h_b \rangle = \D_{\rm liq} - \D_{ab} \geqslant 0$.
For convenience we can also express the entropy in terms of an overlap matrix 
$Q_{ab} = \D_{\rm liq}  - \D_{ab} = \frac{2d}{\s^2} \la x_{a} \cdot x_{b} \ra$.
The matrix $\hat Q$ is determined by $\partial\SS/\partial Q_{ab}=0$ for $a<b$ (the matrix is symmetric and the diagonal elements are $Q_{aa}=\D_{\rm liq}$).
However we assume a general form for $\hat Q$ and add a Lagrange multiplier term $\wh\mu\sum_a Q_{aa}$ to $\SS$ in order
to impose the constraint on the diagonal elements.
We obtain a very simple expression:
\beq\label{eq:addmu}
\begin{split}
\SS(\hat Q) & =
\frac{d}2  \log\det(\hat Q) 
-\frac{d}{2} \wh \f \FF(\hat Q)  - \frac{d}2 \b \wh\m \sum_a Q_{aa}
 \ ,
\hskip20pt
 \FF(\hat Q) = - e^{-\D_{\rm liq}/2}  \int  \DD_{\hat Q} \bar h \, \Psi(\bar h)  \ .
\end{split}\eeq
The matrix $\hat Q$ is now determined by $\forall (a,b),~\partial\SS/\partial Q_{ab}=0$.
Using the relations $\frac{\partial}{\partial Q_{ab}} \log\det\hat Q = Q^{-1}_{ba}$ 
and $\frac{\partial Q^{-1}_{cd}}{\partial Q_{ab}} = - Q^{-1}_{ca} Q^{-1}_{bd}$, we obtain
\beq\label{eq:MCTreplica}
0 = Q_{ab}^{-1} 
- \b \wh\mu \d_{ab} - \wh\f \frac{\partial \FF}{\partial Q_{ab}} \ ,
\hskip20pt
\frac{\partial\FF}{\partial Q_{ab}} =
  \frac12 Q_{ab}^{-1} e^{-\D_{\rm liq}/2} \int \DD_{\hat Q} \bar h \,  \Psi(\bar h)
- \frac12  e^{-\D_{\rm liq}/2} \sum_{cd} Q^{-1}_{ac} \int \DD_{\hat Q} \bar h \,  \Psi(\bar h) \,  h_c h_d  \, Q^{-1}_{db}  \ .
\eeq
This equation can be simplified by observing that, by integration by parts:
\beq\begin{split}
&\sum_{cd} Q^{-1}_{ac} \int \DD_{\hat Q} \bar h \,   h_c h_d e^{-\b\sum_{a=1}^n \redv( h_a + \l ) } \, Q^{-1}_{db}
= \int  \de \bar h\, \frac{e^{-\b\sum_{a=1}^n \redv( h_a + \l ) }}{(2\pi)^{n/2} \sqrt{\det(\hat Q)}}  \left( \frac{\partial^2}{\partial h_a \partial h_b} + Q^{-1}_{ab} \right) e^{  -\frac{1}2 \bar h^T \hat Q^{-1} \bar h} 
 \\
&= \int \DD_{\rm \hat Q} \bar h\, \left( \frac{\partial^2}{\partial h_a \partial h_b} + Q^{-1}_{ab} \right)e^{-\b\sum_{a=1}^n \redv( h_a + \l ) } \ .
\end{split}\eeq
Using this relation (and the same relation with $\redv=0$) we obtain
\beq
\sum_{cd} Q^{-1}_{ac} \int \DD_{\rm \hat Q} \bar h \,  \Psi(\bar h) \,  h_c h_d  \, Q^{-1}_{db} = 
Q^{-1}_{ab} \int \DD_{\rm \hat Q} \bar h\, \Psi(\bar h) + \int \de \l\, e^\l \int \DD_{\rm \hat Q}\bar h\, \frac{\partial^2}{\partial h_a \partial h_b} e^{-\b\sum_{a=1}^n \redv( h_a + \l ) }
\eeq
and defining $f_\l(h) = - \redv'( h + \l )$ we get:
\beq\label{eq:medvdef}
\begin{split}
&\frac{\partial\FF}{\partial Q_{ab}} =  - \frac12  \la \b^2 f_\l(h_a) f_\l(h_b) + \b f_\l'(h_a) \d_{ab} \ra_v 
= \frac1{\wh \f}\left[- \b^2 M_{ab} + \b \d \m_a \d_{ab} \right] \ ,
\\
&\la \OO \ra_v =  \int \de\l \, e^{\l-\D_{\rm liq}/2} \, \int \DD_{\rm \hat Q} \bar h \, e^{-\b\sum_{a=1}^n \redv( h_a + \l ) } \, \OO \ ,
\end{split}\eeq
where we defined
\beq\label{eq:Mabdef}
M_{ab} = \frac{\wh\f}2   \la  f_\l(h_a) f_\l(h_b)  \ra_v \ ,
\hskip20pt
\d\m_a =   -\frac{\wh\f}2  \la  f_\l'(h_a) \ra_v \ .
\eeq
Then Eq.~\eqref{eq:MCTreplica} takes the form
\beq\label{eq:MCTreplica2}
0 = Q_{ab}^{-1}  + \b^2 M_{ab}(\hat Q)- \b (\wh\mu + \d\m_a) \d_{ab} \ ,
\hskip30pt \Leftrightarrow\hskip30pt
0 = \d_{ab} + \b^2 \sum_c M_{ac}(\hat Q) Q_{cb} -  \b (\wh\mu + \d\m_a) Q_{ab} \ .
\eeq
Written in this form, the saddle-point equation for $Q$ is manifestily similar to the exact dynamic
equations that are
the basis of 
Mode-Coupling Theory~\cite{Go09}: roughly, one has $\hat Q \sim \hat M(\hat Q) \hat Q$, where $\hat M(\hat Q)$ is the analog of
the memory kernel (this will appear more clearly in Section~\ref{sec:dynamics}). 
Mode-Coupling Theory amounts to a polynomial approximation $M_{ab}(\hat Q)\sim Q_{ab}^2$, 
which is exact for some spin glass models~\cite{CC05};
while
here we obtain a more complicated form for $\hat M$.

\subsection{A microscopic expression of the memory kernel: force-force, stress-stress correlations, and the shear modulus}
\label{sec:VD}

We now provide a microscopic interpretation of the memory kernel.

\subsubsection{Force-force correlation}
\label{sec:VD1}

First, we wish to show that $M_{ab}$ is related to the correlation of inter-particle forces.
Using Eq.~\eqref{eq:twoO}, we have:
\beq\begin{split}
F_{ab} &= \frac{\s^2}{2d^3N} \sum_{i \neq j} \la  \nabla v(|x^a_i - x^a_j|) \cdot \nabla v(|x^b_i - x^b_j|) \ra 
= \frac{N\s^2}{2d^3}  \int \de\bar x \de\bar y\, \r(\bar x)\r(\bar y) \prod_c e^{-\b v(|x^c - y^c|)} \nabla v(|x^a - y^a |) \cdot \nabla v(|x^b - y^b |) 
\ .
\end{split}\eeq
For large $d$ we have $x^a - y^a = X + \OO(1/d)$ with $|X|= \s$. Also when we use
Eq.~\eqref{eq:Iffinal} we have to compute the function $f$ in $|x^a - y^a| = \s(1+h_a/d+\l/d)$.
Thus at leading order 
$\frac{(x^a - y^a )\cdot (x^b - y^b)}{ |x^a - y^a ||x^b - y^b |} = \frac{X \cdot X}{\s^2} =  1$,
and $v'(|x^a - y^a |) = (d/\s) \redv'(\l+h_a)$ according to Eq.~\eqref{eq:redv}. We obtain
\beq\label{eq:app44889}
\nabla v(|x^a - y^a |) \cdot \nabla v(|x^b - y^b |) = v'(|x^a - y^a |)v'(|x^b - y^b |) \frac{(x^a - y^a )\cdot (x^b - y^b)}{ |x^a - y^a ||x^b - y^b |}
\sim \left(\frac{d}\s\right)^2 \, \redv'(\l+h_a) \redv'(\l + h_b)
 \ .
\eeq
Finally
using Eq.~\eqref{eq:Iffinal} with $A = \Dl$ we obtain
\beq
F_{ab} =  \frac{\wh\f}2 e^{-\D_{\rm liq}/2}  \int  \DD_{\hat Q} \bar h \, \de\l \, e^\l  \prod_c e^{-\b \redv(\l + h_c)} \redv'(\l+h_a) \redv'(\l + h_b) 
= \frac{\wh\f}2  \la f_\l(h_a) f_\l(h_b) \ra_v = M_{ab}
\ .
\eeq
Therefore the kernel that enters in Eq.~\eqref{eq:MCTreplica} is also the microscopic force-force correlation.

\subsubsection{Stress-stress correlation}
\label{sec:stress}

We can take another step and compute the stress-stress correlation, following~\cite{Yo12,YZ14}.
Note that in this derivation we neglect the kinetic component of the stress tensor~\cite{hansen}: we do this to simplify the 
computations, and because this component remains small in the glass transition regime.
According\footnote{Note that there is a typo in the factors of $N$ in~\cite{Yo12}; the correct ones are given here.}
to~\cite[Eq.~(136)-(138)]{Yo12},
we define 
respectively the Born term $B_a$, the replicated stress-stress correlation $\Sigma_{ab}$ and the potential part of the stress tensor at zero wavevector 
$\s_{ij}^a$ evaluated at $x=x_i^a-x_j^a$
\beq
B_a = \frac{1}{d N} \sum_{i<j} \langle b^a_{ij} \rangle \ ,
\hskip30pt
b_{ij}^a = \{ \hat x^2_1 [ |x |^2 v''(|x |)  \hat x_2^2 + |x| v'(|x|) (1 -  \hat x_2^2)   ]  \}_{x = x_{i}^a - x_{j}^a} \ ,
\eeq
and
\beq\begin{split}
\Si_{ab} &= \frac1{dN} \sum_{i<j, k<l} [ \langle \s_{ij}^a \s_{kl}^b \rangle - \langle \s_{ij}^a \rangle \langle \s_{kl}^b \rangle] =  \frac1{dN} \sum_{i<j, k<l} \langle \s_{ij}^a \s_{kl}^b \rangle \sim \frac1{dN} \sum_{i<j} \langle \s_{ij}^a \s_{ij}^b \rangle
 \ , \\
\s_{ij}^a &= [ |x | v'(|x |) \hat x_1 \hat x_2 ]_{x = x_{i}^a - x_{j}^a} \ ,
\end{split}\eeq
where $\hat x = x/|x|$, and $\hat x_\mu$ are its spatial components. By isotropy the stress tensor for two directions $\m\neq\n$ is the same as the one written here for directions 1,2.
Here we used that $\langle \s_{ij}^a \rangle=0$ again by isotropy
and that in $d\to\io$ only the terms with $i=j$ and $k=l$ contribute to $\Si_{ab}$ (see~\cite[Appendix A]{YZ14} and~\cite{MKZ15} for a more detailed discussion). Physically it is related to the 
tree-like structure of the interactions as emphasized in sections~\ref{sec:intro} and \ref{sec:II}.
From $B_a$ and $\Si_{ab}$ we obtain the shear modulus matrix\footnote{The name $\m$ is standard in the literature, and is not to be confused with the Lagrange multiplier introduced in the saddle-point Eq.~\eqref{eq:addmu}.}
\beq
\wh\mu_{ab} = \frac{\mu_{ab}}d = B_a \d_{ab} - \b \Si_{ab} \ .
\eeq

Following the same reasoning that leads to Eq.~\eqref{eq:app44889}, and observing that
on average, $\hat X_1^2 = \hat X_2^2 \sim 1/d$, we obtain
\beq\begin{split}
b_{ij}^a & \sim  \frac1d \left[ \frac{\s^2}d v''(|x^a - y^a |)
+ \s v'(|x^a - y^a |)
\right] = \redv''(\l+h_a) + \redv'(\l+h_a)
\\
\s_{ij}^a \s_{ij}^b &\sim \s^2 \hat X_1^2 \hat X_2^2 v'(|x^a - y^a |)v'(|x^b - y^b |) \sim \frac{\s^2}{d^2} v'(|x^a - y^a |)v'(|x^b - y^b |)
\sim \redv'(\l+h_a) \redv'(\l + h_b) \ .
\end{split}\eeq
Then performing similar steps as in Section~\ref{sec:VD1}, we arrive to
\beq\begin{split}
B_a &= - \frac{\wh\f}2  \la f_\l(h_a) + f_\l'(h_a)  \ra_v = \b \sum_{b} M_{ab} \ ,
\\
\Si_{ab} &=  \frac{\wh\f}2 \la f_\l(h_a) f_\l(h_b) \ra_v = M_{ab} \ ,
\end{split}\eeq
where the relation $B_a = \sum_b M_{ab}$ is obtained through a simple integration by parts on $\l$.
Therefore, the stress-stress correlation coincides, in $d\to\io$, with the force-force correlation, and both coincide with $M_{ab}$.
Finally, for the shear modulus we obtain
\beq
\b \wh\mu_{ab} =  \d_{ab} \sum_{c (\neq a)} \b^2 M_{ac} -  (1-\d_{ab}) \b^2 M_{ab} \ .
\eeq
Recalling that from Eq.~\eqref{eq:MCTreplica} we have for $a\neq b$ that $\b^2 M_{ab} =  \wh \f \frac{\partial\FF}{\partial\D_{ab}}$,
this results coincides\footnote{
The factor of 2 in~\cite{YZ14} is due to the fact that in that paper the derivatives with respect to $\D_{ab}$ are defined
for a symmetric matrix, hence
only for $a<b$ and multiplied by 2.
}
with the one in~\cite[Eq.(15)]{YZ14}. We refer to~\cite{YZ14} for a discussion of the physical consequences of this result.

\subsection{Replica symmetric solution}

\subsubsection{Product measure}

In the liquid phase, the solution to this equation is $Q_{ab} = \D_{\rm liq} \d_{ab} + Q_0 (1-\d_{ab})$ where
$Q_0 = \Dl - \D_0$. We already know that $Q_0$ is exponentially small in the limit $\Dl \to \io$ (Section~\ref{sec:SPrepRS}), 
we therefore consider for simplicity a RS solution with $Q_{ab} = \Dl \d_{ab}$ and $Q^{-1}_{ab} = \d_{ab}/\Dl$.
In this case the measure in Eq.~\eqref{eq:medvdef} becomes a product measure and defining 
\beq
\HH_0(h,\l) =  \redv( h + \l ) + \frac{T h^2}{2\Dl} \ ,
\hskip20pt 
\ZZ_0(\l) = \int \de h \, e^{-\b\HH_0(h,\l)} \ ,
\eeq
we obtain for some observable $\OO$ when $n\to0$:
\beq\label{eq:avRS}
\begin{split}
\la \OO(h_a) \ra_v &= \int \de\l \, e^{\l-\D_{\rm liq}/2} \,\int \de h\, e^{-\b \HH_0} \OO(h) \left( \int \de h\, e^{-\b \HH_0} \right)^{n-1} \\ & = 
\int \de\l \, e^{\l-\D_{\rm liq}/2} \,\frac{1}{\ZZ_0} \int\de h\, e^{-\b \HH_0} \OO(h) =\int \de\l \, e^{\l-\D_{\rm liq}/2} \, \la \OO(h) \ra_{\HH_0} \ .
\end{split}\eeq
For later purposes, it is useful to compute some of these averages.
First of all, it is easy to show through an integration by parts that
\beq\label{eq:IBP}
 \la \frac{d\OO}{dh} \ra_{\HH_0} =  \la \OO \left(-\b f_\l(h) + \frac{h}{\Dl}  \right) \ra_{\HH_0} \ ,
 \hskip20pt  \Rightarrow \hskip20pt
 \la f_\l(h)\ra_{\HH_0} = \frac{T}{\Dl} \la h \ra_{\HH_0} \ .
\eeq
where the second result is obtained by choosing $\OO=1$.
Eq.~\eqref{eq:avRS} is readily generalized for $a\neq b$ by:
\beq\label{eq:av2RS}
\la \OO(h_a)\OO(h_b) \ra_v =\int \de\l \, e^{\l-\D_{\rm liq}/2} \, \la \OO(h) \ra_{\HH_0}^2
\eeq
which will be compared to long-time limits of dynamical quantities later on, in the liquid phase. Indeed, in the replica symmetric language, diagonal elements represent equal-time values of the corresponding dynamical observables, while off-diagonal elements represent long-time limits.

\subsubsection{Averages for large $\Dl$}

We will be particularly interested in computing averages $\la \bullet \ra_v$ for
$\Dl \to \io$.
For an observable $\OO(y,\l)$ that decays quickly to zero for large $y$, we have
\beq\label{eq:114}
\begin{split}
\la \OO(h+\l,\l) \ra_v
 & = \int \de\l \,e^{\l -\Dl/2}\,
\frac{\int \de h\, e^{-\b \redv(h+\l) - \frac{h^2}{2\Dl}} \OO(h+\l,\l)}
{\int \de h\, e^{-\b \redv(h+\l) - \frac{h^2}{2\Dl}} } \\ &=  
\Dl\int \de\a \,e^{ -\frac\Dl2(1-\a)^2}\,
\frac{\int \de y\, e^{-\b \redv(y)+y\a - \frac{y^2}{2\Dl}} \OO(y,\a\Dl)}
{\int \de h\, e^{-\b \redv(h+\a\Dl) - \frac{h^2}{2\Dl}} } \\
&\underset{\Dl\to\io}{\sim }
\int  \de y\, e^{  -\b \redv(y) + y} \, \OO(y,\Dl)
\ .
\end{split}\eeq
The above chain of equalities is based on the following reasoning:
{\it (i)} since for $\Dl\to\io$ the integral over $\l$ is dominated by large values of $\l$, we set $\a=\l/\Dl$;
{\it (ii)} we changed variable from $h$ to $y = h + \l$
in the numerator; because $\OO(y,\bullet)$ decays to zero for large $y$, the term $y^2/2\Dl$ is negligible for $\Dl\to\io$; 
{\it (iii)} we can evaluate the integral over $\a$ by a saddle-point method in $\Dl\to\io$, dominated by $\a=1$;
{\it (iv)} in the denominator, contrary to the numerator, there is no damping function $\OO$, hence $h^2/2\Dl$ is not negligible and
we use $\redv(r\to\io) = 0$ to compute it for large $\Dl$.
Note that the factor $\de y\,e^{  -\b \redv(y) + y}$ corresponds to the $d\to\io$ limit of $\de r\,r^{d-1}g(r)$ with $r=1+y/d$.

From Eq.~\eqref{eq:114} we obtain several useful relations. We specialize for simplicity on the Hard Sphere potential,
which we consider as the limit of a soft potential, e.g. $\redv(y) = - \ee y \th(-y)$ for $\ee\to\io$.
We get, for example:
\beq\label{eq:mediev}
\begin{split}
&\frac{T}{\Dl}  \la \l^n h \ra_v = 
\la \l^n f_\l(h) \ra_v = - \Dl^n \int_{-\io}^0  \de y\, e^{  -\b \redv(y) + y} \redv'(y) = 
 T \Dl^n \ , \\
& \la (h+\l)^n f_\l(h) \ra_v =- \int_{-\io}^0  \de y\, e^{  -\b \redv(y) + y} y^n \redv'(y) 
= 0 \ , \hskip30pt \forall n>0 \ ,
\\
&  \la h f_\l(h) \ra_v = \la (h+\l) f_\l(h) \ra_v - \la \l f_\l(h) \ra_v = - \la \l f_\l(h) \ra_v =  -T \Dl \ .
\end{split}\eeq

As an example, from Eqs.~\eqref{eq:MCTreplica2} and \eqref{eq:Mabdef} we obtain the expression of the Lagrange multiplier $\wh\mu$:
\beq\label{eq:mustatic}
\begin{split}
\b\wh\mu     & = \frac1\Dl + \frac{\wh\f}2    
\langle  \b^2 f_\l(h)^2 + \b f_\l'(h) \rangle_{v} 
= \frac1\Dl + \frac{\wh\f}2 \frac{\b}{\Dl} \la h f_\l(h) \ra_v
= \frac1\Dl - \frac{\wh\f}2 
\ .
\end{split}\eeq
where the last equality holds for Hard Spheres using Eqs.~\eqref{eq:IBP} and \eqref{eq:mediev}.

\subsection{1RSB solution}
\label{sec:VC}

We now consider the 1RSB solution. We restrict to the case $m=1$ for simplicity, and once again we consider that $Q_0=0$.
We thus have $Q_{ab} = \D_{\rm liq} \d_{ab} + Q_1 (I^m_{ab}-\d_{ab})$ with $Q_1 = \Dl - \D_1$, and
\beq
Q^{-1}_{ab} = \frac{1}{\Dl} \d_{ab} + \left( \frac{1}{\Dl} - \frac{1}{\D_1} \right) (I^m_{ab}-\d_{ab}) 
= \frac{1}{\Dl} I^m_{ab} - \frac{1}{\D_1}  (I^m_{ab}-\d_{ab}) 
\ .
\eeq
By taking in Eq.~\eqref{eq:MCTreplica} indeces $a \neq b$ that belong to the same block, and using $n\to 0$,
we obtain the equation (where the index $a = 1 \cdots m$ with $m\to 1$)
\beq\label{eq:plateau_static}
\begin{split}
\frac1{\D_1} - \frac1{\Dl} & = \frac{ \wh\f }2  \int \de\l \, e^{\l-\D_{\rm liq}/2}
\frac{
\int \left( \prod_a \de h_a\, e^{-\b \redv(h_a+ \l)} \right) e^{ - \frac{1}{2\D_1}\sum_a h_a^2 +\frac12 \left( \frac1{\D_1} - \frac1{\Dl} \right)\left( \sum_{a} h_a \right)^2 } \b f_\l(h_1) \b f_\l(h_2) 
}
{
\int \left( \prod_a \de h_a \,e^{-\b \redv(h_a+ \l)} \right) e^{ - \frac{1}{2\D_1}\sum_a h_a^2 +\frac12 \left( \frac1{\D_1} - \frac1{\Dl} \right) \left( \sum_{a} h_a \right)^2} 
}
\\
& = \frac{ \wh\f }2  \int \de\l \, \frac{e^{\l-\D_{\rm liq}/2}}{
  \int \de h \, e^{-\b \redv(h+ \l) - \frac{h^2}{2\Dl}  } 
}
 \int \DD\h\,
\frac{\left[ \int \de h\, e^{-\b \redv(h+ \l) - \frac{h^2}{2\D_1} + \h h \sqrt{ \frac1{\D_1} - \frac1{\Dl} } } \b f_\l(h)   \right]^2}
{  \int \de h\, e^{-\b \redv(h+ \l) - \frac{h^2}{2\D_1} + \h h \sqrt{ \frac1{\D_1} - \frac1{\Dl} } } }
 \\
& = \frac{ \wh\f }2  \int \de\l \, e^{\l-\D_{\rm liq}/2}
\frac{1}{\ZZ_0}
 \int \DD\h  \, \ZZ_1 \,
\la \b f_\l(h) \ra_{\HH_1}^2 \ ,
\end{split}\eeq
where we defined
\beq\label{eq:H1}
\HH_1(h,\l) = \redv(h+ \l) + \frac{T h^2}{2\D_1} - \h h T \sqrt{ \frac1{\D_1} - \frac1{\Dl} } \ ,
\hskip20pt 
\ZZ_1(\l) = \int \de h\, e^{-\b\HH_1(h,\l)} \ .
\eeq

It remains to be checked that Eq.~\eqref{eq:plateau_static} is equivalent to the one derived in Section~\ref{sec:1RSB_C}
for $\Dl \to \io$.
From the second line of Eq.~\eqref{eq:plateau_static}, shifting in all the integrals $h + \l \to h$ and
$\h + \l \sqrt{ \frac1{\D_1} - \frac1{\Dl} } \to \h$, we obtain
\beq
\begin{split}
\frac1{\D_1} - \frac1{\Dl} & = \frac{ \wh\f }2  \int \de\l \, \frac{e^{ - \frac{1}{2\Dl} (\l - \Dl)^2 }}{
  \int \de h \, e^{-\b \redv(h) - \frac{(h-\l)^2}{2\Dl}  } 
}
 \int \DD\h\,
\frac{\left[ \int \de h\, \left(\frac{\de}{\de h}e^{-\b \redv(h)}\right) e^{ - \frac{h^2}{2\D_1} + \frac{h\l}{\Dl} + \h h \sqrt{ \frac1{\D_1} - \frac1{\Dl} } }    \right]^2}
{  \int \de h\, e^{-\b \redv(h) - \frac{h^2}{2\D_1} + \frac{h\l}{\Dl} + \h h \sqrt{ \frac1{\D_1} - \frac1{\Dl} } } } \ ,
\end{split}\eeq
From this form one sees as in the replica symmetric case that for large $\Dl$ the integral over $\l$ is strongly peaked on $\l = \Dl$. With this choice at leading order in large
$\Dl$ we have
\beq
\begin{split}
\frac1{\D_1}  & = \frac{ \wh\f }2   \int \DD\h
\frac{\left[ \int \de h\, \left(\frac{\de}{\de h}e^{-\b \redv(h)}\right) e^{ - \frac{h^2}{2\D_1} + h + \h h \sqrt{ \frac1{\D_1} } }    \right]^2}
{  \int \de h\, e^{-\b \redv(h) - \frac{h^2}{2\D_1} + h + \h h \sqrt{ \frac1{\D_1}  } } } \ .
\end{split}\eeq
Specializing to hard spheres, we obtain
\beq
\begin{split}
\frac1{\D_1}  & = \frac{ \wh\f }2   \int \DD\h\,
\frac{\left[ \int_0^\io \de h\, \frac{\de}{\de h} e^{ - \frac{h^2}{2\D_1} + h + \h h \sqrt{ \frac1{\D_1} } }    \right]^2}
{  \int_0^\io \de h\, e^{ - \frac{h^2}{2\D_1} + h + \h h \sqrt{ \frac1{\D_1}  } } }
= \frac{ \wh\f }2   \int \DD\h\,
\frac{ e^{-\frac12 (\h + \sqrt{\D_1})^2} }{\sqrt{2\pi \D_1}}
\frac{1}
{  \Th[ (\h + \sqrt{\D_1})/\sqrt{2} ]} \ ,
\end{split}\eeq
which is equivalent to Eq.~\eqref{eq:D1}.

%
%
%
%

\section{Dynamics through a thermodynamic analogy}
\label{sec:dynamics}

In this section we derive the dynamics of the system, drawing a formal analogy between time dependence of observables in the dynamics and replica index
of the corresponding observables in the statics, as discussed in~\cite{MPV87,CK93,CC05} and illustrated in Fig.~\ref{fig:delta}.
Indeed, although dynamics is formally more difficult to handle than the statics, it is not needed to resort to replicas
in order to average over the disorder, which is a conceptual and technical advantage~\cite{DD78,CC05}. 
This is a consequence of the observation that the dynamic partition function is 1 by probability conservation if one considers all possible paths, hence independent of the Hamiltonian of the system.

We consider a Langevin dynamics
\beq\label{eq:HSLang}
m \ddot x_i(t) + \g \dot x_i(t) = -\nu_i(t) x_i(t) - \nabla_{x_i}H+ \h_i(t) 
\hskip30pt
\langle \h^\mu_i(t) \h^\nu_j(t') \rangle = 2 T \g \d_{ij} \d_{\mu\nu} \d(t-t')
\eeq
where $\mu,\nu=1,\cdots,d$ are the spatial indices and $H = \sum_{i < j} v(x_i - x_j)$.
In order to impose the constraint $x_i \cdot x_i = R^2$,
we introduce a Lagrange multiplier $\n_i$ per particle, whose value is determined by the constraint.
The inertial term $m \ddot x_i(t)$ will be dropped for simplicity, but as it will be clear in the following, it can be 
reinserted at any time.

Another possible strategy is to keep the inertial term, let the system equilibrate, and then remove the friction and noise terms.
One may do so directly assuming the Gibbs-Boltzmann distribution for the initial condition. 
Remarkably enough, nothing dramatic happens with the equation and its solution in the limit $\gamma \rightarrow 0$. The external noise is absent,
but the one induced on a particle by the others is still here, just as the induced friction term.
It is tempting to think that we have thus ``proven'' chaoticity for a particle system, but a  caveat is in order. Our path integrals are defined
for finite noise level, which we are taking to zero {\em after} the limit of large particle number and of large dimension.  
Thus, we are ``proving chaos'' with some level of coarse-graining, which we are taking to zero after all other parameters have gone to infinity.

\subsection{Lagrange multiplier}
\label{sec:Lagdyn}

Let us compute the value of $\n_i$ in $d\to\io$.
We can discretize Eq.~\eqref{eq:HSLang} as follows 
(in the It\^o sense):
\beq\label{eq:langevin}
x_i(t+\de t) = x_i(t) -\frac1\g \nu_i(t) x_i(t) \de t - \frac1\g \nabla_{x_i}H \de t + \frac1\g\h_i \ ,
\hskip30pt
\langle \h_i^\m \h^\n_j \rangle = 2 T \g \de t \d_{ij} \d_{\m\n} \ .
\eeq
We impose the spherical constraint. At order $\de t$, using that for large $d$
one has $A \cdot \h_i \sim 0$ (for any vector $A$ uncorrelated with $\h_i$ in the It\^o sense) and $\h_i \cdot \h_i \sim 2 d T \g \de t$
due to the central limit theorem, giving
\beq
R^2 = x_i(t+\de t) \cdot x_i(t+\de t) = x_i(t)\cdot x_i(t) - \frac{2 \de t}\g x_i(t) \cdot \left[
\n_i(t) x_i(t) + \nabla_{x_i}H  \right] +\frac{2d T \de t}{\g} \ ,
\eeq
and therefore
\beq\label{eq:HSmu}
\n_i(t) = - \frac1{R^2} x_i \cdot  \nabla_{x_i}H + \frac{d \, T}{R^2} \ .
\eeq
We have a general relation\footnote{
This relation can only be derived for a confined system in a (cubic) volume $V$. Here instead 
we are considering a system on the hypersphere. However, we can take Eq.~\eqref{eq:pdef}
as a {\it definition} of the pressure for any system. In the limit $R\to\io$, the system on the hypersphere
is equivalent to a system in Euclidean space with periodic boundary conditions, which is also equivalent
to a confined system when $V\to\io$, because in the liquid phase the boundary conditions are irrelevant.
In that case therefore the pressure $p$ is the same for all systems and is given by Eq.~\eqref{eq:pdef}.
}~\cite[Eq.(2.2.10)]{hansen} 
\beq\label{eq:pdef}
p = \frac{\b P}{\r} = 1 - \frac{\b}{d \, N} \la \sum_i x_i \cdot  \nabla_{x_i}H \ra
= 1 - \frac{\b}{2 d \, N} \la \sum_{i\neq j} |x_i-x_j|  v'(|x_i - x_j|)   \ra \ .
\eeq
For $d\to\io$ the fluctuations vanish because we average over $d$ dimensions, we thus have
$ x_i \cdot  \nabla_{x_i}H \sim \la x_i \cdot  \nabla_{x_i}H \ra = d\, T (1-p)$,
and plugging this in Eq.~\eqref{eq:HSmu} we obtain 
that all $\nu_i(t)$ are equal and constant in time, and given by
\beq
\nu_i(t) \sim \nu =  \frac{d \, T}{R^2} p \ ,
\hskip30pt
\wh\n \equiv \frac{\s^2}{2d^2} \n = 
\frac{T}{\Dl} p \ .
\eeq
We recall that in $d\to\io$ we have, in the liquid phase of hard spheres, 
$p = 1 + d\wh\f/2$, as it can be easily derived from Eq.~\eqref{eq:Sliq}~\cite{PZ10,MK11}.
The same result can also be obtained more directly as follows:
\beq\begin{split}
\n &= \frac1N \sum_i \la \n_i \ra = \frac{d \, T}{R^2}- \frac1{N R^2} \sum_i \la x_i \cdot  \nabla_{x_i}H \ra
=  \frac{d \, T}{R^2} - \frac1{2 N R^2} \sum_{i\neq j} \la |x_i-x_j|  v'(|x_i - x_j|) \ra \\
&=\frac{d \, T}{R^2} -
 \frac{N}{2 R^2}  \int \de\bar x \de\bar y \r(\bar x)\r(\bar y) \prod_c e^{-\b v(|x^c - y^c|)} |x^a-y^a| v'(|x^a - y^a|) \\
&=\frac{d \, T}{R^2} - 
  \frac{d \wh\f}{2 R^2} e^{-\D_{\rm liq}/2}  \int  \DD_{\hat Q} \bar h \, d\l \, e^\l \prod_c e^{-\b \redv(\l + h_c)}\, \s (1+ h_a/d + \l/d )\frac{d}{\s} \redv'(\l+h_a) \\
&=\frac{d \, T}{R^2} +
  \frac{d \wh\f}{2 R^2} \la f_\l(h) (d+ h+\l) \ra_v 
  =\frac{d \, T}{R^2} \left( 1 + \frac{d \wh \f}2 \right) \ . 
\end{split}\eeq
where we used Eq.~\eqref{eq:twoO}, \eqref{eq:Iffinal} and~\eqref{eq:mediev}.

\subsection{Path integral and supersymmetry}

The supersymmetric (SUSY) formulation of the dynamics can be found in~\cite{Cu02,Ku92,K03}, but here we use slightly different conventions and do not introduce fermionic fields\footnote{They 
are only necessary in case of possible ambiguities in the discretization of the stochastic Eq.~\eqref{eq:langevin}, that are irrelevant here.}.
 Within the Martin-Siggia-Rose formalism~\cite{Cu02,CC05} we introduce a dynamic partition function of the form
\beq\begin{split}
\overline{Z} &= \overline{ \int \De\eta \int_{\SSS} \De X \, \d( \g \dot x_i(t) + \nu_i(t) x_i(t) + \nabla_{x_i}H - \h_i(t) ) } \\
& =
\int_{\SSS} \De X \int_{\partial \SSS} \De \hat X\,
e^{   - \int \de t \sum_i [ T \g \hat x_i \cdot\hat x_i +  i \hat x_i \cdot \g\dot x_i + i \hat x_i \cdot \n_i x_i ]} 
\,\overline{
e^{ -\int \de t \sum_i  i\hat x_i \cdot \nabla_i H } 
} 
\end{split}\eeq
Here $\De X$ denotes a functional integral over trajectories $X(t) = \{ x_i(t) \}$.
Note that here we assume to start in a random configuration at time $t=-\io$, hence all integrals
over $t$ extend from $-\io$ to $\io$. Also, note that $x_i(t) \in \SSS$ while $\hat x_i(t)$ is introduced
to exponentiate the delta function of $\dot x_i$. From the spherical constraint $\dot x_i\cdot x_i=0$, therefore $\hat x_i$ is orthogonal to $x_i$ and belongs
to the tangent plane to $\SSS$, which we call $\partial \SSS$.
We introduce SUSY fields defined in terms of Grassmann variables $\th$ and $\bth$ as follows:
\beq\begin{split}
a &= \{ t, \th, \bth \} \ , \\
\bm x_i(a) &= x_i(t)  + i \hat x_i(t) \th\bth \ , \\
\int \de a f[\bm x_i(a) ] &= \int \de t \de \bth \de \th f[x_i(t) + i \hat x_i(t) \th\bth]
= \int \de t\,  i\hat x_i(t) f'[x_i(t)]  \ , \\
\partial^2_a &= 2T\g \frac{\partial^2}{\partial\th \partial\bth} - \g\th \frac{\partial^2}{\partial\th\partial t}+\g\frac{\partial}{\partial t}  \ , \\
\bm\d(a,b) &= \d(t_a - t_b) (\th_a\bth_a + \th_b \bth_b) \ , \\
\frac12 \int \de a \,\bm x_i(a)  \cdot \partial^2_a \bm x_i(a) &= \int \de t [ T \g \hat x_i(t) \cdot \hat x_i(t) +  i\hat x_i(t) \cdot \g\dot x_i(t) ]  \ .
\end{split}\eeq
Using these notations we can write the partition function in a very compact form:
\beq\label{eq:olZdyn}
\overline{Z} = \int_{\SSS} \De \bm X\,
e^{   - \frac12  \int \de a \sum_i\bm x_i(a) \cdot [\partial^2_a +\n_i ]\bm x_i(a)
} 
\overline{
e^{ -\int \de a H[\bm x(a)] } 
} \  .
\eeq
Here, the constraints that $x_i\in \SSS$ and $\hat x_i \in \partial \SSS$ are equivalently encoded in a single constraint
$\bm x_i(a) \cdot\bm x_i(a) = R^2$.
Here $\De\bm X \equiv \De X \De \hat X$.

The formal analogy between Eqs.~\eqref{eq:olZdyn} and \eqref{eq:olZ}, apart from the single-particle kinetic term which is easily dealt with as an additive contribution 
to the exponent, is evident. The replica index $a=1,\cdots,n$ becomes 
the SUSY variable $a = \{t,\th,\bth\}$. 
Except that, the structure of the dynamical partition function $\overline{Z}$ is identical to the one of the replicated partition 
function, see~\cite{Ku92,K03,PR13} for a general discussion.
We can thus repeat all the steps that we performed with replicas (details can be found in~\cite{MKZ15})
and we arrive to a similar result. The dynamic partition function can be expressed as an integral over a dynamical order parameter
$Q(a,b)$ as follows:
\beq\begin{split}
\overline{Z} &= \int \de \bm Q(a,b)\, e^{N \SS(\bm Q)} \ , 
\hskip30pt \bm Q(a,a) = R^2  \ ,
\\
\SS(Q) &= - \frac{d}2 \int \de a \de b\, \bm\KK(a,b)\bm Q(a,b) + \frac{d}2  \log\det(\bm Q) +  
\frac{d}{2} \wh\f e^{-\D_{\rm liq}/2}  \int  \DD_{\bm Q}\bm h \,  \Psi(\bm h) 
 \ ,
 \\
 \DD_{\bm Q} \bm h & \propto \De\bm h\,  e^{  -\frac{1}2 \int \de a \de b \,\bm h(a)\bm Q^{-1}(a,b)\bm h(b)} \ , \\
 \Psi(\bm h) & = \int \de\l \, e^\l \, \left(- 1 + e^{- \int \de a \, \redv(\bm h(a) + \l )} \right)  \ .
\end{split}\eeq
Here $\De\bm h = \De h \De \hat h$
is the usual measure (no constraint) for path integrals of a scalar function of time
and the measure $\DD_{\bm Q}\bm h$ is normalised to 1. 
We recall that $\bm Q(a,b) = (2d/\s^2) \la\bm x(a) \cdot\bm x(b) \ra$, and we use that in 
large $d$ all $\n_i \sim \n$ as discussed in Section~\ref{sec:Lagdyn}.
The operator $\bm\KK(a,b)$ is just a rewriting of the kinetic term, defined by the equality
\beq
\frac12  \int \de a\,\bm x(a) \cdot [\partial^2_a + \nu ]\bm x(a)
= \frac12  \int \de a \de b\, [\partial^2_a + \nu ] \bm\d(a,b) \frac{\s^2}{2d}\bm Q(a,b)
= \frac{d}2 \int \de a \de b\, \bm\KK(a,b)\bm Q(a,b) 
\eeq
which gives
\beq
\bm\KK(a,b) = \frac{\s^2}{2d^2} [\partial^2_a + \nu ] \bm\d(a,b) =  [\wh\partial^2_a + \wh\nu ]\bm \d(a,b) \ ,
\eeq
where $\wh \partial^2_a$ is identical to $\partial^2_a$ but with a rescaled coefficient $\wh \g = \frac{\s^2}{2d^2}\g$.

\subsection{Saddle-point equation for the dynamical order parameter}

To obtain the saddle-point equation for $\bm Q(a,b)$ we must impose the condition $\d\SS(\bm Q)/\d\bm Q(a,b)=\bm 0$ together
with $\bm Q(a,a)=R^2$. As in the replica calculation, to impose the constraint we introduce a Lagrange multiplier
and optimize $\SS(\bm Q) - \frac{d}2 \int \de a\,\bm{\d\wh\n}(a) \bm Q(a,a)$. This amounts to sum the microscopic multiplier $\wh\n$ with an
additional term $\bm{\d\wh\n}(a)$. We will simply call $\bm{\wh\m}$ the sum of the two terms and we thus 
redefine the kinetic operator as
\beq
\bm\KK(a,b) \equiv   [\wh\partial^2_a + \bm{\wh\mu}(a) ] \bm\d(a,b) \ .
\eeq
The saddle-point equation is therefore identical to the one obtained in the replica calculation:
\beq\label{eq:HSSUSY}
\bm 0 = - \bm\KK(a,b) + 
\bm Q^{-1}(a,b) \left[ 1  -\frac12 \wh\f e^{-\D_{\rm liq}/2} \int \DD_{\bm Q}\bm h \,  \Psi(\bm h) \right]
+ \frac12 \wh\f e^{-\D_{\rm liq}/2} \int \de c \de e\,\bm Q^{-1}(a,c) \int \DD_{\bm Q}\bm h \,  \Psi(\bm h) \, \bm h(c)\bm h(e)  \, \bm Q^{-1}(e,b)  \ .
\eeq
As for replicas, we have
\beq\begin{split}
\int \DD_{\bm Q}\bm h& \,
e^{  - \int \de c \, \redv(\bm h(c) + \l )}
 \int \de c \de e\,\bm Q^{-1}(a,c)\bm h(c)\bm h(e)  \,\bm Q^{-1}(e,b) \\ 
 &= \int \De\bm h \,
 e^{- \int \de c \, \redv(\bm h(c) + \l )} \left[ \frac{\d^2}{\d\bm h(a) \d\bm h(b)} + \bm Q^{-1}(a,b) \right]
e^{  -\frac{1}2 \int \de c \de e \,\bm h(c)\bm Q^{-1}(c,e)\bm h(e)} \\
&=\int \DD_{\bm Q}\bm h \,
\left[ \frac{\d^2}{\d\bm h(a) \d\bm h(b)} +\bm Q^{-1}(a,b) \right] e^{- \int \de c \, \redv(\bm h(c) + \l )} \ .
\end{split}\eeq
Introducing $f_\l(\bm h) = -\redv'(\bm h+ \l)$, we get
\beq\label{eq:app555}
\begin{split}
\int \de c \de e\,& \bm Q^{-1}(a,c) \int \DD_{\bm Q}\bm h \, \Psi(\bm h) \, \bm h(c)\bm h(e)  \,\bm Q^{-1}(e,b)
=\bm Q^{-1}(a,b) \int \DD_{\bm Q}\bm h\, \Psi(\bm h) + \int \de \l\, e^\l
\int \DD_{\bm Q}\bm h \,
 \frac{\d^2}{\d\bm h(a) \d \bm h(b)}  e^{- \int \de c \, \redv(\bm h(c) + \l )} \\
 &=\bm Q^{-1}(a,b) \int \DD_{\bm Q}\bm h \Psi(\bm h) +  
 \int \de \l\, e^\l \int \DD_{\bm Q}\bm h \, e^{- \int \de c \, \redv(\bm h(c) + \l )}
[ f_\l(\bm h(a)) f_\l(\bm h(b))  + f_\l'(\bm h(a)) \bm\d(a,b) ] \ .
\end{split}
\eeq
Let us define the dynamical measure
\beq\label{eq:app634}
\la \OO \ra_v = \int \de \l \, e^{\l - \Dl/2} \int \De \bm h \,
e^{  -\frac{1}2 \int \de a \de b \,\bm h(a)\bm Q^{-1}(a,b)\bm h(b) - \int \de a \, \redv(\bm h(a) + \l )} \OO \ ,
\eeq
and
\beq\label{eq:MABSUSYf}
\bm M(a,b) = \frac{\wh\f}2    \la f_\l(\bm h(a)) f_\l(\bm h(b))  \ra_v \ ,
\hskip20pt
\bm{\d\mu}(a) = -\frac{\wh\f}2   \la  f_\l'(\bm h(a))   \ra_v \ .
\eeq
Plugging Eq.~\eqref{eq:app555} and Eq.~\eqref{eq:MABSUSYf} into Eq.~\eqref{eq:HSSUSY} we obtain
\beq\label{eq:HSSUSY2}
\bm0 = - \bm\KK(a,b) + 
\bm Q^{-1}(a,b) +\bm M(a,b)
- \bm{\d\m}(a) \bm\d(a,b) \ ,
\eeq
which can also be written in the equivalent form:
\beq\label{eq:HSSUSY3}
\begin{split}
\bm0 &= - [\wh\partial^2_a + \bm{\wh\mu}(a) + \bm{\d\mu}(a) ]\bm Q(a,b)  + 
\bm\d(a,b) +  \int \de c\,\bm M(a,c)\bm Q(c,b)  \ . \\
\end{split}\eeq
Defining $\bm\LL(a,b) = \bm\KK(a,b) + \bm{\d\m}(a) \bm\d(a,b) = [\wh\partial^2_a + \bm{\wh\mu}(a) + \bm{\d\m}(a) ] \bm\d(a,b)$ and
using the first of Eqs.~\eqref{eq:HSSUSY2} in Eq.~\eqref{eq:app634} one has
\beq\label{eq:HSSUSYmeasure}
\la \OO \ra_v = \int \de \l \, e^{\l - \Dl/2} \int \De\bm h \,
e^{  -\frac{1}2 \int \de a \de b \, \bm h(a) [ \bm\LL(a,b) -\bm M(a,b)  ] \bm h(b) - \int \de a \, \redv(\bm h(a) + \l )} \OO
\eeq
Eqs.~\eqref{eq:MABSUSYf}, \eqref{eq:HSSUSY3} and \eqref{eq:HSSUSYmeasure} provide a simple and compact
expression for the saddle point equation in SUSY form.

\subsection{Equilibrium dynamics}

From the SUSY saddle point equations one can in principle derive the dynamical equation in full generality, i.e. without assuming
equilibrium~\cite{Cu02}. The resulting equations can describe all the dynamical regimes of the system including transient regimes
and the long time aging behaviour. Here however, in order to simplify the derivation, we specialize to the equilibrium regime. 

\subsubsection{Equation for the position time-autocorrelation in terms of the memory kernel}

We assume that the system is in equilibrium at all times.
The SUSY correlators have thus the equilibrium form~\cite{Cu02}:
\beq\label{eq:SUSYeq}
\begin{split}
\bm Q(a,b) &= C(t_a-t_b) + \th_a\bth_a R(t_b-t_a) + \th_b\bth_b R(t_a - t_b) \ , \\
\bm M(a,b) &= M_C(t_a-t_b) + \th_a\bth_a M_R(t_b-t_a) + \th_b\bth_b M_R(t_a - t_b) \ ,
\end{split}\eeq
and the fluctuation-dissipation theorem (FDT) further gives $R(t) = -\b \th(t) \dot C(t)$. 
With these hypotheses, Eq.~\eqref{eq:HSSUSY3} shows that $\bm{\wh\mu}(a) + \bm{\d\mu}(a)=\wh\mu + \d\mu$ are real numbers, independent of time at equilibrium\footnote{
This is because, if both $\bm M$ and $\bm Q$ do not have a $\th_a\bth_a\th_b\bth_b$ term, then either does $\int \de c\,\bm M(a,c)\bm Q(c,b)$. 
Besides, $\bm\d(a,b)$ has no $\th_a\bth_a\th_b\bth_b$ term. Thus, to fulfill Eq.~\eqref{eq:HSSUSY3}, the first term must not have this component either. 
This implies that $\bm{\wh\mu}(a) + \bm{\d\mu}(a) = \wh\mu(t) + \d\mu(t)$. Then, we know that in equilibrium one-time quantities are constant in time.
}.
Therefore the real component (with no Grassmann variables) of Eq.~\eqref{eq:HSSUSY3} becomes, using Eq.~\eqref{SUSYproduct}:
\beq\label{eq:HSfinal0}
\wh \g \dot C(t) = - [\wh\m+ \d\m - \b M_C(0)]C(t) - \b \int_0^t \de s\, M_C(t-s) \dot C(s) \ .
\eeq
This equation gives the correlation $C(t)$ in terms of the memory kernel $M_C(t)$. It has the form of a MCT equation~\cite{Go09}; note
that the integral is restricted in $[0,t]$ so the previous history disappears.

To compute $\wh\m + \d\mu$, we observe that Eq.~\eqref{eq:HSfinal0} at $t=0$ gives
\beq
\wh\g \dot C(0) = - [\wh\mu + \d\m - \b M_C(0)] \Dl \ .
\eeq
$\dot C(0)$ can be computed using the non-interacting dynamics because interaction effects are irrelevant at very short times.
From Eq.~\eqref{eq:HSLang} and \eqref{eq:HSmu} we have $\g \dot x_i = -(dT/R^2) x_i + \h_i$ which gives
$x_i(t) = x_i(0) \exp\left[ - \frac{dT}{\g R^2} t \right] + \{\text{a linear term in the noise}\}$ and thus
\beq
C(t) = C(0) \exp\left[ - \frac{dT}{\g R^2} t \right] = \Dl \exp\left[ - \frac{T}{\wh\g \Dl} t \right] 
\ .
\eeq
Hence $-\wh\g \dot C(0) = T$ and we get
\beq\label{eq:HSmurel}
\frac{T}{\Dl} = \wh\mu + \d\m - \b M_C(0)
\eeq
Plugging this in Eq.~\eqref{eq:HSfinal0} we get the final result:
\beq\label{eq:HSfinalC}
\wh \g \dot C(t) = - \frac{T}{\Dl} C(t) - \b \int_0^t \de s M_C(t-s) \dot C(s) \ .
\eeq

\subsubsection{Self-consistent equation for the memory kernel}

The self-consistent equation for $M_C(t)$ follows from Eqs.~\eqref{eq:MABSUSYf}
and \eqref{eq:HSSUSYmeasure}.
The average over $h$ that appears in Eq.~\eqref{eq:HSSUSYmeasure} corresponds~\cite{Cu02} to an average over the effective Langevin process
\beq\label{eq:HSeffh_app}
\begin{split}
\wh\g \dot h(t) &= -(\wh\m +\d\m) h(t) + \int_{-\io}^t \de s M_R(t-s) h(s) + \xi(t) -   \redv'(\l + h(t)) \\
&= - [\wh\mu + \d\m -\b M_C(0)] h(t) - \b \int_{-\io}^t \de s M_C(t-s) \dot h(s) + \xi(t) -  \redv'(\l + h(t)) \ , \\
\langle \xi(t) \xi(t') \rangle & = 2 T \wh \g \d(t-t') + M_C(t-t') \ .
\end{split}
\eeq
Taking in Eqs.~\eqref{eq:MABSUSYf}
the terms without any $\th$ variable, $f_\l(h(a))$ is simply the force
$f_\l(t) = - \redv'(\l + h(t))$, and we thus get the self-consistent equation for the memory kernel:
\beq\label{eq:HSfinal2}
M_C(t-t')  = \frac{ \wh\f}2 e^{-\D_{\rm liq}/2}\int \de\l\, e^\l  
 \langle f_\l(t) f_\l(t') \rangle_h \ ,
\eeq
where the average $\langle \bullet \rangle_h$ is over the stochastic process in Eq.~\eqref{eq:HSeffh_app}.
Using the argument of Appendix~\ref{app:B}, starting from
any configuration at time $-\io$ is equivalent to starting at equilibrum at $t=0$ for the purpose of computing
equilibrium correlations at positive times. Therefore Eq.~\eqref{eq:HSeffh_app} is equivalent to,
using Eq.~\eqref{eq:HSmurel}:
\beq\label{eq:HSfinal1}
\begin{split}
\wh\g \dot h(t) 
&= - \frac{T}{\Dl} h(t) - \b \int_{0}^t \de s M_C(t-s) \dot h(s) + \xi(t) - \redv'(\l + h(t)) \ , \\
\langle \xi(t) \xi(t') \rangle & = 2 T \wh \g \d(t-t') + M_C(t-t') \ , \\ 
\PP_{0}(h_0,\l) & = \frac1{\ZZ_0(\l)} e^{-\b  \redv(h_0 + \l) - \frac{h_0^2}{2\Dl}  } =\frac1{\ZZ_0(\l)} e^{-\b \HH_0(h_0,\l)}  \ ,
\end{split}
\eeq
where $\PP_0$ is the equilibrium probability measure with which the initial ``position'' $h_0$ is picked at $t=0$.
These are our final expressions for the dynamical equations.
A numerical procedure (\ie the logical steps) to obtain the memory kernel could be
\begin{itemize}
\item Start with a guess for $M_C(t)$
\item Solve the process in Eq.~\eqref{eq:HSfinal1} to compute the correlation that
appear in Eq.~\eqref{eq:HSfinal2}.
\item Use Eq.~\eqref{eq:HSfinal2} to obtain a new guess for $M_C(t)$
\item Iterate until convergence
\item Use Eq.~\eqref{eq:HSfinalC} to obtain $C(t)$ from the memory kernel.
\end{itemize}

\subsection{Thermodynamic limit: dynamical equations in terms of the mean square displacement}

Because we know that $\Dl \to \io$ in the thermodynamic limit, 
it is interesting to eliminate it from the dynamical equations. This is possible if we consider finite times $t$ for which the system
cannot explore the whole volume.
First we note that the equal time value of $M_C$ is given by
\beq\label{eq:Mequaltimes}
M_C(0)  = 
 \frac{\wh\f}2 \int \de\l \,e^{\l - \Dl/2} \,
\frac{\int \de h\, e^{-\b \HH_0} \redv'(h+\l)^2}
{\int \de h\, e^{-\b\HH_0} } 
=  \frac{\wh\f}2 \langle f_\l(h)^2 \rangle_v
\sim 
 \frac{ \wh\f}2
\int  \de y\, e^{  -\b \redv(y) + y} \, \redv'(y)^2
\ ,
\eeq
where we used Eq.~\eqref{eq:114}.
Therefore, $M_C(0)$ is finite (for a non-singular potential\footnote{For Hard Spheres, 
$M_C(0)$ is divergent, but here we are interested in the behavior of $M_C(t)$ in the thermodynamic limit; 
the Hard Sphere limit can be taken after the thermodynamic limit, and in this case
the divergence of $M_C(0)$ gives rise to a short-time singularity in the memory that can be treated without problems.
}) 
in the thermodynamic limit $\Dl \to \io$; 
we expect that $M_C(t)$ is a monotonically decreasing function
and we will see that $M_C(t\to\io)$ is also finite (actually, it vanishes in the liquid phase).
We thus conclude that
$M_C(t)$ is finite at all times for $\Dl \to \io$.
We then consider the self-consistent equations~\eqref{eq:HSfinal2} and \eqref{eq:HSfinal1} for $M_C(t)$, in which
we change variables to $y = h+\l$ and
we introduce the average $\la \OO \ra_{y_0}$ over the dynamics constrained to the initial condition $y_0$. We obtain
\beq\begin{split}
M_C(t-t')  &=  \frac{\wh\f}2 \int \de \l\, e^{-\frac{\Dl}{2}\left(1-\frac{\l}{\Dl}\right)^2} \int \de y_0\, \frac1{\ZZ_0(\l)} e^{-\b  \redv(y_0)+\frac\l\Dl y_0 - \frac{y_0^2}{2\Dl}  }  
 \langle \redv'(y(t)) \redv'(y(t')) \rangle_{y_0} \ , \\
\wh\g \dot y(t) 
&= - \frac{T}{\Dl} (y(t) - \l) - \b \int_{0}^t \de s\, M_C(t-s) \dot y(s) + \xi(t) - \redv'(y(t)) \ , \\
\langle \xi(t) \xi(t') \rangle & = 2 T \wh \g \d(t-t') + M_C(t-t') \ , 
\end{split}
\eeq
Now following the same reasoning as in Eq.~\eqref{eq:114} we see that only finite values of $y_0$ can contribute
to the average in the first line. Then the integral over $\l$ is dominated by $\l = \Dl$ and terms in $y/\Dl$ are negligible. 
We obtain\footnote{Note that in Eq.~\eqref{eq:MCnoDl} $\langle \redv'(y(t)) \redv'(y(t')) \rangle_{y_0}$ is small if $y_0$ is large,
because the potential falls quickly to zero for positive $y_0$, hence the integral over $y_0$ is convergent.}
\beq\label{eq:MCnoDl}
\begin{split}
M_C(t-t')  &=  \frac{\wh\f}2 \int \de y_0\, e^{-\b  \redv(y_0) + y_0  }  
 \langle \redv'(y(t)) \redv'(y(t')) \rangle_{y_0} \ , \\
\wh\g \dot y(t) 
&= T - \b \int_{0}^t \de s\, M_C(t-s) \dot y(s) + \xi(t) - \redv'(y(t)) \ , \\
\langle \xi(t) \xi(t') \rangle & = 2 T \wh \g \d(t-t') + M_C(t-t') \ , 
\end{split}
\eeq
Finally, we can define the time-dependent mean square displacement $\D(t) = \Dl - C(t)$.
For finite times, $\D(t)$ is finite and from
Eq.~\eqref{eq:HSfinalC} we get
\beq\label{eq:HSfinalD}
 \wh \g \dot \D(t) =  T - \b \int_0^t \de s\, M_C(t-s) \dot \D(s) \ .
\eeq
Eqs.~\eqref{eq:MCnoDl} and \eqref{eq:HSfinalD} are the final dynamical equations for $\Dl\to\io$, written in terms of $\D(t)$.

To conclude this section, let us note that 
an interesting alternative self-consistent equation for $M_C(t)$ is obtained from Eq.~\eqref{eq:MCnoDl} if we choose $t'=0$, which is possible because
the dynamics starts in equilibrium. Let us also assume that $\redv(y)=0$ for $y\geqslant 0$.
We obtain
\beq\begin{split}
M_C(t)  &= \frac{\wh\f}2 \int_{-\io}^0 \de y_0\, e^{-\b  \redv(y_0) + y_0  }  \redv'(y_0)
 \langle \redv'(y(t))  \rangle_{y_0} \\
 & = 
-\frac{\wh\f T}2 \left\{ \left[ e^{-\b  \redv(y_0) + y_0  }\langle \redv'(y(t))  \rangle_{y_0}  \right]_{-\io}^0
- \int_{-\io}^0 \de y_0\, e^{-\b  \redv(y_0)}  \frac{\partial}{\partial y_0} \left[ e^{y_0  }  
 \langle \redv'(y(t))  \rangle_{y_0} \right] 
 \right\} \\
  & = 
-\frac{\wh\f T}2 \left\{  \langle \redv'(y(t))  \rangle_{y_0=0} 
- \int_{-\io}^0 \de y_0\, e^{-\b  \redv(y_0)}  \frac{\partial}{\partial y_0} \left[ e^{y_0  }  
 \langle \redv'(y(t))  \rangle_{y_0} \right] 
 \right\}  
 \ .
\end{split}\eeq
For Hard Spheres we have $\redv(y) \to\io$ for $y<0$ and the second term in the last line can be neglected,
so we obtain a very simple expression: $M_C(t) =- \frac{\wh\f T}2 \langle \redv'(y(t))  \rangle_{y_0=0} $.
Note that $ \langle \redv'(y(t))  \rangle_{y_0}$ is not independent of time, 
firstly because the dynamics in Eq.~\eqref{eq:MCnoDl} has a drift term proportional to $T$, and secondly because
if we impose a fixed initial condition $y_0=0$ and consider finite times the
system is not in stationary state anyway.

\subsection{Diffusion coefficient, viscosity, and Stokes-Einstein relation}\label{sub:diff}

From Eq.~\eqref{eq:HSfinalD} we obtain an expression for the diffusion coefficient.
Let us assume that $\D(t)$ has some structure at short times followed by a linear regime
$\D(t) \sim \wh D t$ at large times, while $M_C(t)$ decays to zero over a finite time. Then
we have $\underset{t\to\io}{\lim}\, \int_0^t \de s\, M_C(t-s) \dot \D(s) = \wh D \int_0^\io \de s\, M_C(s)$
and we obtain from Eq.~\eqref{eq:HSfinalD}:
\beq
\wh\g \wh D = T - \b \wh D \int_0^\io \de s\, M_C(s) 
\hskip20pt
\Rightarrow
\hskip20pt
\wh D = \frac{T}{\wh\g + \b \int_0^\io \de s\, M_C(s)} \ .
\eeq
At low density $M_C =0$ and we recover the diffusion coefficient $\wh D = T/\wh\g$ of the free dynamics.
Upon increasing density, $M_C$ increases and the diffusion coefficient decreases. At the dynamical transition,
where a finite plateau of $M_C$ emerges, $\int_0^\io \de s M_C(s)$ diverges and the diffusion coefficient vanishes.
Going back to non-scaled variables, we have for the diffusion coefficient\footnote{
Again, recovering the diffusive behaviour of the liquid phase might seem surprising 
since our scaling hypotheses in the derivation of Section~\ref{sec:III} constrain a trajectory to vibrate 
around an initial position with amplitude $O(1/d)$ and two trajectories to be almost at contact, 
which is not the case in the liquid phase. 
The crucial point is that diffusive behavior sets in when $\D$ is still of order one, or equivalently, the mean-square displacement
is of the order of $1/d$.
More details are given in Appendix~\ref{app:MKHS}.}:
\beq\label{eq:diffusion}
D = \underset{t\to\io}{\lim}\, \frac{ \DE(t) }{2 d t} = \underset{t\to\io}{\lim}\, \frac{ \s^2 \D(t) }{2 d^2 t} = \frac{\s^2 \wh D}{2 d^2} 
= \frac{T}{\g + (2 d^2 /\s^2) \b \int_0^\io \de s\, M_C(s)}
\ .
\eeq

The viscosity can be deduced from the auto-correlation function of the stress~\cite{hansen}. 
We have seen in Section~\ref{sec:stress} that this quantity actually coincides with $M_C(t)$ in $d\to\io$.
Here we follow the conventions of~\cite{Yo12}, hence we neglect the kinetic term of the stress tensor
(i.e. we neglect the contribution of the ideal gas, which is irrelevant in the glassy regime) and define a viscosity $\h$ 
as follows\footnote{Following the convention of~\cite{Yo12}, here $\h$ is the mass times the kinematic viscosity, $\h = m \h_K$, 
or the shear viscosity divided by the number density, $\h = \h_S / \r$; 
i.e. it has units of kg m$^2/$s.}:
\beq\label{eq:viscosity}
\h = \b \int_0^\io \de s\, N \la \s(s) \s(0) \ra = d \, \b \int_0^\io \de s\, M_C(s) \ .
\eeq
Putting together Eq.~\eqref{eq:diffusion} and Eq.~\eqref{eq:viscosity} we obtain
\beq
D = \frac{T}{\g + \frac{2 d}{\s^2} \h} \ .
\eeq
This relation is interesting. At low densities, $\h\to 0$ while $D \to T/\g$. Upon approaching the glass transition,
$D \to 0$ and $\h \to\io$ with constant $D\h = \frac{T \s^2}{2 d} = \frac{T}{\z}$. Hence, the Stokes-Einstein
relation is satisfied with an apparent Stokes drag $\z = \s^2/(2d)$. 
Note that expressing the Stokes-Einstein relation in terms of the shear viscosity $\h_S = \r \h$ we obtain
\beq
D \h_S = \frac{T \s^2 \r}{2d} = \frac{T}{\z_S} \ ,
\hskip20pt
\z_S = \frac{2d}{\r\s^2} = \frac{\s^{d-2}}{\wh\f} \frac{2 \pi^{d/2}}{\G(d/2+1)} \ .
\eeq
This scaling of the Stokes drag is very close to the hydrodynamic one~\cite{CCJPZ13}.
Also, the prediction that $D \h_S \propto \r$ is very well satisfied by the data of~\cite[Fig.7b]{CCJPZ13}
for high densities and large dimension.

\section{Connection between statics and dynamics}
\label{sec:VII}

In this section, we show that the dynamical equations give the same result as the replica
equations.
From replicas we can compute three reference values: the equal-time value $M_C(0)$, the
long time limit $M_C(\io)$, and the plateau value in the glass phase. We show here
that dynamics gives the same values~\cite{Ku92,PR13}.

\subsection{Equal time limit}

We first discuss the equal time limit.
Eq.~\eqref{eq:HSmurel} gives
\beq
\begin{split}
 \wh\mu - \frac{T}\Dl & = - \d\m +
\b M_C(0)  = 
 \frac12 \wh\f e^{-\D_{\rm liq}/2}  \int \de\l\, e^\l \,  \la  f_\l'(h)   \ra_{\HH_0}
+ \frac{\b}2 \wh\f e^{-\D_{\rm liq}/2}\int \de\l\, e^\l 
 \langle f_\l(h) \, f_\l(h) \rangle_{\HH_0} \ ,
\end{split}
\eeq
which is equivalent to the static result in Eq.~\eqref{eq:mustatic}.

\subsection{Long time limit in the liquid phase}
\label{sec:LTLLP}

Second, we consider the long time limit of the memory function in the liquid phase, where
we assume a complete decorrelation of the system.
From Eq.~\eqref{eq:HSfinalC} we have
\beq\label{eq:pMC}
-\frac{T}{\Dl} C(\io) - \b M_C(\io) [ C(\io) - \Dl ] =0 
\hskip5pt
\Rightarrow
\hskip5pt
C(\io) = \frac{\b^2 \Dl^2 M_C(\io)}{1 + \b^2 \Dl M_C(\io)}
\hskip5pt
\Leftrightarrow
\hskip5pt
\b^2 M_C(\io) = \frac1\Dl \frac{C(\io)}{\Dl - C(\io)}
\ .
\eeq
Because we assume a complete decorrelation, at long times 
the correlation $\langle f_\l(t) f_\l(0) \rangle$ that appears in Eq.~\eqref{eq:HSfinal2}
becomes the product of equilibrium averages:
$\langle f_\l(t) \rangle \langle f_\l(0) \rangle = \langle f_\l(h) \rangle_{\HH_0}^2$.
We thus have:
\beq
M_C(\io) = \frac12 \wh\f e^{-\D_{\rm liq}/2}\int \de\l\, e^\l  \langle f_\l(h) \rangle_{\HH_0}^2 =
\frac{T^2}{2 \Dl^2} \wh\f e^{-\D_{\rm liq}/2}\int \de\l\, e^\l  \langle h \rangle_{\HH_0}^2
 \ , 
\eeq
which can be directly identified with the static expression~\eqref{eq:av2RS}.
We now focus for simplicity on Hard Spheres, for which
\beq
\la h \ra_{\HH_0} = \frac{
\int_{-\l}^\io \de h \, e^{- \frac{h^2}{2\Dl}  } \, h
}{
\int_{-\l}^\io \de h \, e^{- \frac{h^2}{2\Dl}  } 
} \ ,
\eeq
and after a short computation we find
\beq
\b^2 M_C(\io) =  \frac{ \wh\f}{2 \sqrt{\Dl}} e^{-\D_{\rm liq}/4}
\int_{-\io}^\io \de t \, \frac{e^{-t^2}}{ \left( \int_{-t-\Dl/2}^\io \de s\, e^{-s^2/2} \right)^2 }
=\frac{\wh\varphi}{4\sqrt\pi}\frac{e^{-\D_{\rm liq}/4}}{\sqrt{\Dl}}
\eeq
which shows that both $M_C(\io)$ and $C(\io)$ go exponentially to zero for $\Dl \to \io$,
consistently with the static result of Sec.~\ref{sec:SPrepRS}, 
see Eqs.~\eqref{eq:75} and~\eqref{eq:MCTreplica2} for $a\neq b$, which corresponds to a long-time limit in the replica symmetric language.

\subsection{Calculation of the plateau}

Finally, we consider the ``plateau'' value of the correlation function. 
In the liquid phase close to the glass transition, we are in a 
situation where there is a strong separation of time scales between the fast motion (with characteristic time scale $\tau_f$) 
and the slow diffusive motion (with characteristic time scale $\t_s \sim 1/D$)~\cite{CK00}.
The plateau corresponds to an intermediate regime of time difference $\t_f \ll t \ll \t_s$: $t$ is much larger than that the fast vibrational dynamics, 
but much shorter than the diffusive regime (Fig.~\ref{fig:delta}). 
In the glass phase, the diffusion is arrested and the plateau now corresponds to the long time limit of the memory function.
To compute the plateau, 
we split the correlation and the memory function as 
\beq
M_C(t-t') = M^f_C(t-t') + M^s_C(t-t') \ ,
\hskip20pt
C(t-t') = C^f(t-t') + C^s(t-t') \ ,
\eeq
that decay on time scales $\t_f \ll \t_s$, respectively.
We define $C_{\rm P} = C^s(0) = \Dl - \D_1$ where $\D_1$ is the plateau of the mean square displacement. 
Eq.~\eqref{eq:pMC} also holds for the plateau values, hence we obtain
\beq\label{eq:MCp}
\b^2 M_{\rm P} = \frac1\Dl \frac{C_{\rm P}}{\Dl - C_{\rm P}} =  \frac1{\D_1} - \frac1{\Dl}  \ .
\eeq
Following~\cite[Sec.4.1]{CK00} (see also~\cite{MKZ15}), one can show that on time scales $\t_f \ll t - t' \ll \t_s$, one has a quasi-stationary regime of Eq.~\eqref{eq:HSfinal1} 
described by a probability distribution $\PP(h|\h)$ where $\h$ is a slow variable with distribution $\PP(\h)$. We have
\beq
\PP(h | \h) = \frac{1}{\ZZ(\h)} e^{-\b (\HH_0 + \b M_{\rm P} \frac{h^2}2 - \sqrt{M_{\rm P}} \h h)} \ ,
\hskip15pt \ZZ(\h) =\int\de h\,e^{-\b (\HH_0 + \b M_{\rm P} \frac{h^2}2 - \sqrt{M_{\rm P}} \h h)} =e^{-\b \afunc{F}(\h)} \ , \hskip15pt
\PP(\h) = \frac{1}{\ZZ} e^{-\b \afunc{F}(\h) - \frac{\h^2}{2}} \ .
\eeq
Note that
\beq\begin{split}
&\ZZ = \int \de\h\, e^{ - \frac{\h^2}{2 }} \int \de h\, e^{-\b (\HH_0 + \b M_{\rm P} \frac{h^2}2 -\sqrt{M_{\rm P}} \h h)} = \sqrt{2\pi} \ZZ_0(\l) \ , \\
&\la \OO(h) \ra_{\rm P} = \int \de\h \de h\, \PP(\h) \PP(h | \h) \OO(h) = \frac1{\ZZ_0(\l)} \int \de h\, e^{-\b\HH_0(h,\l)} \OO(h) = \la \OO(h) \ra_{\HH_0} \ ,
\end{split}\eeq
which is consistent with the stationary measure of Eq.~\eqref{eq:HSfinal1}.
Also, recalling Eq.~\eqref{eq:H1} and \eqref{eq:MCp}, we get
\beq\begin{split}
& \HH_1 = \HH_0 + \b M_{\rm P} \frac{h^2}2 - \sqrt{M_{\rm P}} \h h \ , \hskip20pt
\PP(h | \h) = \PP_1(h) \ ,
\hskip20pt
\PP(\h) = \frac{\ZZ_1}{\ZZ_0} \frac{e^{- \frac{\h^2}{2}}}{\sqrt{2\pi}} \ . 
\end{split}\eeq
Correlation functions in the plateau regime when $\t_f \ll t-t' \ll \t_s$ are given by
\beq
\la f_\l(t) f_\l(t') \ra_h = \int \de \h \,\PP(\h) \left( \int \de h\, \PP(h | \h) f_\l(h) \right)^2 \ .
\eeq
Thus we get from Eq.~\eqref{eq:HSfinal2}
\beq\begin{split}
\b^2 M_{\rm P}  &= \frac1{\D_1} - \frac1{\Dl} =  \frac12 \wh\f e^{-\D_{\rm liq}/2}\int \de\l\, e^\l  
\int \DD \h\, \frac{ \ZZ_1}{ \ZZ_0} 
\left( \frac{1}{\ZZ_1} \int \de h\,  e^{-\b \HH_1} \b f_\l(h) \right)^2
 \end{split}
\eeq
which coincides exactly with the static result given in Eq.~\eqref{eq:plateau_static}.

Note that in the glassy regime, both the present solution $M_{\rm P}$ and the solution $M_C(\io)$ discussed
in the previous Section~\ref{sec:LTLLP} formally exist as solutions for the long-time limit of $M_C(t)$; however,
the dynamics always selects the solution with the largest value of $M$, which is $M_{\rm P}$ (see e.g. the discussion in~\cite{CC05}).

\section{Conclusions}

In this paper we presented a parallel derivation of the glassy thermodynamics, using replicas, and of the dynamics, using supersymmetry, of an infinite-dimensional
particle system. 
We introduced an irrelevant quenched disorder~\cite{Mo95} that was helpful to derive the entropy functional without having to justify a truncation of the virial 
expansion, as originally done in~\cite{FRW85,WRF87,FP99,EF88} for liquids and in~\cite{PZ10,MKZ15} for glasses. 
We discussed a derivation of the replicated thermodynamics that is simpler but equivalent to previous ones~\cite{KPZ12,KPUZ13,CKPUZ13}, and
that, contrary to previous ones, can be easily generalised to the supersymmetric formalism. In this way we can derive dynamical equations along the same lines
and straightforwardly show the equivalence of thermodynamic and dynamic results in the glassy regime.

In previous papers~\cite{PZ10,KPZ12,KPUZ13,CKPUZ13,nature,RUYZ14,MKZ15}, focusing in particular on the hard sphere potential, these equations have been used to
derive many observables characterising the glassy regime, namely:
\begin{enumerate}
\item The full time-dependent correlations in the liquid phase, and in principle also the out-of-equilibrium correlations in the aging
regime~\cite{MKZ15}.
\item The dynamical  
transition density~\cite{KPUZ13,MKZ15}, at which the liquid phase becomes infinitely viscous and ergodicity is broken,
and the so-called MCT parameter $\l$ that controls dynamic criticality at the transition~\cite{Go09}.
\item The Kauzmann transition~\cite{PZ10}, where the number of metastable states becomes sub-exponential, giving rise
to an ``entropy crisis" and a second order equilibrium phase transition\footnote{Note that in the MK model 
the additional term $N!$ due to particle distinguishability (see Section~\ref{sec:disting}) induces an additional term $\log N$ in the entropy
of metastable states, which
shifts the Kauzmann transition to infinite density.
In the normal system (consider e.g. Hard Spheres) this factor is replaced by a $\log d$ term, which shifts the Kauzmann transition to
values of $\wh \f_{\rm K} \sim \log d  \gg\wh \f_{\rm d}$, where $\wh\f_{\rm d}\approx 4.8$ is the dynamical transition scaled density~\cite{PZ10}.
}.
\item The Gardner transition line, that separates a region where glass basins are stable from a region where they are broken in a complex structure of metabasins~\cite{KPUZ13,CKPUZ13,RUYZ14}.
\item The density region where jammed packings exist (also known as ``jamming line'' or ``J-line''~\cite{PZ10}), 
which is delimited by the threshold density and the glass close packing density~\cite{PZ10}.
\item The equation of state of glassy states, computed by compression and decompression of equilibrium glasses~\cite{RUYZ14}.
\item The response of the glass state to a shear strain~\cite{YZ14,RUYZ14}.
\item The long time limit of the mean square displacement in the glass 
(the so-called Edwards-Anderson order parameter)~\cite{CKPUZ13}.
\item The behaviour of correlation function, structural $g(r)$ and non-ergodicity factor of the glass~\cite{PZ10,CKPUZ13,MKZ15}.
\item The probability distribution of the forces in a packing, and the average number of particle contacts~\cite{CKPUZ13}.
\end{enumerate}
In the present paper, the equations have been obtained for a generic scaled inter-particle potential $\redv(h)$, therefore similar computations can be done
for infinite-dimensional versions of, e.g. the Lennard-Jones or WCA potentials. Some results in this directions have already been obtained for a square-well
potential in~\cite{SZ13}. 
To sum up, one can compute almost every interesting observable of almost every potential of interest in the infinite-dimensional limit.
Hopefully, this will also be helpful to start investigating $1/d$ corrections systematically.

\acknowledgments

We wish to thank C.~Rainone, G.~Szamel and P.~Urbani for many useful discussions on the content of this manuscript. T. M. acknowledges funding from a fondation CFM grant.


\clearpage

\appendix

\section{Useful mathematical formulae}
\label{app:formulae}

We collect here a few mathematical definition and general formulae that are used throughout the paper.

\subsection{Definition of basic quantities of the model}
\label{eqapp:liquid}

\subsubsection{Static quantities}

\begin{center}
\begin{tabular}{| l | l |}
\hline
$d$ & Dimension of space \\
$N$ & Number of particles \\
$\SSS$ & A $d+1$-dimensional hypersphere of radius $R$ \\
$x \in \SSS$ & Position of a particle \\
$\Omega_d = \frac{2 \pi^{d/2}}{\G(d/2)} $ & $d$-dimensional solid angle \\
$V = $vol$(\SSS) = \Omega_{d+1} R^{d}$ & Volume of $\SSS$ in $\RRR^{d+1}$ \\
$\s$ & Particle diameter \\
$\VV_d(\s) = V \int \de\RR\, \th(\s - | x - \RR x |) = \int \de x\, \th(\s - | x |)$ & Volume excluded by a particle on the surface of $\SSS$ \\
$\f = N \VV_d(\s)/(2^d V) $ & Packing fraction \\
$\wh \f = 2^d \f /d $ & Scaled packing fraction \\
$r = |x - y|$ & Euclidean distance between two particles $x$ and $y$ \\
$v(r)$ & Interaction potential energy between two particles \\
$h = d (r - \s)$ & Scaled ``interparticle gap'' \\
$\redv(h) = \underset{d\to\io}{\lim}\, v[ \s (1 + h /d) ]$ & Scaled interaction potential \\ 
$f_\l(h) = -\redv'(h+\l)$ & Force, shifted by $\l$ \\
\hline
\end{tabular}
\end{center}

\subsubsection{Dynamic quantities}

\begin{center}
\begin{tabular}{| l | l |}
\hline
$x(t)$ & Time-dependent particle position \\
$\De x$ & Functional integration measure over $x(t)$ \\
$\g$ & Friction coefficient of the Langevin equation \\
$\wh\g = \frac{\s^2}{2d^2} \g$ & Scaled friction coefficient \\
$D$ & Diffusion coefficient \\
$\wh D = \frac{2 d^2}{\s^2} D $ & Scaled diffusion coefficient \\
\hline
\end{tabular}
\end{center}

\subsection{Replica coordinates}
\label{eqapp:MSD}

\begin{center}
\begin{tabular}{| l | l | l |}
\hline
$\bar x = \{ x_1, \cdots, x_n \}$ & Coordinates of a replicated atom & \\
$\hat M = \{ M_{ab} \}$ & $n\times n$ replica matrix & \\
$q_{ab} = x_a \cdot x_b$ & Matrix of scalar products, or overlaps & $q_{aa} = R^2$ \\
$\DE_{ab} = (x_a - x_b)^2 $ & Matrix of mean square displacements & $\DE_{aa} =0 $ \\
$Q_{ab} = 2d \, q_{ab} / \s^2 $ & Scaled overlaps  & $Q_{aa} = 2d R^2 /\s^2 = \Dl$ \\
$\D_{ab} = d \DE_{ab} / \s^2$ & Scaled mean square displacements & $\D_{aa} = 0$ \\
$\Dl = 2 d R^2 / \s^2$ & Scaled overlap of the liquid phase & \\
$v = \{ 1,\cdots,1\}$ & All-ones vector in replica space & $v_a = 1 \, , \, \forall a$ \\
\hline
\end{tabular}
\end{center}

\subsection{Gaussian integrals}
\label{app:Gauss}

\begin{center}
\begin{tabular}{| l | l |}
\hline
$\g_a(x) = \frac{e^{-\frac{x^2}{2a}}}{\sqrt{2\pi a}}$ & Gaussian kernel \\
$\DD_a\l = \de \l \frac{e^{-\frac{\l^2}{2a}}}{\sqrt{2\pi a}}$ & Gaussian integration measure \\
$\DD\l = \de \l \frac{e^{-\frac{\l^2}2}}{\sqrt{2\pi}}$ & Gaussian measure of unit variance \\
$\g_a \star f(x) = \int \de z \g_a(z) f(x-z)  = \int \DD_a z f(x-z)$ & Convolution product \\ 
$\Th(x) = \frac12 [1+\erf(x)] = \int_{-x}^\io \DD_{1/2} \l = \int_{-x}^\io  d\l \frac{e^{-\l^2}}{\sqrt{\pi}} = \g_{1/2} \star \th(x) $
& Smoothed theta function \\
$\Th\left(\frac{x}{\sqrt{a}}\right) = \int_{-x}^\io \DD_{a/2} \l = \g_{a/2} \star \th(x)$ &
Smoothed theta function of width $a$ \\
$ \DD_{\hat\D} \bar h  = \de \bar h \frac{1}{(2\pi)^{n/2} \sqrt{\det(\hat\D)}} e^{ - \frac{1}2 \bar h^T \hat\D^{-1} \bar h} $ &
Gaussian measure for replicated variables \\
$ \DD_{\bm Q} \bm h  \propto \De\bm h \, e^{  -\frac{1}2 \int \de a \de b \, \bm h(a)\bm Q^{-1}(a,b)\bm h(b)}$ &
Dynamical SUSY Gaussian measure \\
\hline
\end{tabular}
\end{center}

\subsection{Averages}
\label{app:ave}

\begin{center}
\begin{tabular}{| c | l |}
\hline
$\la \bullet \ra$ & Usually denotes the thermal average \\
$\overline{\bullet}$ & Average over random rotations\\
\hline
$\la \OO \ra_v =  \int \de\l \, e^{\l-\D_{\rm liq}/2} \, \int \DD_{\rm \hat Q} \bar h \, e^{-\b\sum_{a=1}^n \redv( h_a + \l ) } \, \OO$ &
Replica average over the scaled potential \\
\hline
$\HH_0(h,\l) =  \redv( h + \l ) + \frac{T h^2}{2\Dl}$  & RS effective Hamiltonian \\
$\ZZ_0(\l) = \int \de h \, e^{-\b\HH_0(h,\l)} $  & RS partition function \\
$\la \OO(h) \ra_{\HH_0} = \frac{1}{\ZZ_0} \int\de h\, e^{-\b \HH_0} \OO(h)$ & RS average \\
\hline
$\HH_1(h,\l) = \redv(h+ \l) + \frac{T h^2}{2\D_1} - \h h T \sqrt{ \frac1{\D_1} - \frac1{\Dl} }$ & 1RSB effective Hamiltonian \\
$\ZZ_1(\l) = \int \de h\, e^{-\b\HH_1(h,\l)} $ & 1RSB partition function \\
$\la \OO(h) \ra_{\HH_1} = \frac{1}{\ZZ_1} \int\de h\, e^{-\b \HH_1} \OO(h)
$ & Average over $\HH_1$ \\
\hline
$\la \OO \ra_v = \int \de \l \, e^{\l - \Dl/2} \int \De\bm h \,
e^{  -\frac{1}2 \int \de a \de b \,\bm h(a)\bm Q^{-1}(a,b)\bm h(b) - \int \de a \, \redv(\bm h(a) + \l )} \OO $ & Dynamical SUSY average \\
$\langle \bullet \rangle_h$ & Average over the effective dynamics in Eq.~\eqref{eq:HSeffh_app} \\
$\langle \bullet \rangle_{y_0}$ & Average over the effective dynamics in Eq.~\eqref{eq:MCnoDl} \\
& with fixed initial condition $y_0$ \\
\hline
\end{tabular}
\end{center}

\section{Equivalence between MK model and hard spheres without random shifts}
\label{app:MKHS}

We consider the original MK model~\cite{MK11} in $d$ dimensions. It is the $R\to\io$ version of the spherical model presented in the rest of the paper; 
introducing the hypersphere is an irrelevant complication 
for the purpose of this Appendix. The rotations $\RR_{ij}$ are thus replaced by $d$-dimensional shifts $A_{ij}$. Though these random rotations were picked with an infinite variance, here we go back to 
the original model with a variance $\l^2$ of the distribution of the shifts, which is taken to be Gaussian centered. 
We show that this model is described by the entropy functional in Eq.~\eqref{eq:repent} when
\begin{itemize}
 \item one fixes $\l\in\RRR^+$ and takes the limit $d\to\io$
 \item or one fixes $d$ and takes the limit $\l\to\io$
\end{itemize}
except for the additive constant due to discernability in the MK model. We only focus on the hard spheres potential; any short-ranged 
potential can be treated similarly, the conclusions with respect to the scalings are not changed. This implies, quoting~\cite{MK11}, that the limits $\l\to\io$ and $d\to\io$ are of the same nature.

When computing the replicated entropy, we introduced in Section~\ref{sub:repf} the functions
\beq
\begin{split}
 \bar\chi(\bar x,\bar y)&=\int\DD_{\l^2} A\,\prod_{a=1}^n\th(|x^a-y^a+A|-\s)=\int\DD_{\l^2} A\,\th(\underset{a}{\min}|x^a-y^a+A|-\s)=1+\bar f(\bar x,\bar y) \ , \\
 \bar f(\bar x,\bar y)&=\bar f(\bar u\equiv\bar x-\bar y)=-\int\DD_{\l^2} A\,\th(\s^2-\underset{a}{\min}|u^a+A|^2) \ ,
\end{split}
\eeq
where $\bar f$ is the replicated Mayer function. To show the equivalence, following the derivation in Section~\ref{sub:repf} one sees that we only need to prove that it vanishes in these limits. 
Indeed, if so, the leading order of the entropy functional is obtained by keeping the first term in the expansion of $\log(1+\bar f)\sim \bar f$, the other terms giving vanishingly small contributions.
\begin{enumerate}
 \item If $\l=0$, $\DD_{\l^2} A=\d(A)\de A$ and $\bar f=\bar f_{\rm HS}$ as in~\cite{KPZ12}. A computation similar to the one below gives the following conclusions in $d\to\io$:
\beq\label{eq:l0}
 \begin{split}
  \bar f(\bar u)&=-V_d(\s)\FF\left(\frac{\sqrt{d-n}}{\s}\bar u\right)\\
  \FF(\bar x)&=\int \frac{\de^n\e}{(2\p)^{n/2}}\,e^{-\frac12\underset{a}{\min}|x^a+\e|^2} 
   \end{split}
\eeq
Note that $1\leqslant\FF(\bar x)\leqslant\sum_{a=1}^n\int \frac{\de^n\e}{(2\p)^{n/2}}\,e^{-\frac12|x^a+\e|^2}=n$. The minimum 1 is obtained when all $x^a$ are equal and the integrand is a Gaussian centered at 
this position; as soon as they differ, several Gaussians peaked at the different positions contribute to the integral, giving additional contributions. Therefore the prefactor $V_d(\s)$ makes $\bar f$ tend exponentially to zero when $d\to\io$.

\item For $\l>0$, we look at the general case. However, we assume that the $n$ vectors in $\bar u$ are linearly independent. If they are not\footnote{In high dimension, 
the most likely configuration is that all the vectors in $\bar u$ are orthogonal to each other.}, it suffices to reduce the number of components of $A_\sslash$ introduced below to the rank 
of the vectors in $\bar u$ instead of $n$. Using the fact that the Gaussian measure is rotation invariant, we can build an orthonormal basis for $A=A_\sslash+A_\perp$ where $A_\sslash\in\mathrm{Span}(u^1,\dots,u^n)$ 
and $A_\perp$ lies in the orthogonal subspace. Then, with a change of variables, we get
\begin{equation}
\begin{split}
 \bar f(\bar u)&=-\int\DD_{\l^2}^n A_\sslash\DD_{\l^2}^{d-n}A_\perp\,\th(\s^2-\underset{a}{\min}|u^a+A_\sslash|^2-A_\perp^2)\\
 &=-\frac{\Omega_{d-n}}{2(2\p\l^2)^{\frac{d-n}{2}}}\int\DD_{\l^2}^n A_\sslash\,\th(\s^2-\underset{a}{\min}|u^a+A_\sslash|^2)\int_0^{\s^2-\underset{a}{\min}|u^a+A_\sslash|^2}\de r\, r^{\frac{d-n}{2}+1}e^{-r/2\l^2}\\
 &=-\frac{\Omega_{d-n}}{2(2\p\l^2)^{\frac{d-n}{2}}}\int\DD_{\l^2}^n A_\sslash\,\th(\s^2-\underset{a}{\min}|u^a+A_\sslash|^2)(2\l^2)^{\frac{d-n}{2}}\g\left(\frac{d-n}{2},\frac{\s^2-\underset{a}{\min}|u^a+A_\sslash|^2}{2\l^2}\right)\\
 \end{split}
\end{equation}
where the incomplete Gamma function is $\g(\a,z)=\int_0^z\de t\, t^{\a-1}e^{-t}$ defined for $\Re(\a)>0$. For $\Re(z)>0$, we have the relation $\g(\a,z)=\frac{1}{\a} z^\a e^{-z}F_1(1,\a+1,z)$, where $F_1$ is Kummer's function of the first kind which can be given by a hypergeometric series
\beq\label{eq:F_1} 
F_1(a,b,z)=1+\sum_{p=1}^\infty \frac{\prod_{j=0}^{p-1}(a+j)}{\prod_{j=0}^{p-1}(b+j)}z^p\,, \hskip15pt b\notin \mathbb{Z}^-
\eeq
whose radius of convergence is infinite. Then, 
\beq\label{eq:fu}
\bar f(\bar u)=-e^{-\frac{\s^2}{2\l^2}}V_{d-n}(\s/\l\sqrt{2\p})\int\DD_{\l^2}^n A_\sslash\,\GG_{d-n}\left(1-\underset{a}{\min}\abs{\frac{u^a+A_\sslash}{\s}}^2\right)e^{-\underset{a}{\min}|u^a+A_\sslash|^2/2\l^2}\left[1+O\left(\frac{1}{d(\l/\s)^2}\right)\right]
\eeq
where $\GG_\a(x)=x^{\a/2}\th(x)$. A very rough bound on the leading order of the integrand is 1, which gives
\begin{equation}
 |\bar f(\bar u)|\leqslant e^{-\frac{\s^2}{2\l^2}}V_{d-n}(\s/\l\sqrt{2\p})
\end{equation}
So $\bar f$ tends to zero when either $d$ goes to infinity at fixed $\l>0$ or conversely, when $\l$ goes to infinity at finite $d>n$.
\end{enumerate}

Let us make a comment on the values of the replicated Mayer function. In the case $\l=0$ and $d\to\io$, one can easily show~\cite{KPZ12} that
\begin{itemize}
 \item If all $u^a$ are zero, $\bar f(\bar 0)=-V_d(\s)$.
 \item If $\forall a\neq b$, $|u^a-u^b|\gg\s$, $\bar f(\bar u)=-nV_d(\s)$, independent of $\bar u$.
\end{itemize}
Using similar arguments, one can obtain these simplified expressions when $\l>0$ and $d\to\io$:
\begin{equation}
\begin{split}
\bar f(\bar 0)&=-e^{-\frac{\s^2}{2\l^2}}V_d(\s/\l\sqrt{2\p})\\
 \bar f(\bar u)&\sim-e^{-\frac{\s^2}{2\l^2}}V_{d}(\s/\l\sqrt{2\p})\sum_{a=1}^n e^{-\frac{(u^a)^2}{2\l^2}}\underset{\l\to\io}{\sim}-nV_{d}(\s/\l\sqrt{2\p})\,,\hskip15pt {\rm if,}~ \forall a\neq b,~ |u^a-u^b|\gg\s
\end{split}
\end{equation}
that can be compared to the $\l=0$, $d\to\io$ formulas given above. These results are of course compatible with the reasoning of Section~\ref{sub:repf} that for an infinite range of shifts, 
$\bar f=O\left(V_d(\s)/V\right)$ where $V\sim \l^d$ is the volume of the system.\\
From these observations, one concludes that the averaged Mayer function $\bar f(\bar u)$ has a trivial behaviour close to zero and as soon as the different replicas (respectively different trajectories for the dynamics) 
wander away from more than a particle diameter, its value is essentially a constant. Thus the critical regime where $\bar f$ has a non-trivial behaviour is when $\bar u=\bar x-\bar y$, the relative positions 
of two replicated configurations (respectively relative trajectories of two particles in the dynamics) is of the order the particle diameter (and fluctuations of order $1/d$)~\cite{KPZ12}. 
This critical scaling regime where replicas (respectively trajectories) are close reproduces, for large distances within this scaling, the behaviour of the regime where they are not constrained 
to remain close to contact, \ie the liquid phase. Indeed, the entropy of the liquid phase is recovered in the statics, see~Eq.~\eqref{eq:nsliq}, and respectively diffusive behaviour is recovered in the dynamics
at large times, see~\ref{sub:diff}. The fact that the Mayer function (equivalently $\FF(\bar x)$ of Eq.~\eqref{eq:l0} or $\FF(\hat \D)$ of~\eqref{eq:Iffinal}, which are related) becomes a constant 
for distances larger than the particle diameter elucidates these paradoxes.

\section{Integrals for rotationally invariant functions}
\label{app:A}

Here we prove Eq.~\eqref{eq:J}.
We consider a rotationally invariant function $f(x_1,\cdots, x_n)$ and, setting $q_{aa} = R^2$ and $\hat q$ symmetric\footnote{$q_{aa} = R^2$ due to the constraint $\d(x_a^2 - R^2)$ in the left hand side. We take $\hat q$ symmetric in order to integrate only on its independent variables $a<b$.}, we write
\beq
\int_{\RRR^{d+1}} \prod_{a=1}^n \left[ \de x_a  \d(x_a^2 - R^2) \right]\,
f(x_1,\cdots,x_n)
= \int_{\RRR^{d+1}} \prod_{a=1}^n \de x_a\, \int \prod_{a<b}^{1,n} \left[ \de q_{ab}\, \d(x_a \cdot x_b - q_{ab}) \right] f(\hat q)
\eeq
In Appendix~A of~\cite{KPUZ13} it is shown that
\beq
\int_{\RRR^{d+1}} \prod_{a=1}^n \de x_a\, \prod_{a<b}^{1,n} \d(x_a \cdot x_b - q_{ab})= 
2^{-n} \Omega_{d+1} \cdots \Omega_{d-n+2} [ \det \hat q]^{(d-n)/2} \ ,
\eeq
which proves the first equality in Eq.~\eqref{eq:J}:
\beq
\int_{\RRR^{d+1}} \prod_{a=1}^n \left[ \de x_a\,  \d(x_a^2 - R^2) \right]
f(x_1,\cdots,x_n)
= 2^{-n} \Omega_{d+1} \cdots \Omega_{d-n+2} \int \prod_{a<b}^{1,n} \de q_{ab}\,   [ \det \hat q]^{(d-n)/2}  f(\hat q) \ .
\eeq

To obtain the second equality we change variables from $q_{ab}$ to $\DE_{ab} = (x_a - x_b)^2 = 2 R^2 - 2 q_{ab}$,
or $\hat \DE = 2 ( R^2 v v^{\rm T} - \hat q)$. Then we have 
\beq
\hat q = -\frac{\hat\DE}2 ( I - 2 R^2 \hat\DE^{-1} v v^{\rm T} ) \ .
\eeq
and
\beq
\log \det \hat q = \log\det (-\hat\DE/2) + \Tr \log  ( I - 2 R^2 \hat\DE^{-1} v v^{\rm T} ) =\log\det (-\hat\DE/2) +  \log  ( 1- 2 R^2 v^{\rm T} \hat\DE^{-1} v  ) \ .
\eeq
Note that for $a=b$ we have $\DE_{aa}=0$, and we do not integrate over these variables.
For $a<b$, going from $q_{ab}$ to $\DE_{ab}$ is a simple linear change of variables, so we have
\beq
\int \prod_{a<b}^{1,n} \de q_{ab}  \, [ \det \hat q]^{(d-n)/2}  f(\hat q) =
(-2)^{-n(n-1)/2} \int \prod_{a<b}^{1,n} \de \DE_{ab} \,  e^{\frac{d-n}2 \left[\log\det (-\hat\DE/2) + \log  ( 1 - 2 R^2 v^{\rm T} \hat\DE^{-1} v  )
\right]  }  f(\hat \DE) \ ,
\eeq
which proves the second equality.

\section{Equivalence with previous computations of the ideal gas term}
\label{app:equivD}

We show here that the ideal gas term in Eq.~\eqref{eq:SSfinal} 
is equivalent to the results derived in~\cite{KPZ12,KPUZ13,CKPUZ13} for the replicated entropy,
e.g. \cite[Eq.(1)]{CKPUZ13}, for a general order parameter $\hat\D$ (mean-square displacements matrix).
There is a subtlety because in~\cite{KPZ12,KPUZ13,CKPUZ13} the calculation was restricted to a block of $m$ replicas, following \eg\cite{Mo95},
hence it was assumed that the matrix elements of $\hat\D$ are finite. In Eq.~\eqref{eq:SSfinal} we instead considered
the more general case where some matrix elements can be $\sim R^2$ (\eg in the liquid phase). The two methods are equivalent; this can be seen on the ideal gas term as an example we focus on here (the interaction term can be treated by a similar calculation than the one presented in Section~\ref{sub:1RSBent}).
As shown in Section~\ref{sec:V}, it can be written as 
\beq
\SS_{\rm IG}=\frac d2 n \log(\p e\s^2/d^2) +\frac d2 \log\det\hat Q
\eeq
where $\hat Q=\Dl vv^{\rm T}-\hat\D$ reads:
\begin{equation}
\nonumber
\hat Q=\left(
\begin{array}{ccc}
\left(
\begin{array}{ccc}
 &  &  \\
 & \hat Q_m &  \\
 &  &  \\
\end{array}
\right) & & \Dl-\D_0 \\
 &\ddots &\\
\Dl-\D_0 & &
\left(
\begin{array}{ccc}
&  &  \\
 & \hat Q_m &  \\
 &  &  \\
\end{array}
\right) 
\end{array}
\right) \ ,
\end{equation}
where the suffix $m$ denotes the restriction to the first block, $\hat Q_m=\Dl\hat v\hat v^{\rm T}-\hat\D_m$ can be 
any $m\times m$ matrix ($\hat v$ is the $m$-dimensional vector of all ones). As shown in Section~\ref{sec:hierarchical}, we have $\Dl-\D_0=\OO( \D_0^{3/2}e^{-\D_0/4} )$ which tends exponentially 
to zero in the large $R$ limit. As a consequence, the entropy of the $n$ replicas breaks into $n/m$ times the entropy of the $m\times m$ blocks:
\beq
\SS_{\rm IG}=\frac nm\SS_{\rm IG}^{(m)}=\frac nm \left[\frac{d}2 m \log( \pi e \s^2/d^2) + 
\frac{d}2  \log\det(-\hat \D_m) + \frac{d}2 \log\left(1- \frac{2 d R^2}{\s^2} \hat v^{\rm T} \hat \D^{-1}_m \hat v \right)\right]
\eeq
Therefore, we can go back to the setting of~\cite{KPZ12,KPUZ13,CKPUZ13}. We thus focus on $\SS_{\rm IG}^{(m)}$ and drop the hat on $\hat v$ and the suffix $m$ on matrices, restricting on 
a $m\times m$ block. In this setting, $\hat\D$ is finite (as an example, for a 1RSB glass phase, it is a replica symmetric matrix with parameter $\D_1$ which is finite, as shown in Section~\ref{sec:1RSB} and~\ref{sec:VC}). Then, 
for large $R$ we can approximate $\log\left(1 - \frac{2 d R^2}{\s^2} v^{\rm T} \hat \D^{-1} v \right) \sim \log\left(- \frac{2 d R^2}{\s^2} v^{\rm T} \hat \D^{-1} v \right)$
and the ideal gas term of Eq.~\eqref{eq:SSfinal} becomes, recalling Eq.~\eqref{eq:SIG_liquid},
\beq\begin{split}
\SS_{\rm IG}^{(m)} &= \frac{d}2 m \log(2 \pi e \s^2/d^2) + 
\frac{d}2  \log[ \det(-\hat \D/2) (- v^{\rm T} \hat \D^{-1} v)]
 -
\frac{d}2 \log\left( \frac{\s^2}{2 d}  \right) + d \log R \\
&=  \frac{d}2 (m-1) \log(2 \pi e \s^2/d^2) + 
\frac{d}2  \log[2 \det(-\hat \D/2) (- v^{\rm T} \hat \D^{-1} v)] + \log V \ ,
\end{split}
\eeq
which, recalling that to compare with hard spheres we should replace $\log V \to 1- \log\r$, coincides
with the results of~\cite{KPZ12,KPUZ13,CKPUZ13} (see e.g. \cite[Eqs.(1) and (2)]{CKPUZ13}) provided
\beq\label{eq:B2}
 \log[2 \det(-\hat \D/2) (- v^{\rm T} \hat \D^{-1} v)] = \log\det\hat\a^{m,m} + 2 \log m \ ,
\eeq
where $\a_{ab} = d (x_a \cdot x_b)/\s^2$ is a symmetric and Laplacian matrix ($\sum_b \a_{ab}=0$) that is related
to $\hat\D$ by the relation $\D_{ab} = \a_{aa} + \a_{bb} - 2 \a_{ab}$, and $\hat\a^{a,a}$ is the $(a,a)$-cofactor of $\hat\a$.
The task is then to prove Eq.~\eqref{eq:B2}.

We note first that for Laplacian matrices $\det\hat\a=0$ and $\det\hat\a^{a,a}$ is independent of $a$ (Kirchhoff's matrix tree theorem).
Then
\beq
\det(\e I + \hat \a) = \e \sum_a \det\hat\a^{a,a} + O(\e^2) = \e m \det\hat\a^{m,m} + O(\e^2) 
\ \ \ \ \ 
\Rightarrow
\ \ \ \ \ 
\det\hat\a^{m,m} = \underset{\e\to 0}{\lim}\, \frac{1}{\e m}
\det(\e I + \hat \a)
\ .
\eeq
Then we note that defining $\chi_{a} = \a_{aa}$, we have
\beq
\D_{ab} = \a_{aa} +\a_{bb} - 2\a_{ab} = \chi_a + \chi_b -2\a_{ab} 
\ \ \ \ \ 
\Rightarrow
\ \ \ \ \ 
\a_{ab} = \frac12 \left[ \chi_a+\chi_b -\D_{ab} \right] \ ,
\eeq
which is written in matrix notation as $\hat \a = \frac12 [ \chi v^{\rm T} + v \chi^T - \hat \D ]$.
To determine $\chi$ we impose the Laplacian condition, $\hat\a v =0$,
\beq
0 = 2 \hat \a v = m \chi  + v (\chi \cdot v) - \hat \D v 
\ \ \ \ \ 
\Rightarrow
\ \ \ \ \ 
\chi = \frac1m \left[ \hat \D v - v (\chi \cdot v) \right]
\ .
\eeq
Multiplying the last equation by $v^{\rm T}$ we get
\beq
2 m (\chi \cdot v)  - v^{\rm T} \hat \D v = 0
\hskip20pt
\Rightarrow
\hskip20pt
\chi \cdot v = \frac{v^{\rm T} \hat \D v}{2 m} 
\hskip20pt
\Rightarrow
\hskip20pt
\chi  = 
\frac1m \left[ \hat \D v - v \frac{v^{\rm T} \hat \D v}{2 m} \right]
\ .
\eeq
Finally, defining $u = v/\sqrt{m}$ which is normalized to $u^Tu =1$, we get
\beq\label{eq:aD}
\hat \a = \frac12 \left[ - (u^T \hat \D u) u u^T + \hat \D u u^T + u u^T \hat \D - \hat \D \right] \ .
\eeq
Therefore the matrices $\hat\a$ and $\hat\D$ differ by a projector on a vector space spanned
by $u$ and $\hat\D u$. The proof can be done in general, but let us focus here on the case
(of interest for us) in which $\a_{aa}$ is a constant independent on $a$, or equivalently
$\sum_b \D_{ab}$ does not depend on $a$. In this case $u$ is an eigenvector of $\hat\D$, and
$\hat \D u  = \l u $ with $\l = u^T \hat\D u$. Also, $\hat\D^{-1} u =  u/\l$ and therefore
$ u^T \hat\D^{-1} u = 1/\l$.
We have $\hat\a = ( \l u u^T -\hat \D)/2$ and
\beq
\det(\e I + \hat \a) = \det(\e I - \hat \D/2)
\det\left(
1 + \frac\l2 \frac1{\e - \hat\D/2} u u^T
\right)
= \det(\e I - \hat \D/2)
\left[
1 +  \frac12 \frac\l{\e - \l/2}
\right] =
\det(\e I - \hat \D/2) \frac{2\e}{2\e - \l}
\ ,
\eeq
where we used the relations $\frac1{\e - \hat\D/2} u = \frac{1}{\e - \l/2} u$ and
$\det(1 + A u u^T) = 1 + A$.
 Finally,
 \beq\begin{split}
 \det\hat\a^{m,m} &= \underset{\e\to 0}{\lim}\, \frac{1}{\e m} \det(\e I + \hat \a) = -\frac{2}{ m u^T \hat\D u  }\det(- \hat \D/2) 
 = -\frac{2}m ( u^T \hat\D^{-1} u )\det(- \hat \D/2) \\
 &= -\frac{2}{m^2} ( v^{\rm T} \hat\D^{-1} v )\det(- \hat \D/2) \ ,
\end{split} \eeq
which completes the proof of Eq.~\eqref{eq:B2} and therefore of the equivalence of our results with those
of Refs.~\cite{KPZ12,KPUZ13,CKPUZ13}.

\section{Algebra of hierarchical matrices and of SUSY operators}
\label{app:hierarchical}

Here we discuss some general properties of hierachical replica matrices and of SUSY dynamical operators.
We restrict to matrices $\D_{ab}$ such that $\D_{aa}=0$.
We often use a vector $v$ with all components equal to 1.
We also define a $n\times n$ matrix $\hat I^m$ which has elements $I^m_{ab}=1$ in blocks of size $m$ around the diagonal, and $I^m_{ab}=0$ otherwise.
Note that $\hat I^1 = \hat I$ is the identity matrix with $I_{ab} = \d_{ab}$, and $\hat I^n = v v^{\rm T}$ is the matrix of all ones.
Assuming that $m_1$ is a multiple of $m_2$ (hence $m_1 \geqslant m_2$), we have
\beq\label{eq:IIeqI}
\hat I^{m_1} \hat I^{m_2} = m_2 \hat I^{m_1} \ .
\eeq
This relation holds in particular for $m_1 = n$ or for $m_2 = 1$.

\subsection{RS matrices}
\label{app:RSformula}

For a replica symmetric matrix we have
\begin{eqnarray}
\D_{ab} &=& \D_0 (1-\d_{ab}) \hskip70pt \hat \D = \D_0 (\hat I^n - \hat I^1) \ , \label{eq:RSdelta} \\
\D^{-1}_{ab} &=& \frac{1}{\D_0} \left( \frac{1}{n-1} - \d_{ab}   \right) \hskip30pt
\hat\D^{-1} = \frac{1}{\D_0} \left( \frac{1}{n-1} \hat I^n - \hat I^1   \right) \ , \label{eq:RSdeltam1} \\
\sum_{b} \D^{-1}_{ab} &=&  \frac{1}{\D_0} \frac{1}{n-1} \ , \hskip70pt
\hat \D^{-1} v = \frac{1}{n-1} \frac{1}{\D_0} v \ . \label{eq:RSdeltasum}
\end{eqnarray}
The eigenvectors of $\hat\D$ are $v$, with eigenvalue $\D_0 (n-1)$, 
and $n-1$ orthogonal vectors with eigenvalue $-\D_0$;
hence, 
\beq\label{eq:RSdeltadet}
\det \hat \D = (n-1) (\D_0)^n (-1)^{n-1} \ .
\eeq

\subsection{1RSB matrices}
\label{app:1RSBformula}

A 1RSB matrix has the form
\beq\label{eq:1RSBdelta} 
\hat \D = \D_0 \hat I^n + (\D_1 - \D_0) \hat I^m - \D_1 \hat I^1 \  ,
\eeq
and using Eq.~\eqref{eq:IIeqI} one obtains
that the inverse is
\beq\label{eq:1RSBdeltam1} 
\begin{split}
\hat \D^{-1} &= \D^{-1}_0 \hat I^n + (\D^{-1}_1 - \D^{-1}_0) \hat I^m + (\D^{-1}_{\rm d} - \D^{-1}_1) \hat I^1 \ , \\
\D^{-1}_0 &= - \frac{\D_0}{[\D_1 (m-1) - m \D_0][\D_0 (n-m) + \D_1 (m-1)]} \ , \\
\D^{-1}_1 &= \D^{-1}_0   + \frac{\D_0 - \D_1}{\D_1 [\D_1 + m(\D_0 - \D_1)]} \ , \\
\D^{-1}_{\rm d} &= \D^{-1}_1 - \frac{1}{\D_1} \ ,
\end{split}\eeq
and
\beq\label{eq:1RSBdeltasum} 
\hat \D^{-1} v = \frac{1}{\D_0 (n-m) + \D_1 (m-1)} v \ .
\eeq
Finally, the determinant can be computed in the following way. 
The $n$-dimensional vector space can be decomposed in three subspaces:
\begin{enumerate}
\item The vector $v$ of all ones. It has $\hat I^n v = n v$ and $\hat I^m v = m v$. Hence
\beq
\hat \D v = [\D_0 (n-m) + \D_1 (m-1)] v \ .
\eeq
\item A set of $n/m-1$ independent vectors $w$, such that $w_a$ is constant in each block, and $\sum_a w_a = 0$. These are orthogonal to $v$ and
such that $I^n w = 0$ and $I^m w = m w$. Hence
\beq
\hat \D w = [-m \D_0 + \D_1 (m-1)] v \ .
\eeq
\item A set of $(n/m)(m-1)$ vectors $x$ such that $\sum_{a \in B} x_a=0$ in each block $B$. These are orthogonal to $v$ and all the $w$, and they are such
that $I^n x = I^m x =0$. Hence
\beq
\hat \D x = -\D_1 x \ .
\eeq
\end{enumerate}
Therefore we obtain
\beq\label{eq:1RSBdeltadet} 
\begin{split}
\det\hat\D & = [\D_0 (n-m) + \D_1 (m-1)] \times [-m \D_0 + \D_1 (m-1)]^{n/m-1} \times [ -\D_1]^{n/m(m-1)}  \\
& = \frac{\D_0 (m-n) + \D_1 (1-m)}{m \D_0 + \D_1 (1-m)} \times \left[ \frac{m \D_0 + \D_1 (1-m)}{\D_1} \right]^{n/m} \times [ -\D_1]^n
\ .
\end{split}
\eeq
Note that we recover the RS result for $m=1$, as it should be.

Finally, with a similar procedure, we obtain
\beq\label{eq:1RSBinv2}
[ (\D_1 - \D_0) \hat I^m - \D_1 \hat I^1 ]^{-1} = \frac{\D_1 - \D_0}{\D_1 [- m\D_0 + \D_1 (m-1)]} \hat I^m - \frac1{\D_1} \hat I^1 \ ,
\eeq
and
\beq\label{eq:1RSBdeltadet2}
\begin{split}
\det[  (\D_0 - \D_1) \hat I^m + \D_1 \hat I^1 ]  & =  \left[ \frac{m \D_0 + \D_1 (1-m)}{\D_1} \right]^{n/m} \times [ \D_1]^n
\ ,
\end{split}
\eeq
which can be derived in the same way as Eq.~\eqref{eq:1RSBdeltadet}, but taking into account that for the matrix
$(\D_0 - \D_1) \hat I^m + \D_1 \hat I^1$ the eigenvalues associated to the vectors $v$ and $w$ coincide.

\subsection{FullRSB matrices}
\label{app:fRSBformula}

For fullRSB matrices we restrict ourselves to the limit $n\to 0$.
In the limit $n\to 0$,
a hierarchical matrix $\hat \D$ is parametrized by its diagonal element $\D_{\rm d}$
and by a continuous function $\D(x)$ for $0<x<1$.
The algebra of these matrices is 
described in compact form in~\cite{MP91}, see also~\cite{CKPUZ13}. 
In the case of interest here, $\D_{\rm d}=(x_a - x_a)^2 = 0$.
Also, replicas in the outermost block are described by $\D(0)$ which plays a special role.

We follow the notation of~\cite[Appendix II]{MP91} and introduce $\la \Delta \ra = \int_0^1 \de x\, \Delta(x)$ and
$[\Delta](x) = x \Delta(x) - \int_0^x \de y \,\Delta(y)$.
Then, for the special case $\Delta_{\rm d}=0$ which is of interest here,
the determinant is given in~\cite[Eq.~(AII.11)]{MP91}:
\beq\label{eq:fRSBdeltadet} 
\underset{n\to 0}{\lim}\, \frac1n \log\det(\hat\Delta) = \log\left( -\la \Delta \ra \right) - \frac{\Delta(0)}{\la \Delta \ra} - \int_0^1 \frac{\de x}{x^2}\, \log\left(
1 + \frac{  [\Delta](x)}{\la \Delta \ra}
\right)
\eeq
The formula for the inverse, given in~\cite[Eq.~(AII.7)]{MP91}, and specialized to $\Delta_{\rm d}=0$, gives
\beq\label{eq:fRSBdeltam1} 
\begin{split}
\D^{-1}_{\rm d} &= -\frac{1}{\la \D \ra} \left[ 1 + \int_0^1 \frac{\de x}{x^2}\, \frac{[\D](x)}{\la \D \ra + [\D](x)} + \frac{\D(0)}{\la \D \ra} \right] \ , \\
\D^{-1}(x) &= \frac{1}{\la \D \ra} \left[ -\frac{[\D](x)}{x(\la \D \ra + [\D](x))} - \int_0^x \frac{\de y}{y^2}\, \frac{[\D](y)}{\la \D \ra + [\D](y)} - \frac{\D(0)}{\la \D \ra} \right] \ .
\end{split}\eeq
Using these equations one can easily show that
\beq\label{eq:fRSBdeltasum} 
\sum_b \D^{-1}_{ab} = \D^{-1}_{\rm d} - \int_0^1 \de x \,\D^{-1}(x)  = - \frac{1}{\la \D \ra}  \ , \hskip30pt
\hat \D^{-1} v = \frac{1}{\la \D \ra} v \ .
\eeq

\subsection{Product of SUSY fields}

If two superfields $B(a,b)$ and $C(a,b)$ are both cast as an equilibrium form \eqref{eq:SUSYeq}, and
\beq
\bm A(a,b) = \int \de c\,\bm B(a,c)\bm C(c,b)
\eeq
then also $\bm A(a,b)$ has the same form and for $t_a > t_b$ one has
\beq\label{SUSYproduct}
A_C(t_a-t_b) = \b B_C(0) C_C(t_a - t_b) - \b \int_{t_b}^{t_a} \de t_c\, B_C(t_a-t_c) \dot C_C(t_c -t_b) 
\eeq

\section{Dynamics from equilibrium initial condition}
\label{app:B}

Here we give a simple argument to neglect the past history when one starts from equilibrium in a Langevin equation with memory. This discussion applies to exponentially 
decaying memory kernels only. In this special case, only by adding one additional degree of freedom, one can consider an explicit Markovian evolution of the two-body system. For more general 
memory kernels one has to resort to a coupling with a bath containing many degrees of freedom. The corresponding discussion can be found in~\cite{MKZ15}.

Consider the following Langevin equation:
\beq\begin{split}
\g \dot x & = -\frac{\de U}{\de x} + \xi(t) - x(t) + \eta_1(t)   \ , \\
\g \dot \xi &= - \xi(t) + x(t) + \eta_2(t)  \ ,  \\
\la \eta_1(t) \eta_1(t') \ra &= \la \eta_2(t) \eta_2(t') \ra = 2 T \g \d(t-t') \ .
\end{split}\eeq
This equation is Markovian and it admits a Boltzmann stationary distribution
\beq
P_{\rm eq}(x,\xi) = \frac1Z e^{-\b \left[U(x) + \frac12 (x-\xi)^2 \right] } \ ,
\eeq
moreover the marginal distribution of $x$ is the Boltzmann one with potential $U(x)$:
\beq
P_{\rm eq, x}(x) = \int \de \xi\, P_{\rm eq}(x,\xi) = \frac1Z e^{-\b U(x)} \ .
\eeq
We show in the following that these two correlation functions are identical:
\begin{enumerate}
\item Starting with any initial condition at $t=t_0 \to -\io$, we compute $C(t-t') = \la x(t) x(t') \ra$ for $t,t' > 0$.
\item Starting at $t=t_0=0$ with an initial condition $x_0, \xi_0$ drawn from $P_{\rm eq}(x_0,\xi_0)$, 
we compute $C(t-t') = \la x(t) x(t') \ra$ for $t,t' > 0$.
\end{enumerate}

To proceed, we write the effective Langevin equation for $x$ integrating out $\xi$. We have
\beq
\xi(t) = \xi_0 e^{-(t-t_0)/\g} + \frac{1}\g \int_{t_0}^t \de s \, e^{-(t-s)/\g} [x(s) + \h_2(s)] \ .
\eeq
Substituting in the equation for $x$ we obtain
\beq\begin{split}
\g \dot x & = -\frac{\de U}{\de x} + \xi_0 e^{-(t-t_0)/\g} + \frac{1}\g \int_{t_0}^t \de s \, e^{-(t-s)/\g} [x(s) + \h_2(s)]
- x(t) + \eta_1(t) \\
& = -\frac{\de U}{\de x} -  \int_{t_0}^t \de s \, e^{-(t-s)/\g} \dot x(s) 
+ (\xi_0 - x_0) e^{-(t-t_0)/\g} + \frac{1}\g \int_{t_0}^t \de s \, e^{-(t-s)/\g} \h_2(s)
+ \eta_1(t) 
\end{split}\eeq
We define $\r(t,x_0) = (\xi_0 - x_0) e^{-(t-t_0)/\g} + \frac{1}\g \int_{t_0}^t \de s \, e^{-(t-s)/\g} \h(s)
+ \eta_1(t)$ and we note that for fixed $x_0$ it is a random Gaussian variable (because it is a linear combination of Gaussian variables), 
which depends on $\eta_1(t)$, $\eta_2(t)$ and $\xi_0$. We have
\beq\begin{split}
\g \dot x 
& = -\frac{\de U}{\de x} -  \int_{t_0}^t \de s \, e^{-(t-s)/\g} \dot x(s) 
+ \r(t,x_0) \ , \\ 
\la \r(t,x_0) \r(t',x_0) \ra &= 2 T \g \d(t-t') + T e^{-(t-t')/\g} + [ \langle (\xi_0 -x_0)^2 \rangle -T ]
e^{-(t-t_0)/\g}e^{-(t'-t_0)/\g}
\end{split}\eeq
where here $\langle \bullet \rangle$ is an average over $\eta_1(t)$, $\eta_2(t)$ and $\xi_0$ at fixed $x_0$.

Now, in the two cases outlined above, we obtain:
\begin{enumerate}
\item In case (1), the dependence on the inital condition is lost when $t_0\to -\io$. Therefore we obtain
\beq\label{appB:case1}
\begin{split}
\g \dot x 
& = -\frac{\de U}{\de x} - \int_{-\io}^t \de s \, e^{-(t-s)/\g} \dot x(s) 
+ \r(t) \ , \\ 
\la \r(t) \r(t') \ra &= 2 T \g \d(t-t') + T e^{-(t-t')/\g} \ . 
\end{split}\eeq
\item In case (2), we have $\langle (\xi_0 -x_0)^2 \rangle = T$ due to the form\footnote{
Note that this result is completely
independent on the form of the distribution of $x_0$. It is enough that $P_{\rm init}(x_0, \xi_0) \propto
 p(x_0) e^{- \frac\b2 (x_0-\xi_0)^2  }$.
However, we need to assume that $\xi_0$ is in equilibrium, and
therefore also $x_0$ must be in equilibrium. 
} of $P_{\rm eq}(x,\xi)$. 
Then the dependence on $x_0$ in $\r$ again disappears, and we obtain
\beq\label{appB:case2}
\begin{split}
\g \dot x 
& = -\frac{\de U}{\de x} -  \int_{0}^t \de s \, e^{-(t-s)/\g} \dot x(s) 
+ \r(t) \ , \\ 
\la \r(t) \r(t') \ra &= 2 T \g \d(t-t') + T e^{-(t-t')/\g} 
\end{split}\eeq

\end{enumerate}
We conclude therefore that Eqs.~\eqref{appB:case1} and \eqref{appB:case2} give rise to the same correlation
$C(t-t') = \la x(t) x(t') \ra$ at positive times.

Note that this is a particular instance of a memory kernel $M_C(t) = T e^{-t/\g}$, with corresponding
reponse kernel $M_R(t) = -\b \th(t) \dot M_C(t) = \th(t) e^{-t/\g}/\g$.
The corresponding equation is
\beq\label{appB:final}
\begin{split}
\g \dot x 
& = -\frac{\de U}{\de x} -  \b \int_{t_0}^t \de s \, M_C(t-s) \dot x(s) 
+ \r(t) \ , \\ 
\la \r(t) \r(t') \ra &= 2 T \g \d(t-t') + M_C(t-t') \ . 
\end{split}
\eeq
and this argument shows that starting with any initial condition at $t_0=-\io$ is equivalent to starting in equilibrium at $t_0=0$
for the purpose of computing correlations at positive times.


\end{document}